\documentclass[12pt,preprint]{aastex}
\newcommand{\spitzer}{\emph{Spitzer}}

\newcommand{\MIPS}{\emph{MIPS}}
\newcommand{\IRAC}{\emph{IRAC}}

\shortauthors{}

\begin{document}
\title{Current Star Formation in the Ophiuchus and Perseus Molecular
  Clouds: Constraints and Comparisons from Unbiased Submillimeter and Mid-Infrared Surveys. II.}

\author{Jes K. J{\o}rgensen\altaffilmark{1,2,*}, Doug Johnstone\altaffilmark{3,4}, Helen Kirk\altaffilmark{4,3}, Philip
C. Myers\altaffilmark{1}, Lori E. Allen\altaffilmark{1}, \& Yancy
L. Shirley\altaffilmark{5}}
\altaffiltext{1}{Harvard-Smithsonian Center for Astrophysics, 60 Garden Street MS42, Cambridge, MA 02138, USA}
\altaffiltext{2}{Argelander-Institut f\"{u}r Astronomie, University of Bonn, Auf
  dem H{\"u}gel 71, 53121, Bonn, Germany}
\altaffiltext{3}{Herzberg Institute of Astrophysics, National Research Council of Canada, 5071 West Saanich Road, Victoria, BC V9E 2E7, Canada}
\altaffiltext{4}{Department of Physics \& Astronomy, University of Victoria, Victoria, BC, V8P 1A1, Canada}
\altaffiltext{5}{Bart J. Bok Fellow, Steward Observatory, University of Arizona, 933 N. Cherry
  Ave., Tucson, AZ 85721}
\altaffiltext{*}{Current address: Argelander-Institut f\"{u}r Astronomie. E-mail: {\tt jes@astro.uni-bonn.de}}

\begin{abstract}
  We present a census of the population of deeply embedded young stellar
  objects (YSOs) in the Ophiuchus molecular cloud complex based on a
  combination of Spitzer Space Telescope mid-infrared data from the ``Cores to
  Disks'' (c2d) legacy team and JCMT/SCUBA submillimeter maps from the
  COMPLETE team. We have applied a method developed for identifying embedded
  protostars in Perseus to these datasets and in this way construct a
  relatively unbiased sample of 27 candidate embedded protostars with
  envelopes more massive than our sensitivity limit (about 0.1 $M_\odot$). As
  in Perseus, the mid-infrared sources are located close to the center of the
  SCUBA cores and the narrowness of the spatial distribution of mid-infrared
  sources around the peaks of the SCUBA cores suggests that no significant
  dispersion of the newly formed YSOs has occurred. Embedded YSOs are found in
  35\% of the SCUBA cores - less than in Perseus (58\%). On the other hand the
  mid-infrared sources in Ophiuchus have less red mid-infrared colors,
  possibly indicating that they are less embedded. We apply a nearest
  neighbour surface density algorithm to define the substructure in each of
  the clouds and calculate characteristic numbers for each subregion -
  including masses, star formation efficiencies, fraction of embedded sources
  etc. Generally the main clusters in Ophiuchus and Perseus (L1688, NGC1333
  and IC~348) are found to have higher star formation efficiencies than small
  groups such as B1, L1455 and L1448, which on the other hand are completely
  dominated by deeply embedded protostars. We discuss possible explanations
  for the differences between the regions in Perseus and Ophiuchus, such as
  different evolutionary timescales for the YSOs or differences, e.g., in the
  accretion in the two clouds.
\end{abstract}


\section{Introduction}
Although the most detailed theoretical models of star formation continue to
concentrate on the formation of an isolated star within a relatively pristine
environment, these solutions are far removed from the conditions under which a
vast majority of stars begin life.  Stars form in environments which have a
wide range of physical conditions, including large scale clusters containing
high-mass stars, small groups of predominantly low-mass stars, and even the
occasional isolated single star or binary.  Relating the distribution of the
pre- and protostellar cores to their environment on large scales is an
important step toward understanding the global properties of stars such as the
stellar initial mass function and the evolution process through the different
stages of young stellar objects. It is therefore important to perform unbiased
censuses of objects throughout their evolutionary stages. To date, the hardest
sample to quantify has been the most deeply embedded stages where signatures
of the object are observed in the mid- to far-infrared and submillimeter
wavelengths.

In a previous paper \citep{scubaspitz} we performed a detailed comparison
between the \textit{SCUBA} submillimeter dust continuum and \textit{Spitzer
Space Telescope} mid-infrared observations of sources in the Perseus molecular
cloud in order to identify and examine the properties of starless cores and
deeply embedded protostars.  In this paper we extend the SCUBA and Spitzer
analysis to Ophiuchus.  We then use this analysis for a direct comparison
between the Ophiuchus and Perseus clouds, paying particular attention to the
global star formation and clustering properties of the clouds.

Previous studies of the deeply embedded stages have been hampered by low
sensitivity single element (sub)millimeter receivers and confusion in low
resolution, low sensitivity infrared observations such as those from the IRAS
and ISO satellites. In the last few years systematic, detailed surveys of
larger regions have been made possible using wide area imaging at high
resolution and sensitivity using submillimeter telescopes such as the JCMT
with SCUBA and mid-infrared telescopes including Spitzer and its Infrared
Array Camera (\IRAC) and Multiband Imaging Photometer (\MIPS). The ``Cores to
Disks (c2d)'' Spitzer legacy team \citep{evans03} mapped five of the nearby
star forming clouds, including Perseus \citep{perspitz,rebull07} and Ophiuchus
\citep{padgett08}, using the Spitzer cameras. At submillimeter wavelengths the
``COMPLETE'' team \citep{goodman04,ridge06} created maps of Perseus
\citep{kirk06} and Ophiuchus \citep{johnstone04} at submillimeter
wavelengths. Both these surveys compiled systematic samples of sources in an
unbiased sense and together they provide excellent constraints on the
properties of the starless cores and deeply embedded protostars in these cloud
complexes.

\cite{scubaspitz} discussed how to associate mid-infrared Spitzer sources with
submillimeter SCUBA cores: a characteristic population of mid-infrared sources
were found closely associated with the centers of the SCUBA cores. These
mid-infrared sources were furthermore found to have particularly red colors
while the associated SCUBA cores were found to have high ``concentrations'',
i.e., appear centrally peaked. Determining whether this result is a
discriminant of the earliest stage of star formation in general, or whether it
is region specific, requires exploring additional clouds.

The construction of a relatively unbiased samples of embedded protostars,
furthermore makes it possible to address some of the important questions
concerning low-mass star formation and the relation between protostars and
their environment. Based on the analysis of the Perseus data,
\citeauthor{perspitz} for example found that the Spitzer sources were strongly
centered around the peaks of the submillimeter cores, suggesting that little
dispersal of the newly formed YSOs relative to their parental cores occurs
during the protostellar stages. \citeauthor{perspitz} also found similar
numbers of SCUBA cores with and without associated MIPS sources suggesting
that the timescale for the evolution through the dense pre-stellar stages
(where the cores are recognized in the SCUBA maps) is similar to the timescale
for the evolution through the embedded protostellar stages. It is interesting
to utilize a similar method to explore the similarities and differences
between clouds and shed light onto the recent star formation history in
different cloud environments.

The Ophiuchus molecular cloud complex\footnote{Following the other c2d studies
of Ophiuchus \citep{young06,padgett08} a distance of 125$\pm$25~pc \citep[][see
also discussion in \citealt{ophhandbook}]{degeus89} is adopted throughout this
paper.} provides an excellent comparison to Perseus: Ophiuchus is similar to
Perseus in the sense that it contains both embedded protostars and a similar
number of SCUBA cores. In Perseus the population of young stellar objects can be
divided into three main groups: the IC~348 cluster region mainly containing more
evolved YSOs, the NGC~1333 region which is a more currently active cluster and
the remainder, or extended cloud, which contains a handful of smaller groups,
each with $\sim$~10 members and with a relatively high fraction of deeply
embedded low-mass protostars compared to more evolved YSOs. The Ophiuchus cloud
in contrast is dominated by the L1688 cluster with some, but significantly less,
star formation occurring outside this main region. Comparing the embedded
populations of YSOs, the distribution of cores and dust extinction between these
two clouds, and their subgroups, makes it possible to test scenarios for star
formation, and their robustness across differing environments.

This paper follows the approach of \cite{scubaspitz} and compares SCUBA
submillimeter and Spitzer mid-infrared observations of the Ophiuchus molecular
cloud. The paper is laid out as follows: \S\ref{submmandmidir} presents an
overview of the submillimeter and mid-infrared data which forms the basis for
this analysis. \S\ref{method} applies the method from \cite{scubaspitz} to the
Ophiuchus dataset. \S\ref{comparison} presents an analysis of the substructure
of Ophiuchus and Perseus using a nearest neighbor surface density algorithm
and compares the key numbers for the two clouds and the subregions within
them. \S\ref{discuss} discusses the implications for star formation scenarios
in the two clouds.

\section{Observations: Large-scale maps of Ophiuchus}\label{submmandmidir}
As in \cite{scubaspitz} the bases for this analysis are large scale maps of
Ophiuchus at mid-infrared wavelengths with the \textit{Spitzer Space
Telescope} and at submillimeter wavelengths with the \textit{SCUBA} bolometer
array on the \textit{JCMT}. In this section we present a brief discussion of
the data, in particular highlighting differences with respect to the data
presented in \cite{scubaspitz}.

\subsection{Submillimeter observations from JCMT/SCUBA}
The \emph{Submillimetre Common-User Bolometer Array (SCUBA)} on the
\emph{James Clerk Maxwell Telescope (JCMT)} mapped the cloud in 850~$\mu$m
dust continuum emission at 15$''$ resolution as previously presented by
\cite{johnstone04}. For the present analysis these data were re-reduced and
combined with all other SCUBA archive data for Ophiuchus following the
description in \cite{kirk06} (see also \citealt{difrancesco08}). As with the
Perseus data presented in \cite{kirk06}, the standard SCUBA pipeline reduction
was augmented by a matrix inversion image recreation procedure
\citep{johnstone00maps}. The resulting 3$''$ pixel images were smoothed on
small scales by a Gaussian with dispersion of 3$''$ to remove pixel-to-pixel
noise. Furthermore the maps were flattened by removing a large-scale Gaussian
with dispersion of 90$''$. The resulting RMS of the map was about
0.03~Jy~beam$^{-1}$. From these maps, lists of cores were extracted using the
2D version of the Clumpfind algorithm \citep{williams94} as described by
\cite{kirk06} using a contouring level with a spacing, $\sigma$ =
0.03~Jy~beam$^{-1}$ (0.04~Jy~beam$^{-1}$ in the Western part of the map around
the Oph A core and around L1709). Only peaks above 5$\sigma$ are reported. In
this fashion 66 SCUBA cores were identified across the Ophiuchus field. We
summarize the properties of cores in the appendix to this paper (see also the
comparison to previous studies in Sect.~\ref{submmcomparison}).

\subsection{Mid-infrared observations from \spitzer}
\begin{figure}
\centering
\resizebox{0.65\hsize}{!}{\includegraphics{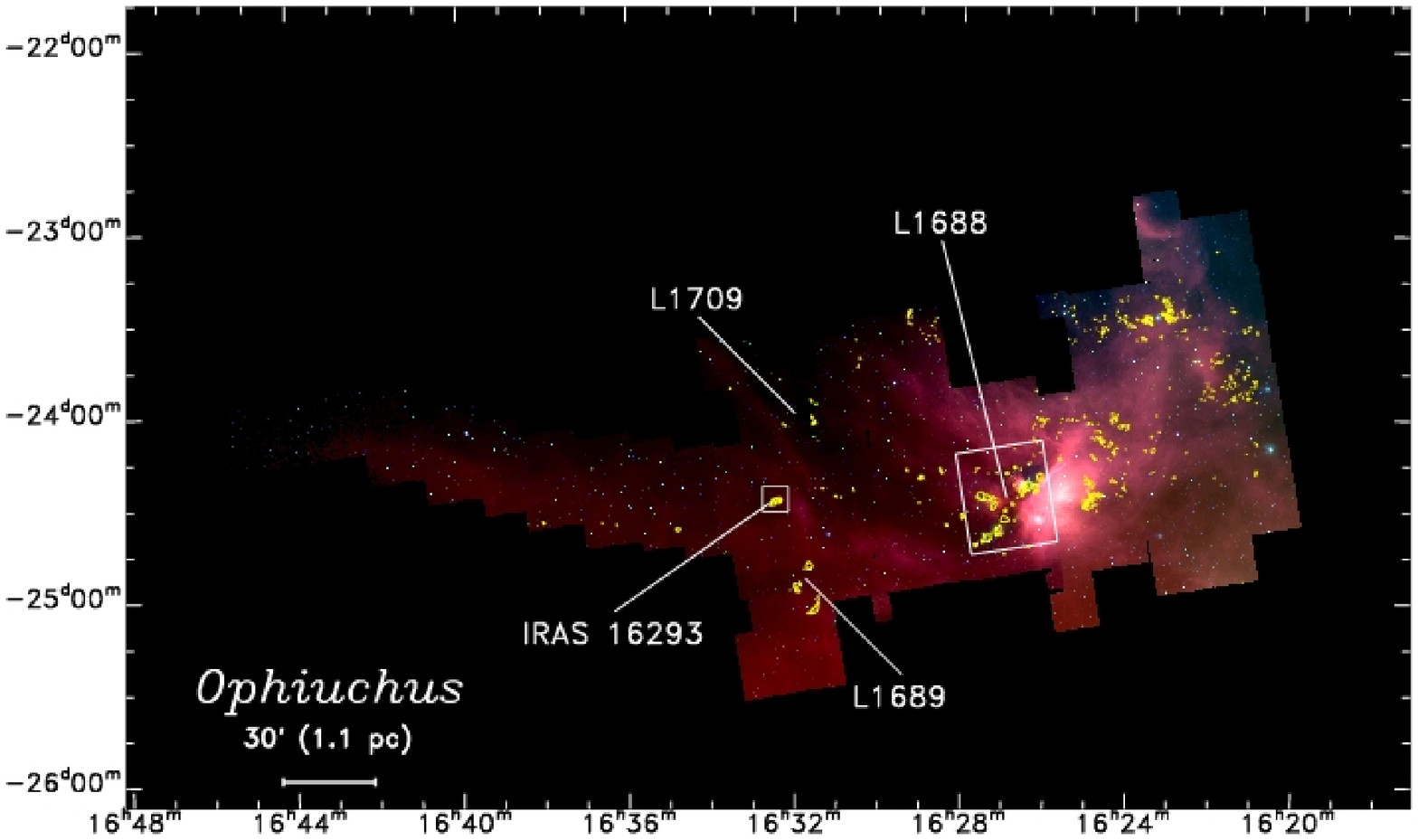}}
\resizebox{0.65\hsize}{!}{\includegraphics{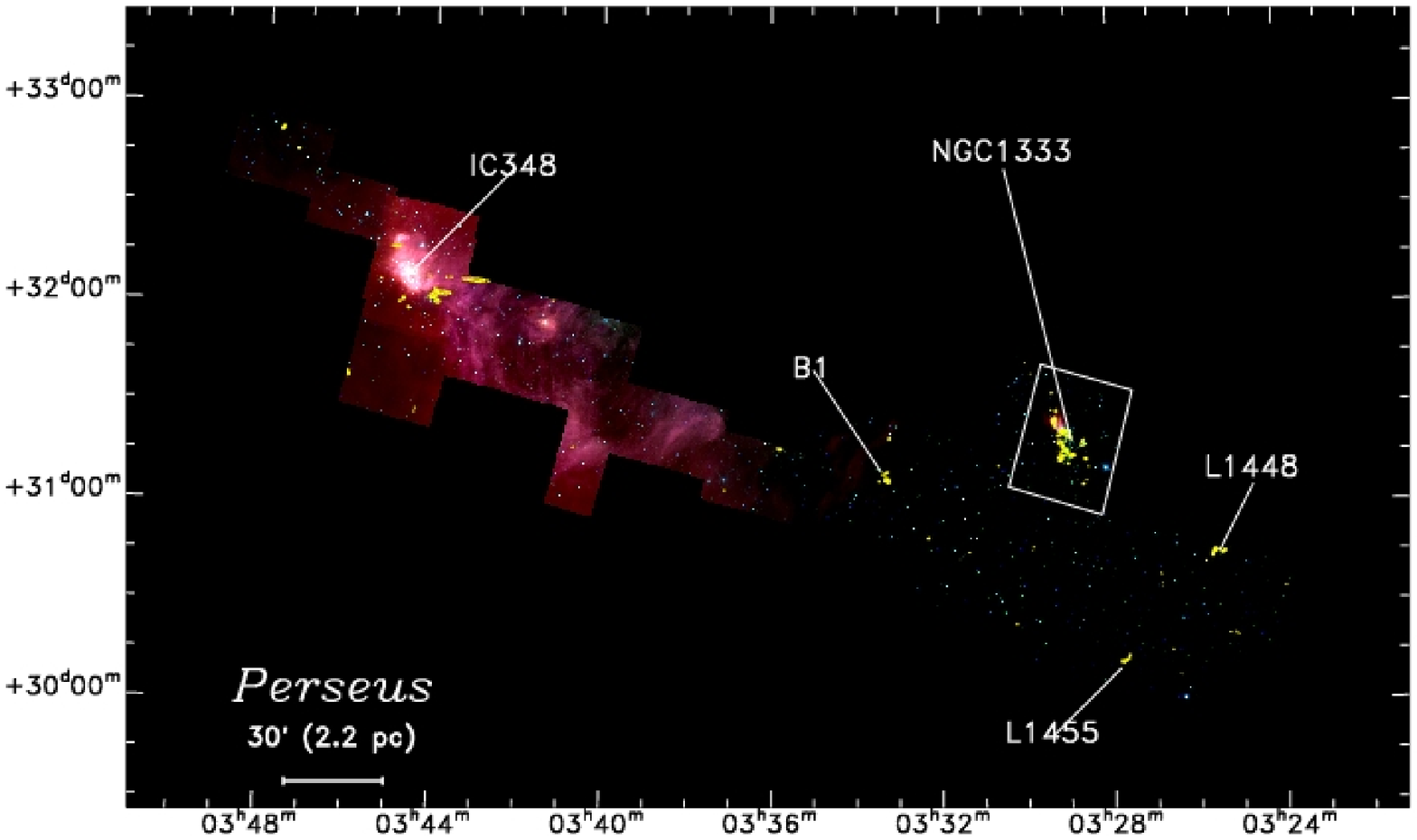}}
\resizebox{0.33\hsize}{!}{\includegraphics{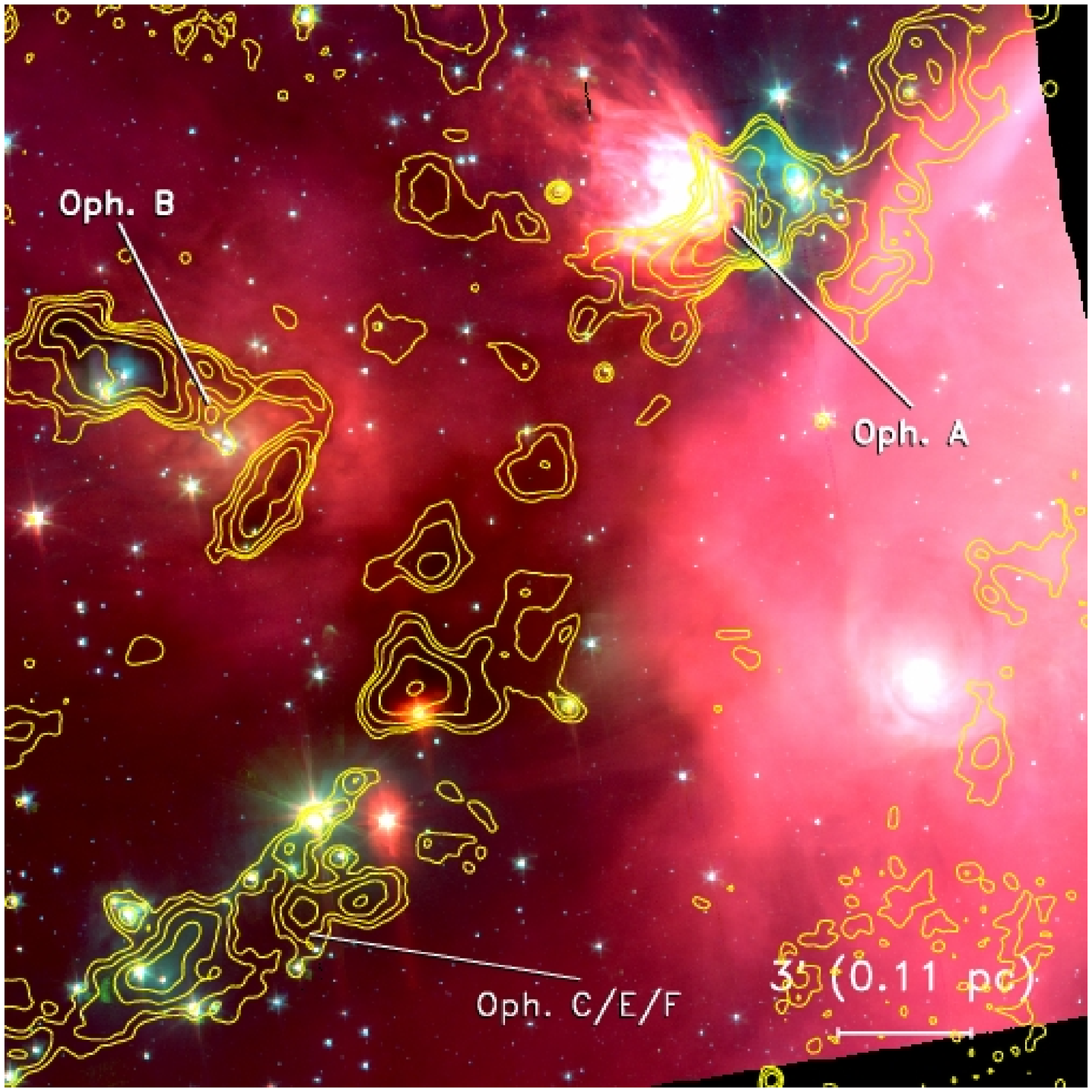}}\mbox{
  }\resizebox{0.33\hsize}{!}{\includegraphics{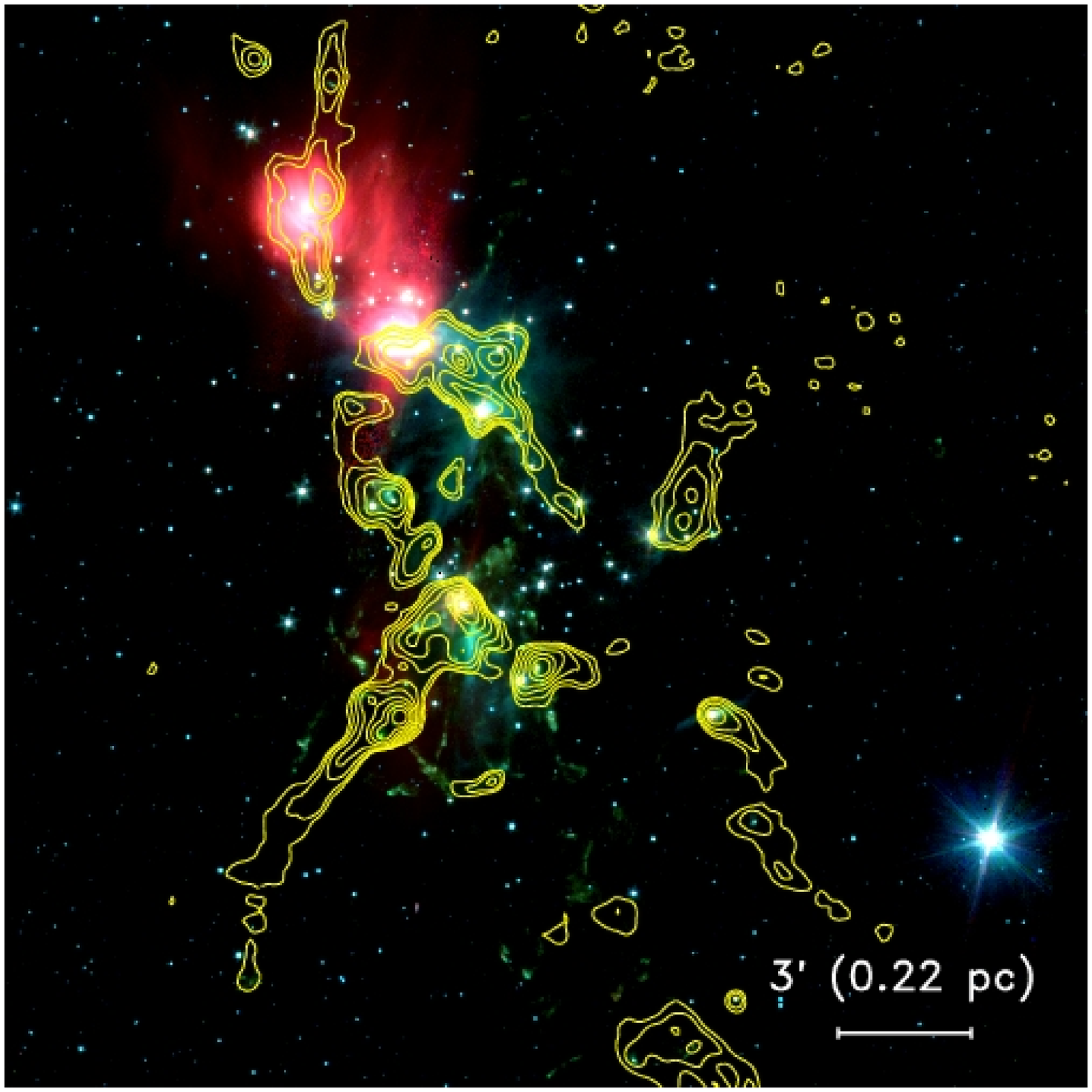}}
  \caption{Three-color image (blue: IRAC1 3.6~$\mu$m, green: IRAC2 4.5~$\mu$m
  and red: IRAC4 8.0~$\mu$m) overviews of the Ophiuchus and Perseus clouds
  with the main regions discussed in the text indicated. The lower panels show
  blow-ups of the L1688 region in Ophiuchus (left) and NGC~1333 region in
  Perseus (right). The yellow contours show the SCUBA 850~$\mu$m emission from
  the maps of \cite{johnstone04} and \cite{kirk06}. The contours are shown at
  9 logarithmically spaced levels between 40~mJy and 6~Jy. For a similar
  blow-up of the IRAS~16293 region see Fig.~\ref{i16293_outflow}. For further
  details about the IRAC observations see \cite{perspitz} and
  \cite{allen08}.}\label{overviewfigure}
\end{figure}

\begin{figure}
\resizebox{0.8\hsize}{!}{\includegraphics{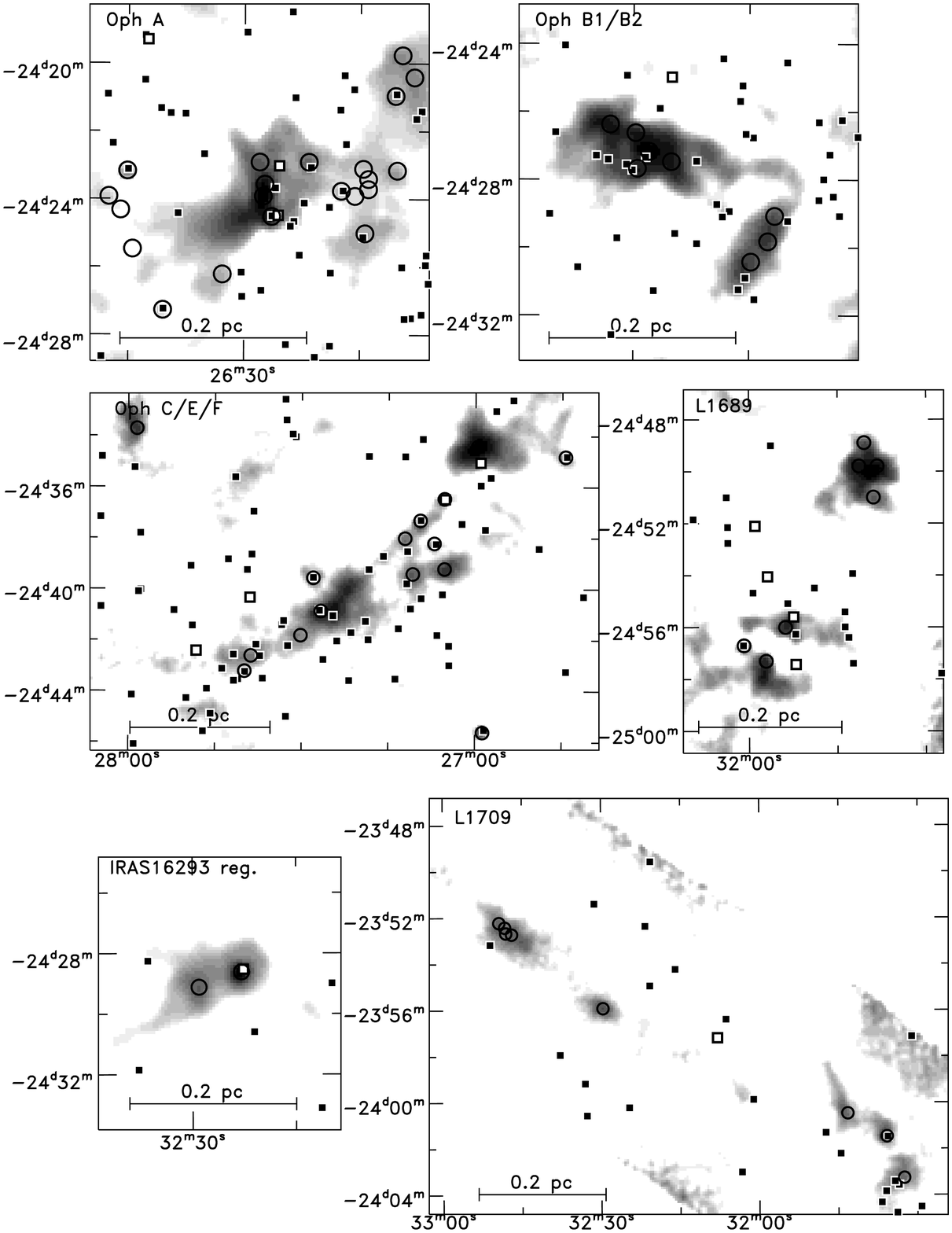}}
\caption{Finding charts for the Ophiuchus cloud. The SCUBA cores (circles) and
  MIPS 24~$\mu$m detected sources (filled squares) are plotted over SCUBA maps
  of six of the most prominent star forming regions indicated in
  Fig.~\ref{overviewfigure}. The sizes of the symbols of the SCUBA cores
  correspond to radii of 15$''$ (twice the SCUBA Beam). The MIPS 24~$\mu$m
  detected sources with $[3.6]-[4.5] > 1.0$ and $[8.0]-[24] > 4.5$ are singled
  out with open squares.}\label{findingchart}
\end{figure}

Together with five other nearby star forming regions, Ophiuchus was mapped at
3.6, 4.5, 5.8, 8.0, 24, 70 and 160~$\mu$m using the \emph{Spitzer Space
  Telescope}'s \IRAC\ and \MIPS\ cameras by c2d \citep{padgett08,allen08}
identifying more than 600,000 sources with a signal-to-noise ratio of 3 or
more in at least one of the 3.6--70~$\mu$m bands over an area of
12.2~degree$^2$. In total the SCUBA map covers approximately 5.7~degree$^{2}$
whereas the overlapping area between the IRAC and MIPS data covers
6.3~degree$^{2}$. In contrast to Perseus where the full area (3.6~degree$^2$)
of the SCUBA map is covered by the Spitzer observations, only about 80\% of
the SCUBA map of Ophiuchus (4.6~degree$^2$) was covered by the Spitzer
observations. No SCUBA cores were found outside this overlap region,
however. Fig.~\ref{overviewfigure} shows the SCUBA dust continuum emission
overlaid as contours on Spitzer images of Ophiuchus and
Perseus. Fig.~\ref{findingchart} shows finding charts with the distribution of
SCUBA cores and Spitzer mid-infrared sources in selected regions of Ophiuchus.
\clearpage
\clearpage

Slight differences exist in the processing of the current (and final) c2d
catalogs compared to the version that was used for analysis of Perseus in
\cite{scubaspitz}. Since that paper, all the c2d data were reprocessed using
the SSC pipeline version S13.  In the earlier catalogs, sources were extracted
from the IRAC and MIPS~24~$\mu$m mosaics individually and the lists of sources
from these catalogs \emph{band-merged} as described in \cite{delivery4}. In
the final catalogs delivered by c2d\footnote{Available at
http://ssc.spitzer.caltech.edu/legacy/c2dhistory.html.}, this procedure has
been improved by \emph{band-filling} individual sources. That is, for sources
without counterparts at one or more wavelengths, those mosaics were reexamined
and a best attempt made to extract a source. However, for detection purposes
sources are potentially suspect: for example outflow knots close to embedded
protostars are easily picked up toward the PSF wings of the MIPS~24~$\mu$m
detections of embedded YSOs and we therefore exclude them in the remainder of
the analysis.

Another addition to the c2d processing, is bandmerging with observations at
70~$\mu$m. In this paper, we include sources that are either detected at
24~$\mu$m (excluding band-filled sources) or 70~$\mu$m as the basis for the
analysis (in the following we refer to those sources as ``MIPS
sources''). This is in contrast to \cite{scubaspitz} where only the 24~$\mu$m
catalogs were utilized. As we return to below, this procedure solves the
problem that some well-known YSOs are saturated at 24$\mu$m, an issue which
was dealt with in \cite{scubaspitz} by including SCUBA cores with high
concentrations.

\subsection{Background contamination}
Using this first cut of MIPS 24~$\mu$m detected sources we can also address
the issue of background contamination: one of the complications in large scale
mappings such as those presented in c2d is separating background contamination
from the sample of candidate YSOs in an accurate statistical manner. This in
particular means separating background galaxies from the sample of YSOs since
both show similar red colors at mid-infrared wavelengths \citep[see,
e.g.,][for an illustrative example and discussion of this issue]{porras07}.
For Perseus, \cite{scubaspitz} found that the density of MIPS 24~$\mu$m
detected sources were low enough that the chance alignment between such a
source and a SCUBA core was statistically small. This fact holds true as well
for Ophiuchus: with 1325~MIPS 24~$\mu$m detected sources over 4.6~degree$^{2}$
we only find 0.08 sources within any 1~arcmin$^{2}$. This translates to a
1.5\% chance of having a MIPS~24~$\mu$m source randomly aligned within a
15$''$ radius (one SCUBA half-power beam-width) of the center a given SCUBA
core (or put in another way, statistically there will be 1 randomly assigned
candidate in total within 15$''$ of the 66 SCUBA cores in Ophiuchus).

\section{Mid-infrared sources associated with dense cores}\label{method}
\subsection{Constructing a sample of embedded protostars}\label{midirsubmm}
\begin{figure}
\resizebox{0.5\hsize}{!}{\includegraphics{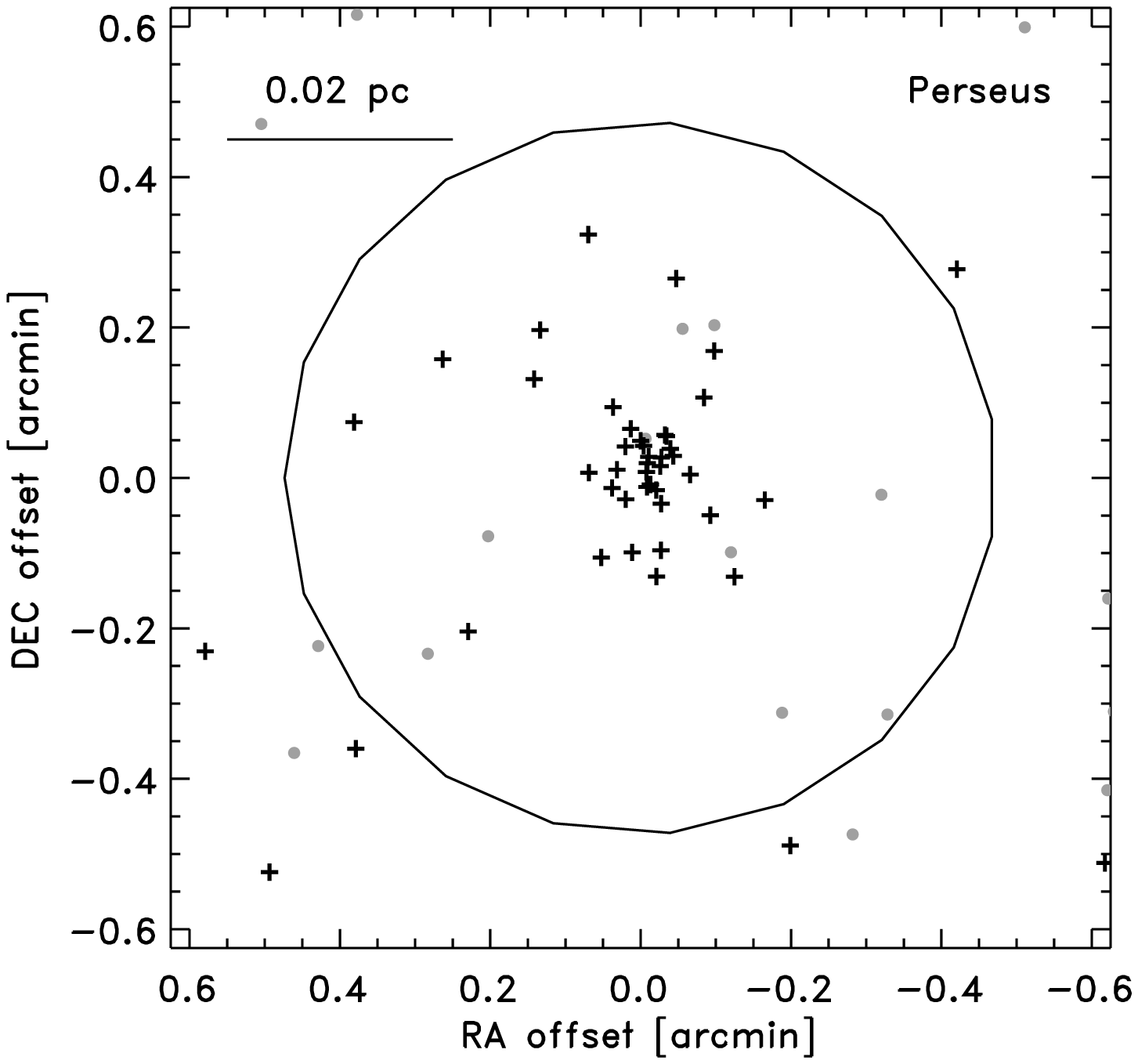}}
\resizebox{0.5\hsize}{!}{\includegraphics{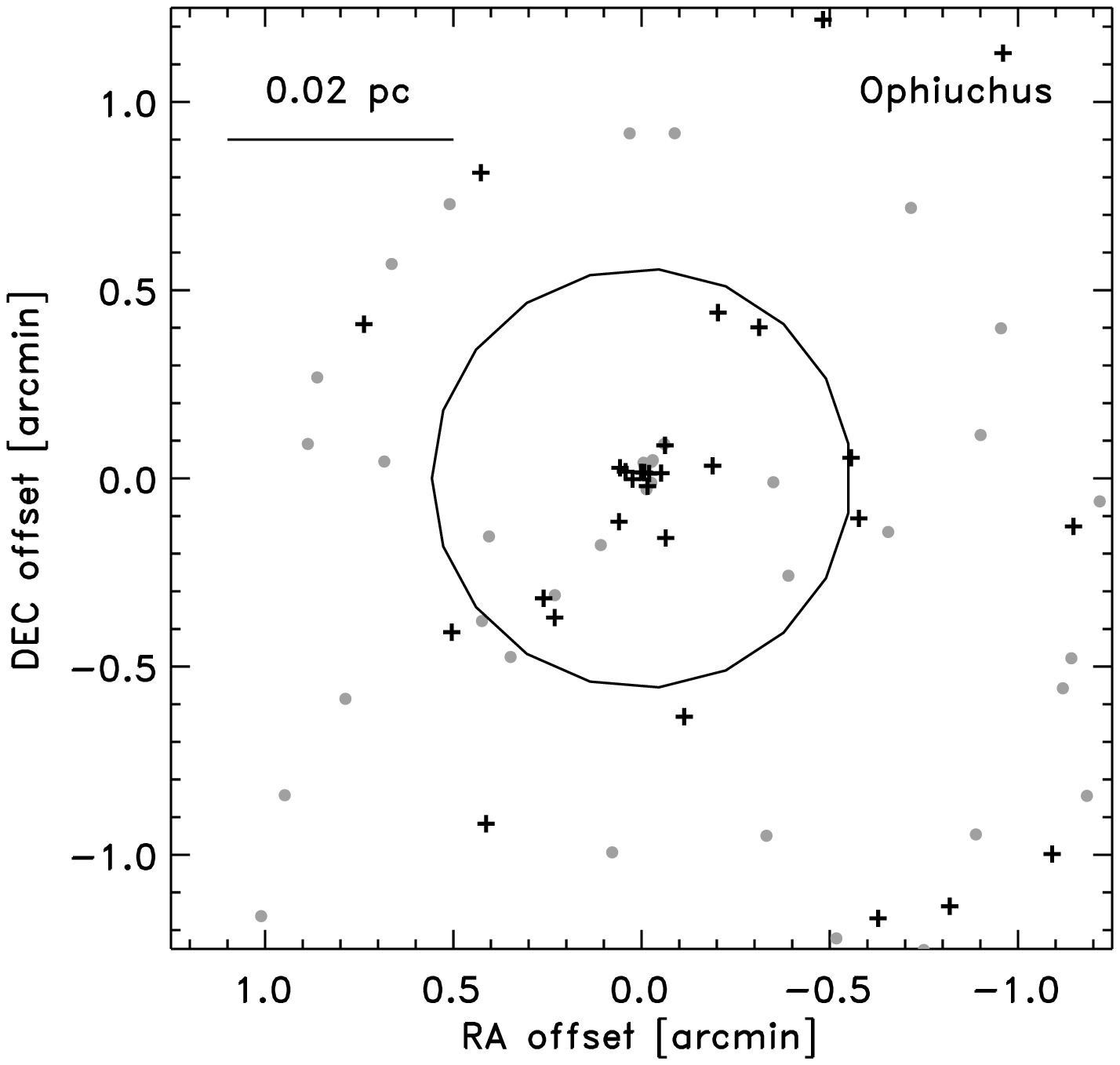}}
\resizebox{0.5\hsize}{!}{\includegraphics{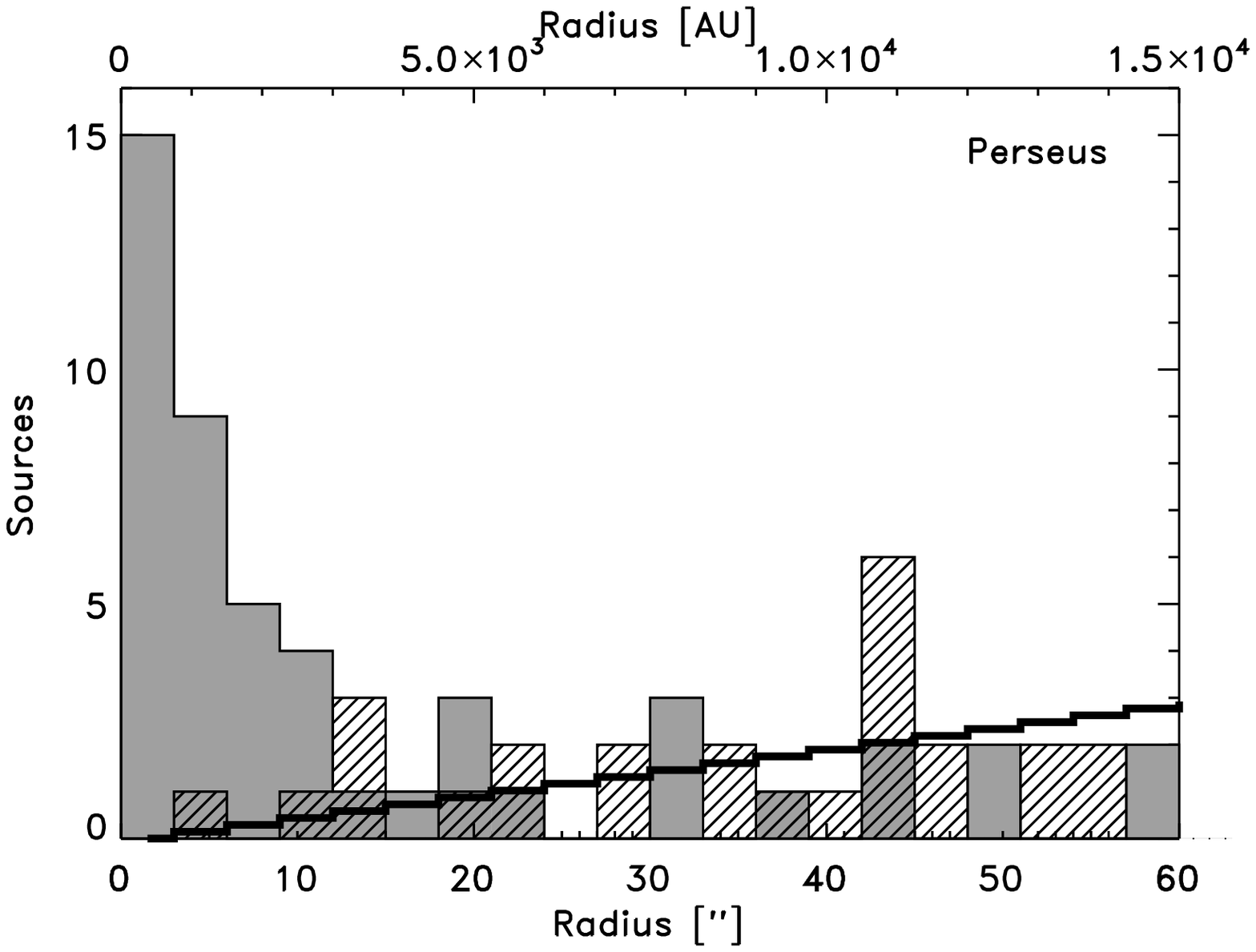}}
\resizebox{0.5\hsize}{!}{\includegraphics{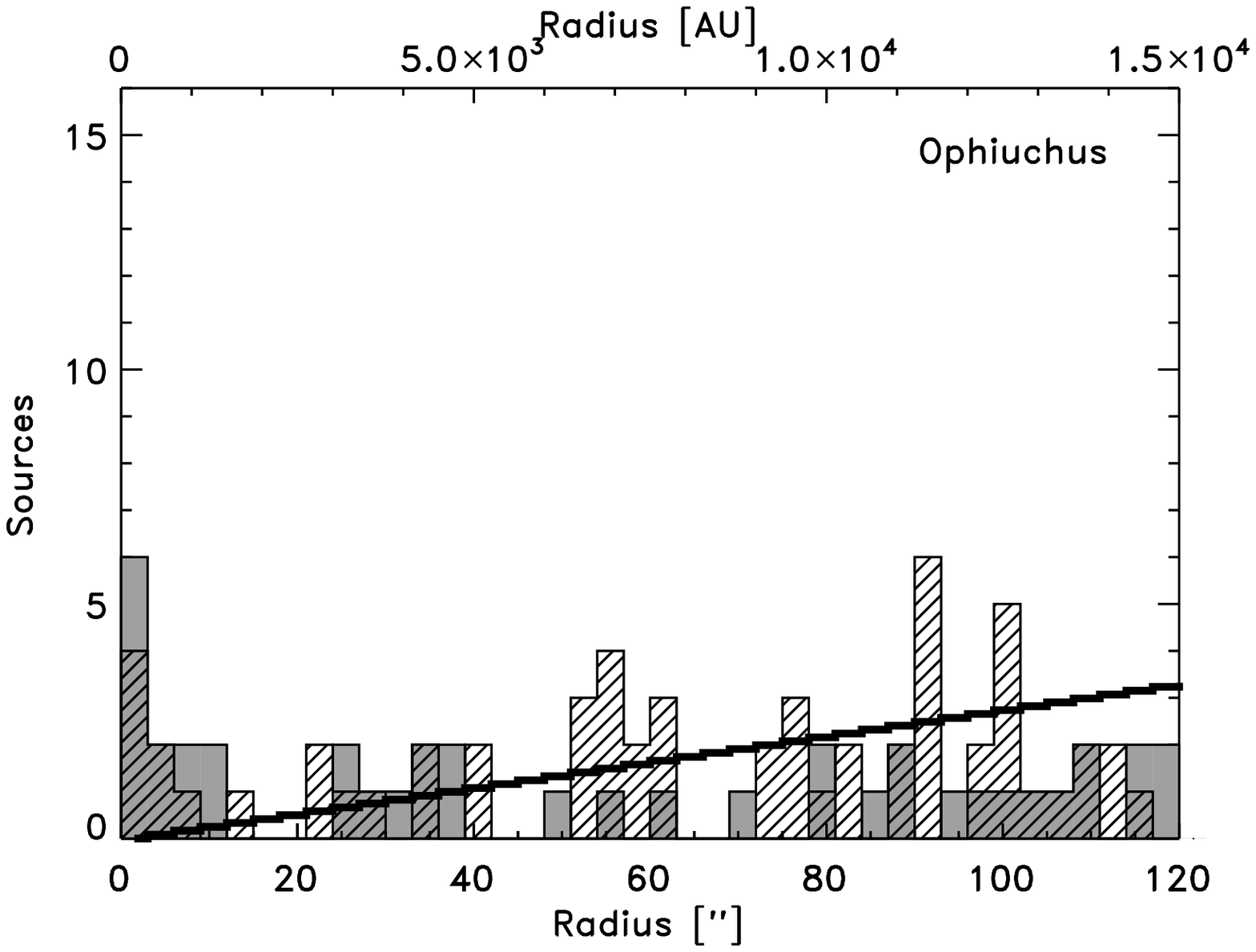}}
\caption{\emph{Upper panels:} Distribution of MIPS sources with red
  mid-infrared colors ($[3.6]-[4.5] > 1.0$ and $[8.0]-[24] > 4.5$; black plus
  signs) compared to other MIPS sources (grey circles) around each SCUBA core
  (shifted to the same center). The size scales in the two plots are stretched
  to represent the same linear scale. The circle indicates the mean SCUBA core
  sizes. \emph{Lower panels:} distribution of number of mid-infrared sources
  as a function of distance to the nearest SCUBA core. Again the filled grey
  histogram indicates the distribution of sources with $[3.6]-[4.5] > 1.0$ and
  $[4.5]-[8.0] > 4.5$ and the hatched histogram the distribution of the other
  sources. The thick black line indicates the prediction from a uniform source
  distribution with the same surface density as the observed
  distribution.}\label{clustplot}
\end{figure}

\begin{figure}
\resizebox{0.8\hsize}{!}{\includegraphics{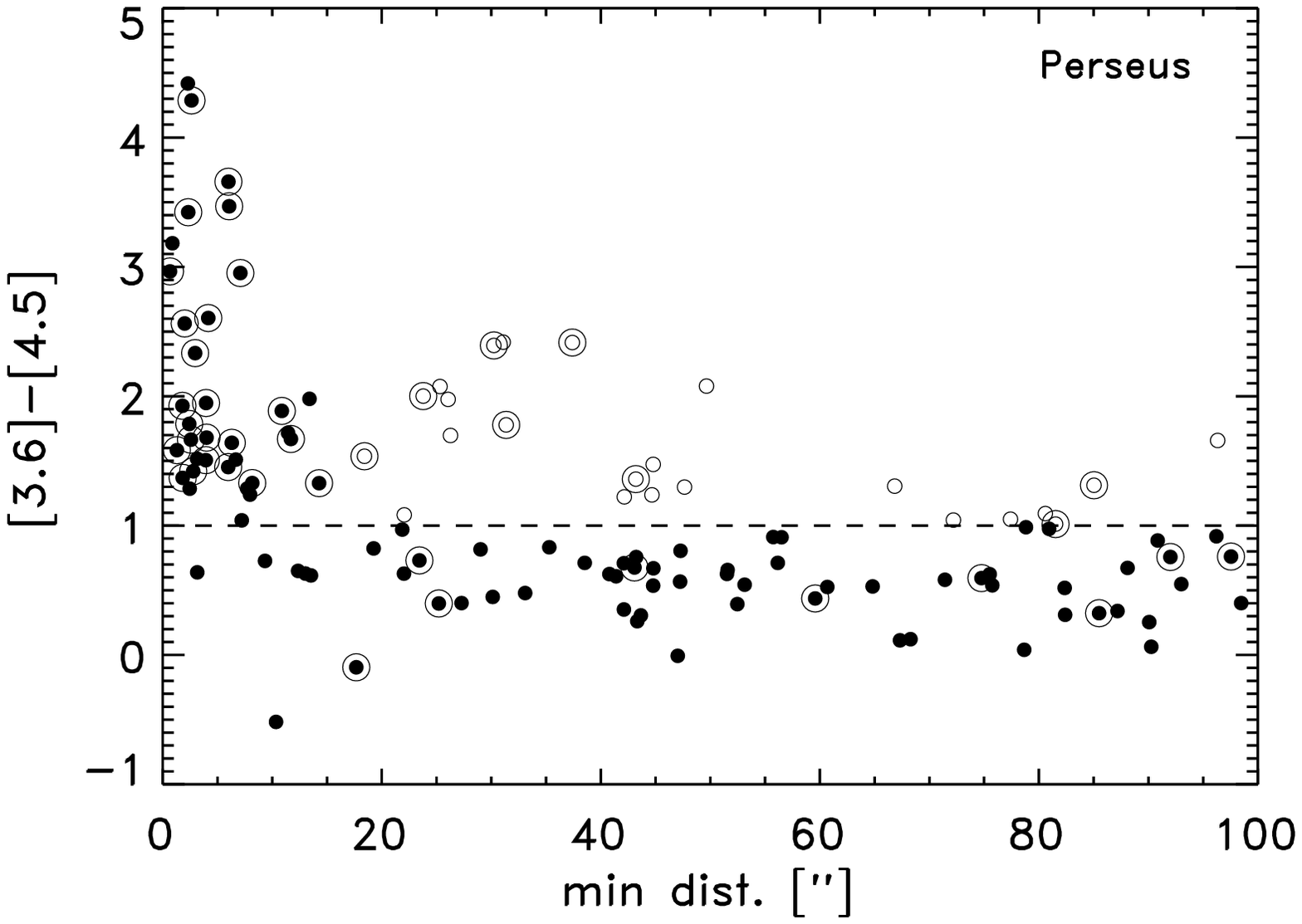}}
\resizebox{0.8\hsize}{!}{\includegraphics{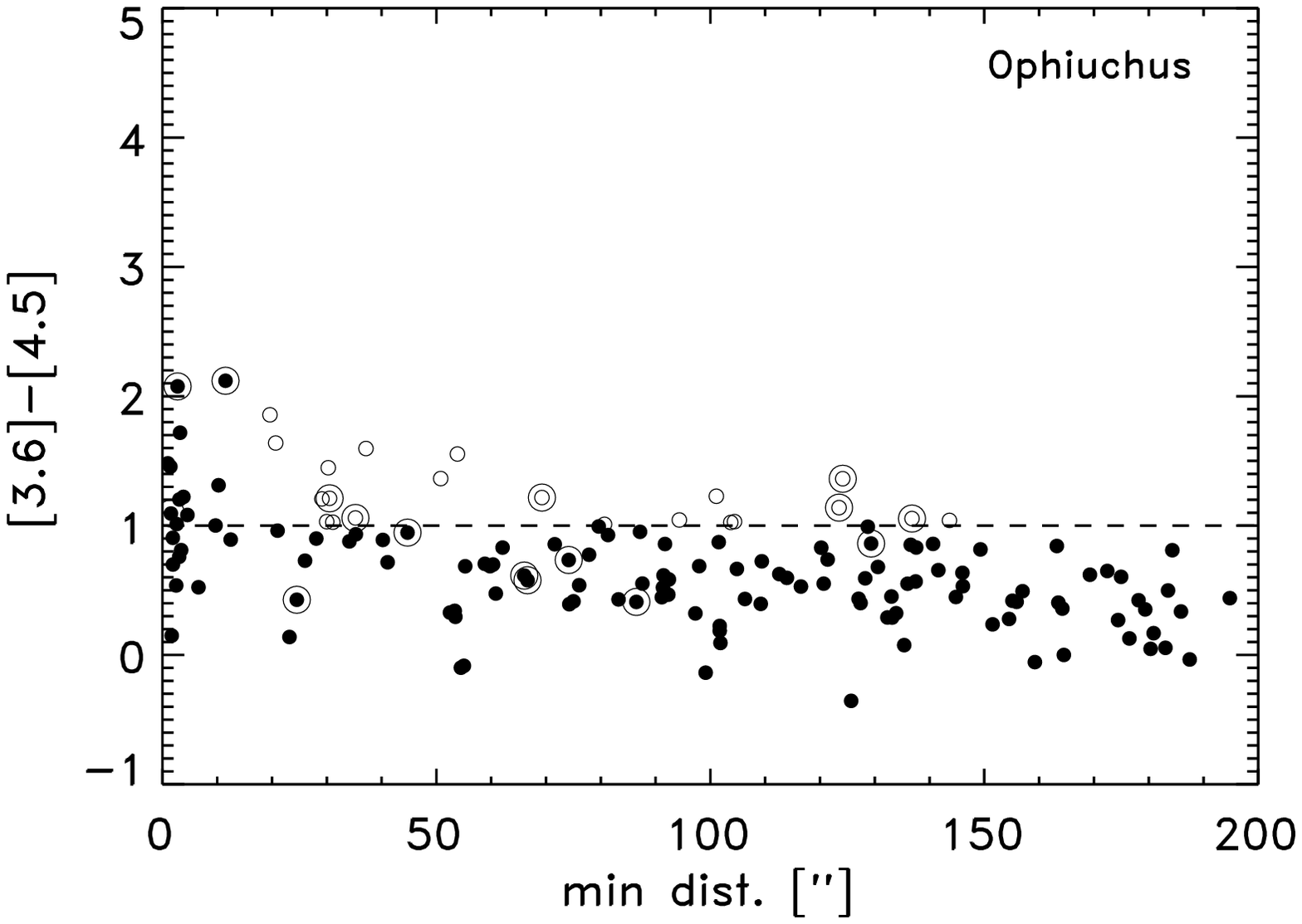}}
\caption{$[3.6]-[4.5]$ color of each MIPS source vs. its distance to the
  nearest submillimeter core. Upper panel for Perseus and lower panel for
  Ophiuchus. Sources are shown with filled circles except those with distances larger than 15$''$ to their nearest cores
  and $[3.6]-[4.5] > 1.0$, which have been indicated by small open circles. Symbols
  (filled or open circles) with an extra larger circle around indicate
  mid-infrared sources with $[8.0]-[24]$ colors greater than
  4.5.}\label{inverse_12}
\end{figure}

\begin{figure}
\resizebox{0.8\hsize}{!}{\includegraphics{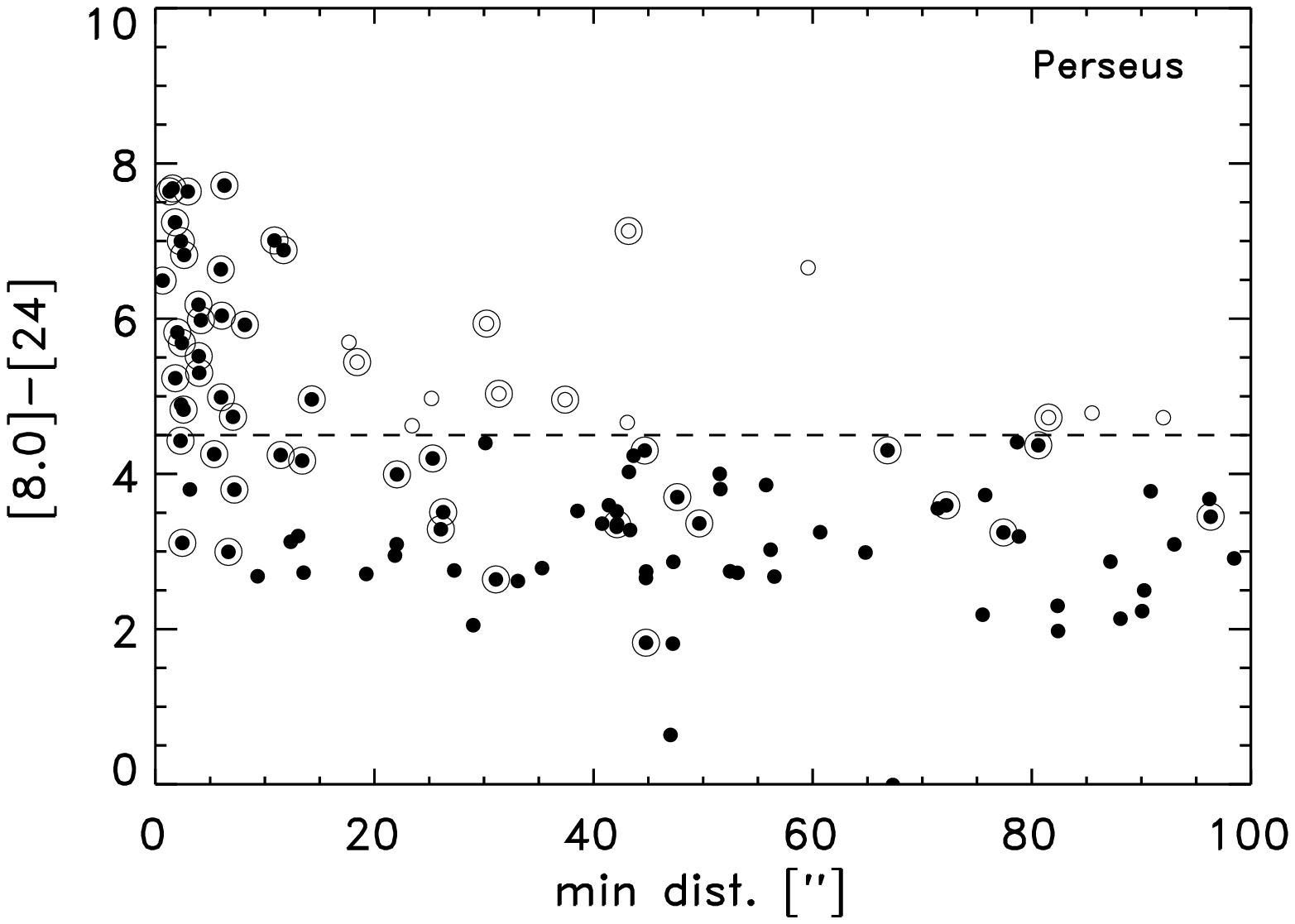}}
\resizebox{0.8\hsize}{!}{\includegraphics{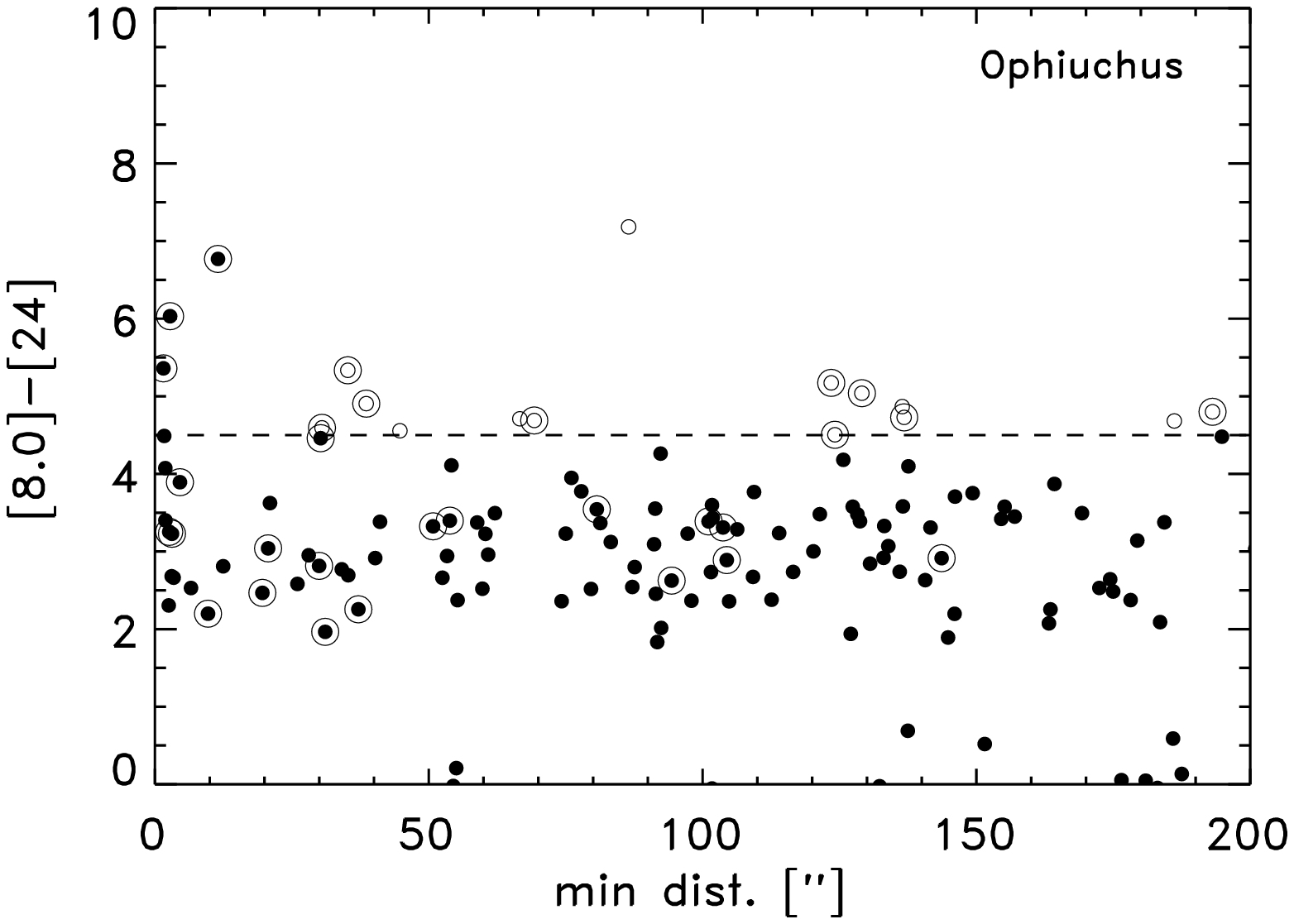}}
\caption{$[8.0]-[24]$ color of each MIPS source vs. its distance to the
  nearest submillimeter core. Upper panel for Perseus and lower panel for
  Ophiuchus. Sources are shown with filled circles except those with distances
  larger than 15$''$ to their nearest cores and $[8.0]-[24] > 4.5$, which have
  been indicated by open circles. Symbols (filled or open circles) with an
  extra larger circle around indicate mid-infrared sources with $[3.6]-[4.5]$
  colors greater than 1.0.}\label{inverse_45}
\end{figure}

Fig.~\ref{clustplot} shows the distribution of MIPS sources with respect to
the nearest submillimeter core for both Ophiuchus and Perseus. Sources with
red mid-infrared colors ($[3.6]-[4.5]>1.0$ and $[8.0]-[24]>4.5$ are
highlighted to reveal their high degree of correlation with the centers of
SCUBA cores. Figs.~\ref{inverse_12} and \ref{inverse_45} show the one
dimensional distributions of $[3.6]-[4.5]$ and $[8.0]-[24]$ colors as a
function of distance to the nearest SCUBA core. Both figures reveal that there
is some clustering of the Ophiuchus sources relative to the core center, but
with less significance than in Perseus. In Perseus the ratio of sources with
$[3.6]-[4.5] > 1.0$ relative to those with with $[3.6]-[4.5] \le 1.0$ is 5.7
within 15$''$ compared to 0.40 at distances of 30$''$--60$''$. In Ophiuchus
the numbers are 1.5 and 0.53, respectively. That is, the number of red sources
in SCUBA cores is about a factor 3.5--4 higher in Perseus than
Ophiuchus. Furthermore, in Perseus, sources with significantly redder colors
are found: 10 sources are found with $[3.6]-[4.5] > 2.5$ with a maximum of 4.4
whereas none this red are found in Ophiuchus. This difference between the
redness of the embedded sources in Perseus vs. Ophiuchus may suggest a
difference in their physical properties (see \S\ref{physorigin}).

\clearpage

To establish samples of embedded YSOs we utilize these results to apply the
same procedure \cite{scubaspitz} used on the Perseus datasets. In this paper
we refer to embedded YSOs as any MIPS source (i.e., a source detected at
either 24 or 70~$\mu$m) which is associated with a submillimeter core as
traced by emission in the SCUBA maps or which fulfills certain color criteria
(see below). Neither the mid-infrared source lists nor the submillimeter core
identifications by themselves can be used unambiguously to define a sample of
embedded objects, but as demonstrated in \cite{scubaspitz} they can be used in
conjunction to form a relatively unbiased list of YSOs complete down to the
resolution of the mid-infrared data. The three criteria developed for
identifying candidate protostars in Perseus were based on the mid-infrared
colors of Spitzer sources, associations between Spitzer sources and
submillimeter cores and the concentration\footnote{The concentration of a
SCUBA core is defined as $C=1-\frac{1.13\, B^2S_{850}}{(\pi R^2_{\rm
obs}f_0)}$, where $B$ is the beam size, $S_{850}$ the total flux from the
SCUBA observations at 850~$\mu$m, $R_{\rm obs}$ the measured radius and $f_0$
the peak flux \citep{johnstone00cores}.} of SCUBA cores. In summary the list
of embedded objects were constructed by selecting:
\begin{description}
\item[A.] MIPS 24~$\mu$m catalog sources with $[3.6]-[4.5] > 1$ and
  $[8.0]-[24] > 4.5$, \emph{or}
\item[B.] MIPS 24~$\mu$m catalog sources with distances less than 15\arcsec\
  to their nearest submillimeter core, \emph{or}
\item[C.] Submillimeter cores with concentrations higher than 0.6.
\end{description}
The first criterion selects the most deeply embedded YSOs with steeply
increasing SEDs but misses sources that are not detected in IRAC bands 1 or 2
- either due to confusion with outflows or simply because they in fact are
very deeply embedded. The second criterion selects all MIPS sources that are
embedded YSOs in the sense that they are located within two SCUBA beam sizes
of a SCUBA core. That list will also include sources which are not necessarily
directly associated with an observed submillimeter core due to confusion in
the submillimeter map, poorly defined core structure, or the limiting
sensitivity of the submillimeter observations. In Perseus it was found that
most of the sources selected according to the second criterion also obeyed the
first (upper panels of Fig.~\ref{inverse_12} and \ref{inverse_45}).

These first two criteria were found to still miss sources saturated at
24~$\mu$m. Such sources were selected by the third criterion, which did not in
itself include all the mid-infrared sources under {\bf A} and {\bf B} (see
\cite{scubaspitz}). On the other hand it was also shown that a number of low
concentration cores in fact had embedded YSOs, so criterion {\bf C} in itself
was also not optimal for picking out candidate embedded YSOs. In contrast to
Perseus, there are a few cores with high concentrations ($C > 0.6$) in
Ophiuchus that in fact are starless. As well, a few sources associated with
lower concentration cores are found to be saturated at 24~$\mu$m and therefore
would not make it onto the list of protostars.

Visual inspection of all SCUBA cores, however, reveals that by adding sources
that are detected at 70~$\mu$m, provides a sample of all the MIPS sources in
Ophiuchus associated with SCUBA cores, even those clearly saturated or
confused at 24~$\mu$m. This method can also be directly applied to Perseus:
the four sources that were associated with high concentration cores all only
have 70~$\mu$m detections, whereas no new sources are added by utilizing this
criterion in Perseus. In this way there is no need to rely on the observed
empirical properties of the SCUBA cores\footnote{The selection according to
the second criterion ``{\bf B}'' of course still relies on the association
between SCUBA cores and MIPS sources.}. Therefore, the two criteria required
to identify candidate protostars in this paper in either Perseus or Ophiuchus
are:
\begin{description}
\item[A.] MIPS 24 or 70~$\mu$m catalog sources with $[3.6]-[4.5] > 1$ and
  $[8.0]-[24] > 4.5$, \emph{or}
\item[B.] MIPS 24 or 70~$\mu$m catalog sources with distances less than 15\arcsec\ to
  their nearest core.
\end{description}

In this way a sample of 27 embedded YSOs are identified: the majority 24 (89\%)
because they are within 15\arcsec\ of a SCUBA core (criterion {\bf
B})\footnote{Note, that we use the same angular scale for the comparison between
the SCUBA cores and MIPS sources in both Ophiuchus and Perseus despite the
factor 2 difference in distance between the two clouds. We return to this
discussion in Sect.~\ref{disteffects}.}  with the remaining three sources due to
their $[3.6]-[4.5]$ and $[8.0]-[24]$ colors (criterion {\bf
A}). Table~\ref{embeddedysolist} summarizes the sample of candidate embedded
YSOs in Ophiuchus. One core, SMM~J162626-24243, is associated with two YSOs (GDS
J162625.6-242429 and VLA~1623) within 15$''$ of its center, although at least
one other, SMM~J162622-24225, is associated with mulitple YSOs (the GSS30-IRS1,
-IRS2, -IRS3 system) where only one is picked up as a separate MIPS 24~$\mu$m
source. This is similar to the analysis from Perseus where only few SCUBA cores
were found to be associated with multiple MIPS sources \citep{scubaspitz}. It is
here worth emphasizing that the resolution of the MIPS observations is only
about 6$''$ (750~AU) and these numbers should therefore not be taken as
statements concerning the close binarity of the embedded YSOs.

Of the 27 embedded YSOs, 16 were also identified as YSO candidates using the
criteria used by c2d to extract YSOs based on their mid-infrared colors
\citep[see discussion in][]{harvey07}, but the remaining 11 were not. This
illustrates the need for the combination of mid-infrared and submillimeter data
when constructing samples of deeply embedded protostars - as it was also
concluded based on the analysis of the Perseus data \citep{scubaspitz}.

\subsection{Distribution of concentrations}
\begin{figure}
\resizebox{0.8\hsize}{!}{\includegraphics{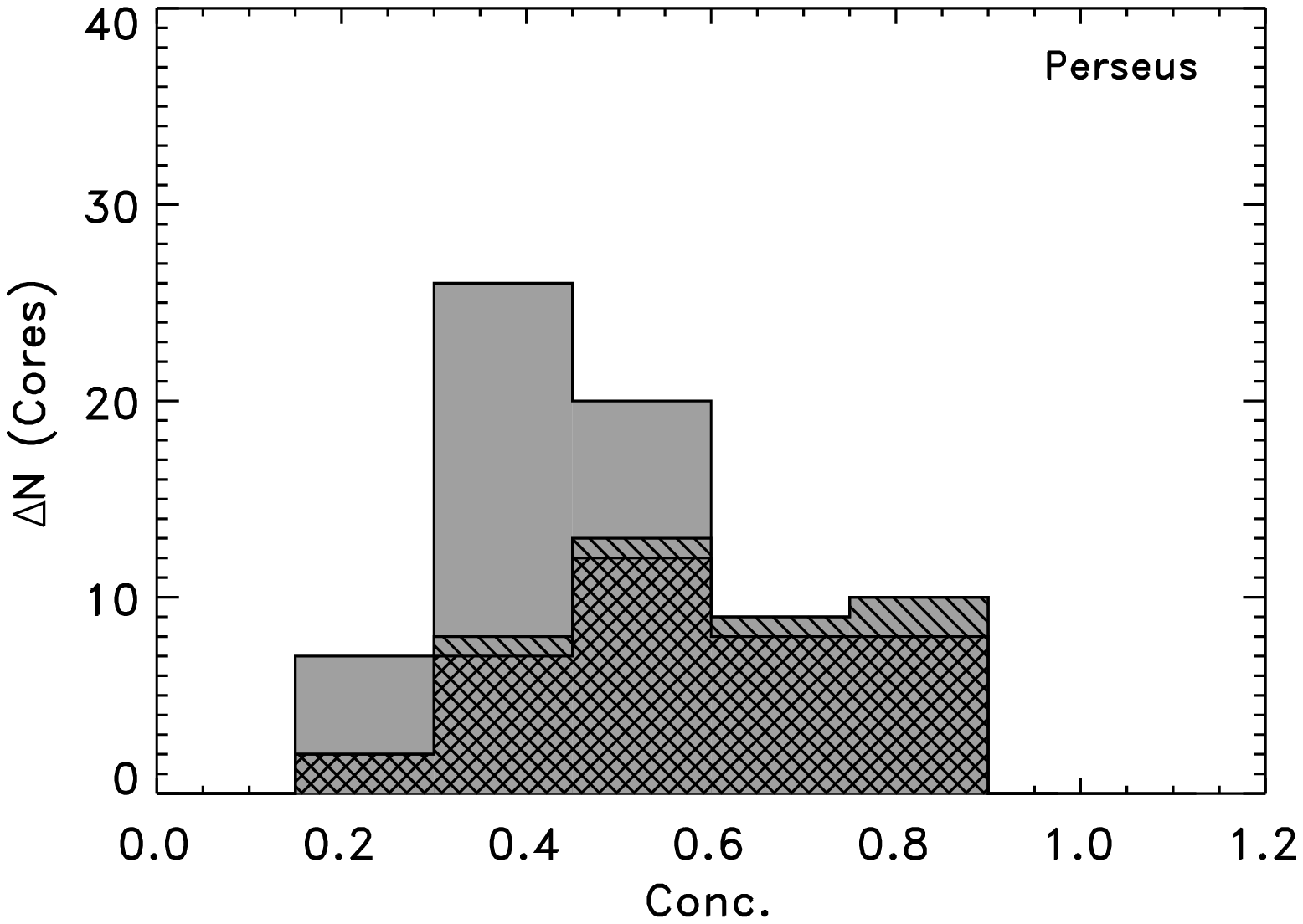}}
\resizebox{0.8\hsize}{!}{\includegraphics{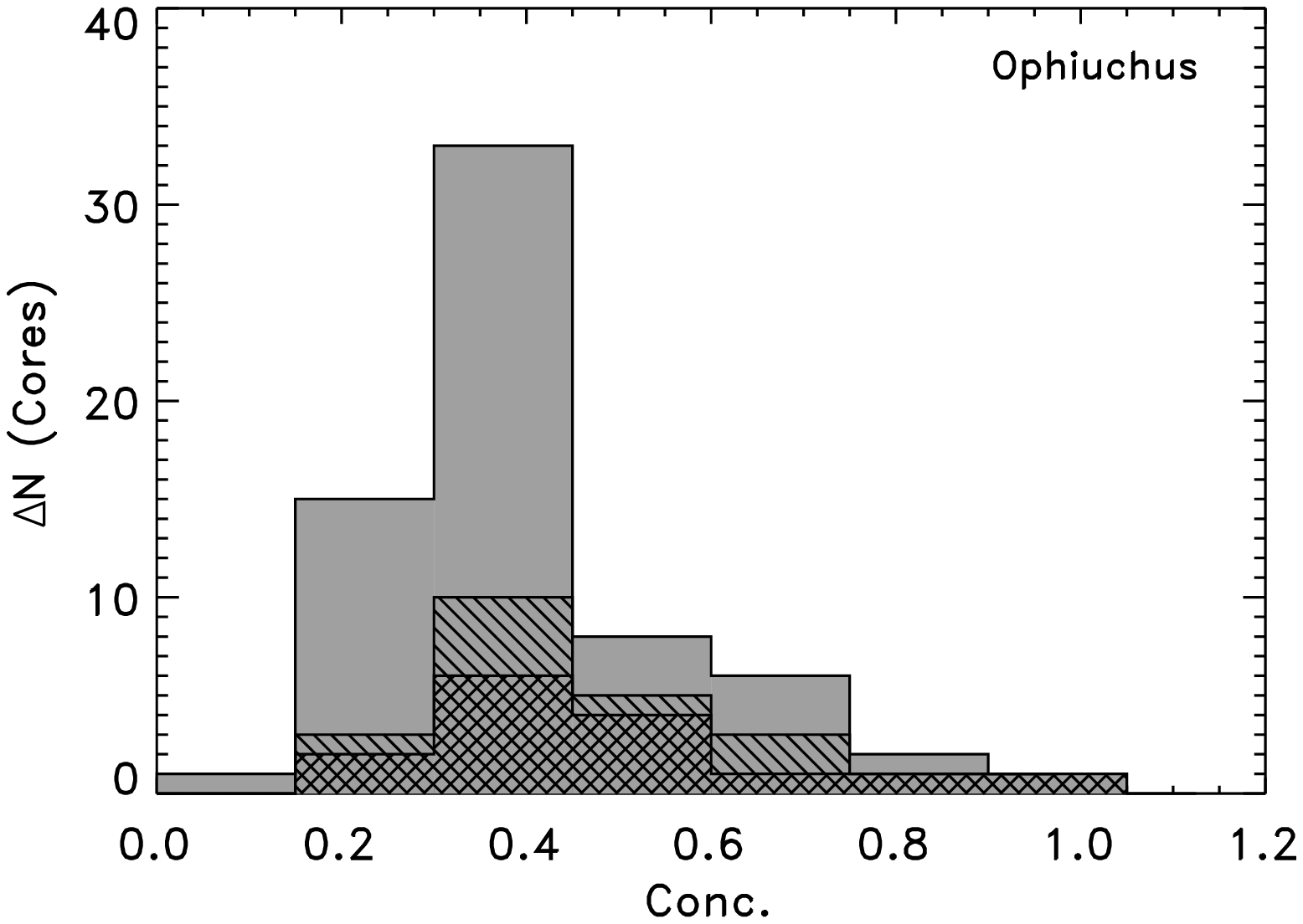}}
\caption{Distribution of ``concentrations'' of SCUBA cores. Those with MIPS
  sources within 15$''$ are hatched (single/double) with the cores with MIPS
  sources with red mid-infrared colors ($[3.6]-[4.5]>1.0$ and
  $[8.0]-[24]>4.5$) within 15$''$ double-hatched.}\label{conc_histo}
\end{figure}

An important finding of \cite{scubaspitz} was that all cores with
concentrations higher than 0.6 had MIPS sources within
15$''$. Fig.~\ref{conc_histo} compares the distributions of the concentrations
of the SCUBA cores with and without embedded MIPS sources in Perseus and
Ophiuchus. The two distributions are quite different: in Perseus the
concentrations of the cores show a very broad distribution with a number of
high concentration cores whereas the Ophiuchus distribution shows a
pronounced peak at concentrations $\approx 0.35$. In \cite{scubaspitz} it was
pointed out that protostellar cores should have higher concentrations than
pre-stellar, simply because of the heating of the dust from the
center. Therefore based on the SCUBA data alone we should expect fewer
embedded protostars in Ophiuchus relative to Perseus. This is directly
confirmed by comparing the number of SCUBA cores with MIPS sources within
15$''$ in the two clouds.  Ophiuchus has embedded YSOs in only 23 out of the
66 SCUBA cores while Perseus has embedded YSOs in 42 out of the 72 SCUBA
cores.  Further, as discussed in \S\ref{disksources} some of the sources in
Ophiuchus are actually disk sources and not deeply embedded.

\clearpage
\clearpage

\subsubsection{Cores with high concentrations and no embedded YSOs}
\begin{figure}
\resizebox{0.7\hsize}{!}{\includegraphics{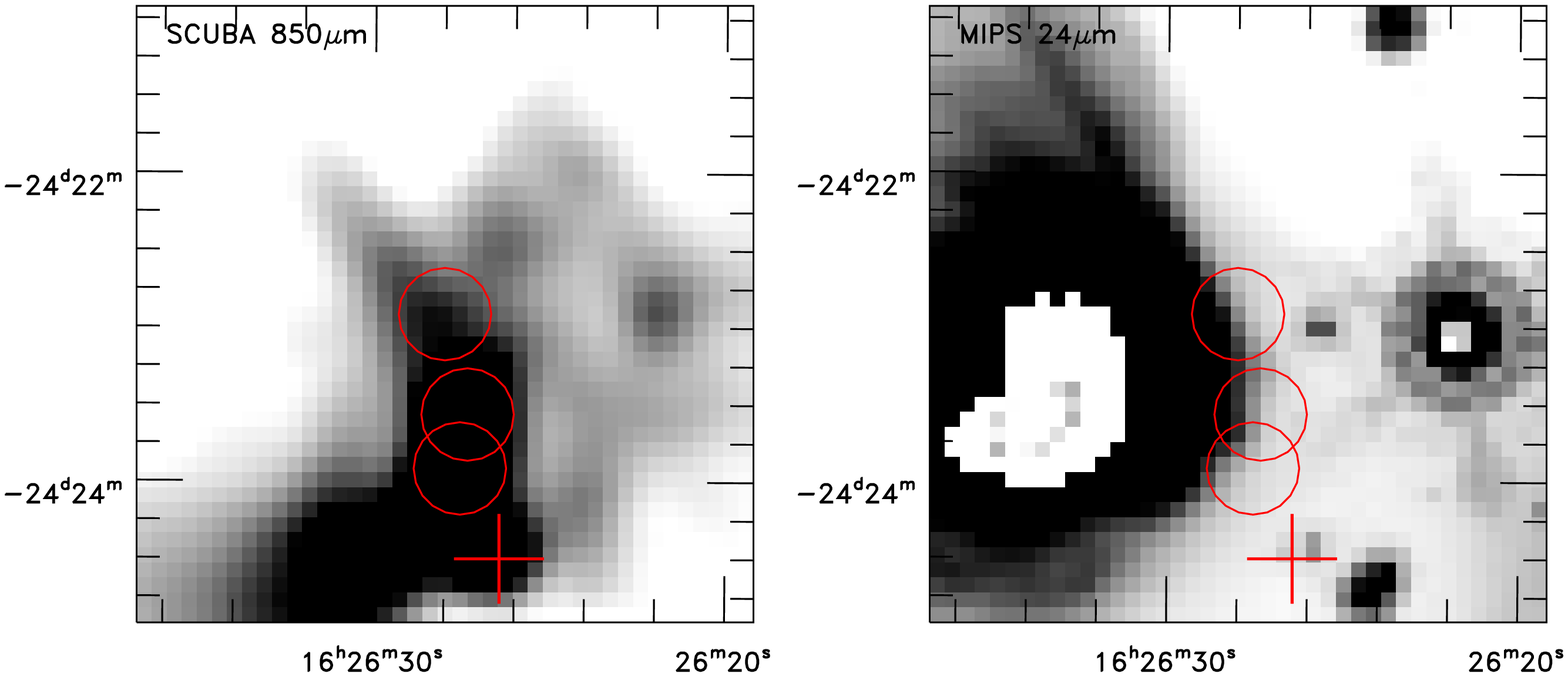}}
\resizebox{0.7\hsize}{!}{\includegraphics{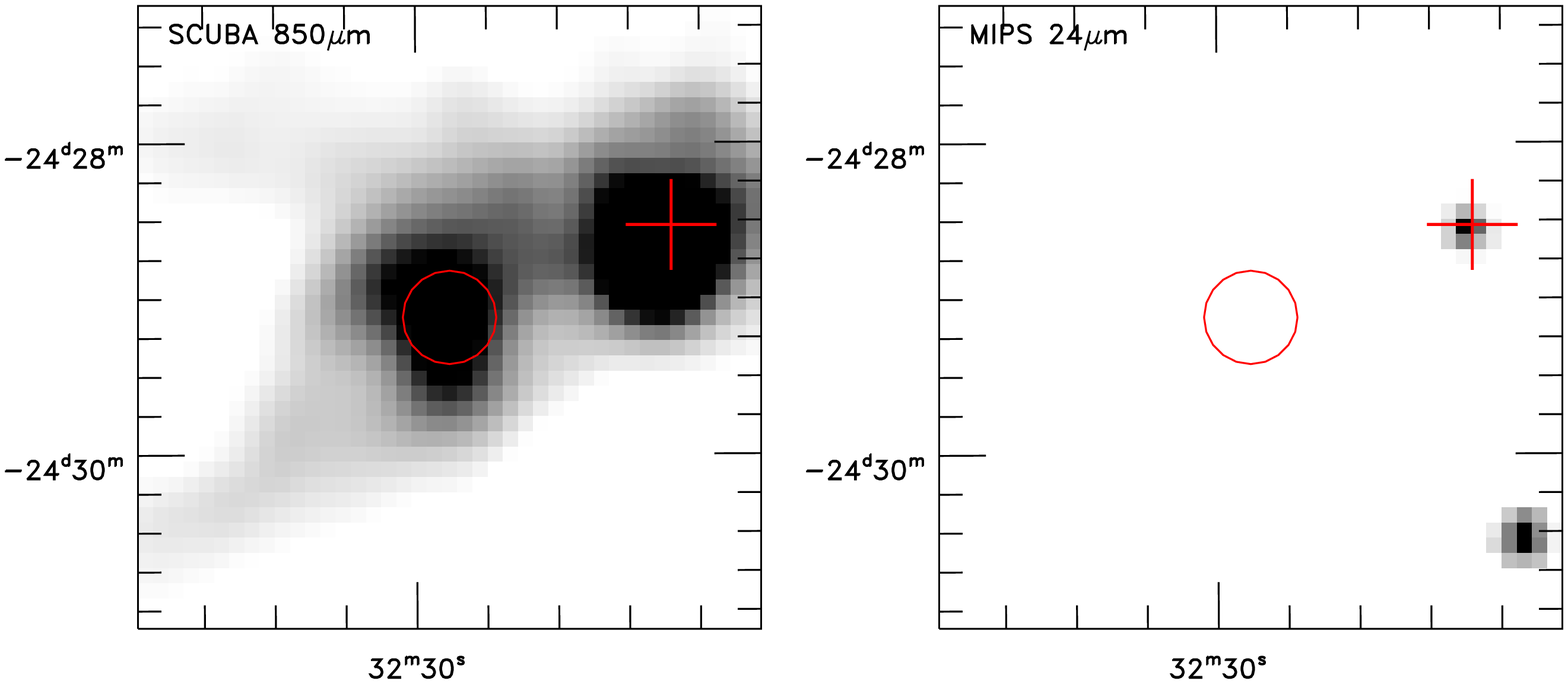}}
\caption{Cores with concentrations larger than 0.6 and no embedded YSOs shown
  on top of the SCUBA 850~$\mu$m image (left panels) and the MIPS~24~$\mu$m
  image (right panels). The upper panels show the Oph.-A ridge and the lower
  panel the region around IRAS~16293-2422. In all panels the circles indicate
  cores with high concentrations but no associated MIPS sources (SMM
  J162627-24233, J162628-24225, J16262628-24235 in the upper panels and SMM
  J163229-24291 in the lower panels) whereas the plus-signs show the location
  of the known Class 0 sources, VLA~1623 (SMM J162626-24243; upper panels) and
  IRAS~16293-2422 (SMM J163223-24284; lower panels).}\label{corenoyso}
\end{figure}

A noteworthy feature of the Ophiuchus SCUBA cores is the existence of four
high concentration cores without embedded YSOs. The four cores in Ophiuchus
are in fact starless as shown in Fig.~\ref{corenoyso}. Three of these cores
are located in the Oph~A ridge, which has previously been the subject of
detailed studies at (sub)millimeter wavelengths
\citep[e.g.,][]{andre93,wilson99}. The Spitzer observations confirm that the
the ridge itself actually appears to be starless but with protostars (e.g.,
VLA~1623 and GSS~30) located in its immediate vicinity. \cite{andre93} (and
likewise \cite{difrancesco04}) found that the cores in the Oph~A ridge were
within a factor two of being in virial equilibrium. Therefore the high
concentrations of the cores in the ridge most likely reflect the extreme local
pressure due to the surrounding environment. \cite{andre93} found that the
millimeter cores in the northern part of the ridge had a high dust temperature
of $27\pm 5$~K, likely due to heating by the nearby B-star [and clearly
heating the dust seen at 24~$\mu$m also (lower right panel of
Fig.~\ref{corenoyso})].

\begin{figure}
\resizebox{0.7\hsize}{!}{\includegraphics{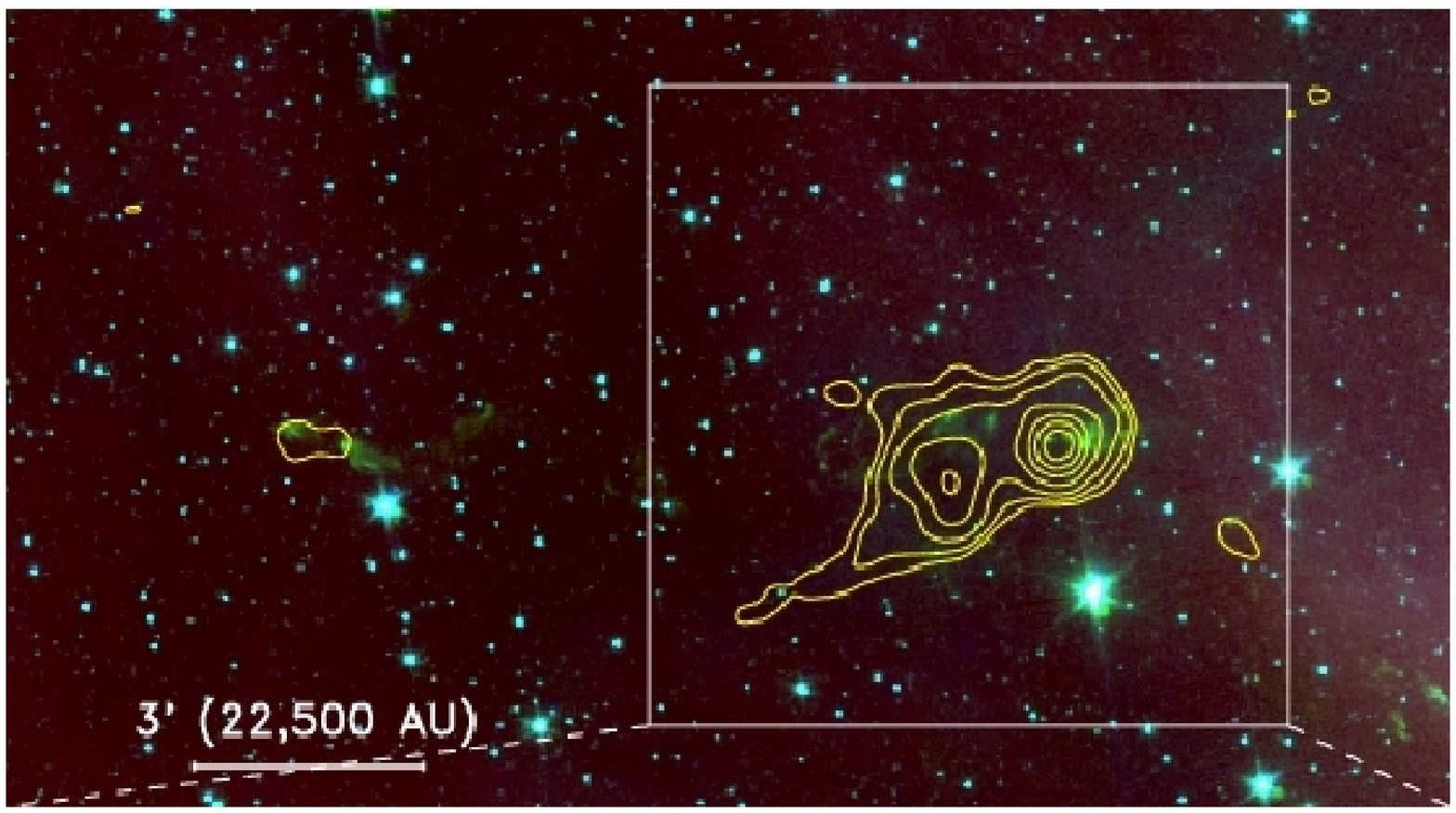}}
\resizebox{0.7\hsize}{!}{\includegraphics{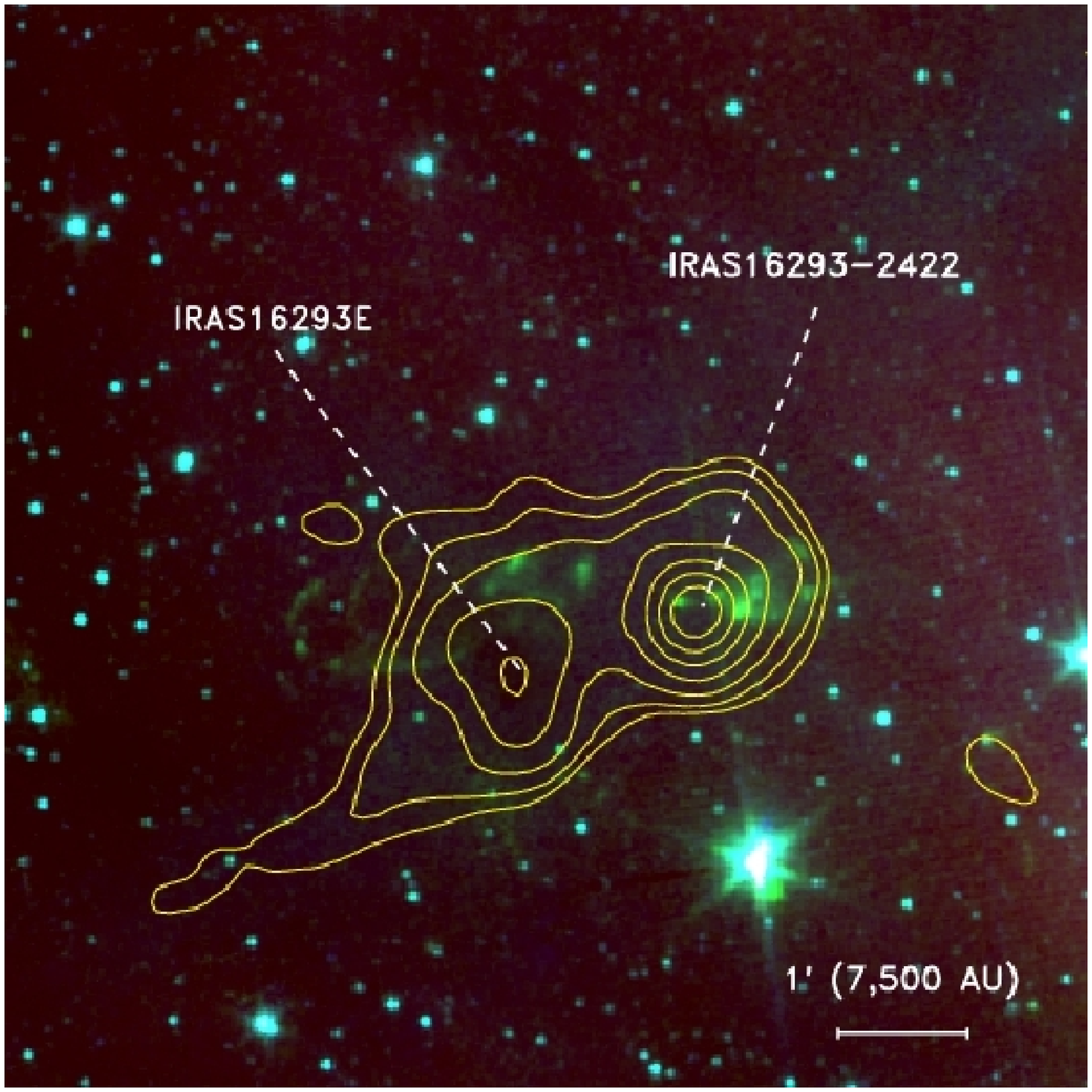}}
\caption{Three-color image (blue: IRAC1 3.6~$\mu$m, green: IRAC2 4.5~$\mu$m
  and red: IRAC4 8.0~$\mu$m) of the region around the Class 0 protostar
  IRAS~16293-2422 and starless core IRAS~16293E. The SCUBA map is overlaid in
  yellow contours. The outflow interaction is predominantly traced by emission from
  rotational transitions of molecular hydrogen picked up in the 
  IRAC2 band (i.e., the green color; see, e.g., \citealt{noriegacrespo04}).}\label{i16293_outflow}
\end{figure}

The fourth high concentration starless core, IRAS~16293E, is a well-studied
submillimeter companion to the IRAS source IRAS~16293-2422 and also has no
counterparts at 24~$\mu$m or 70~$\mu$m (Fig.~\ref{corenoyso}). \cite{castets01}
suggested that IRAS~16293E was a deeply embedded, or ``Class 0''
\citep{andre93}, protostar based on the detection of an outflow inferred from
HCO$^+$. \cite{stark04} presented larger scale maps and showed that this was not
the case, rather the outflowing motions suggested for IRAS~16293E by
\citeauthor{castets01} were a result of the larger scale outflow driven by
IRAS~16293-2422B fanning around IRAS~16293E. The Spitzer images indeed confirm
that the IRAS~16293-2422 outflow extends eastwards of IRAS~16293E
(Fig.~\ref{i16293_outflow}). The absence of a central heating source in
IRAS~16293E is also consistent with the fact that IRAS~16293E shows strong
N$_2$H$^+$ emission relative to the protostellar source in IRAS~16293-2422
itself as reported by \cite{castets01}. N$_2$H$^+$ is found to be prominent in
starless cores with low temperatures where CO has frozen out on dust grains and
thus unavailable to destroy N$_2$H$^+$. In contrast, the conditions within
protostellar cores heat a significant part of the envelope to temperatures where
CO is released into the gas-phase \citep[e.g.,][]{bergin01,paperii}. It is
likely that the submillimeter core IRAS~16293E is affected by the coincident
outflow, e.g., compression and heating possibly increasing its concentration.

\clearpage

\clearpage

\subsection{Distance effects}\label{disteffects}
The most obvious limitation in the direct comparison between Perseus and
Ophiuchus is the factor of two difference in distance between the two clouds
(Perseus at 250~pc; Ophiuchus at 125~pc)\footnote{We note that these distance
estimates in themselves may contain significant uncertainties: a recent
analysis of Hipparcos data for Ophiuchus by \cite{mamajek08} for example
suggests a concensus distance of 139$\pm 6$~pc to this cloud whereas distance
estimates for Perseus range from about 220~pc to 350~pc (see \cite{enoch06}
for a discussion).}. Since the SCUBA and Spitzer surveys are observed to the
same absolute depth for each cloud, we would expect intrinsically weaker
sources to be detected in Ophiuchus. For example, if we assume that each of
the SCUBA cores are isothermal at 15~K the typical 5$\sigma$ sensitivity of
0.15~Jy~beam$^{-1}$ in the SCUBA maps corresponds to total masses within the
15$''$ arcsecond beam (dust+gas) of 0.08 and 0.02~$M_\odot$ in Perseus and
Ophiuchus, respectively.

\begin{figure}
\resizebox{\hsize}{!}{\includegraphics{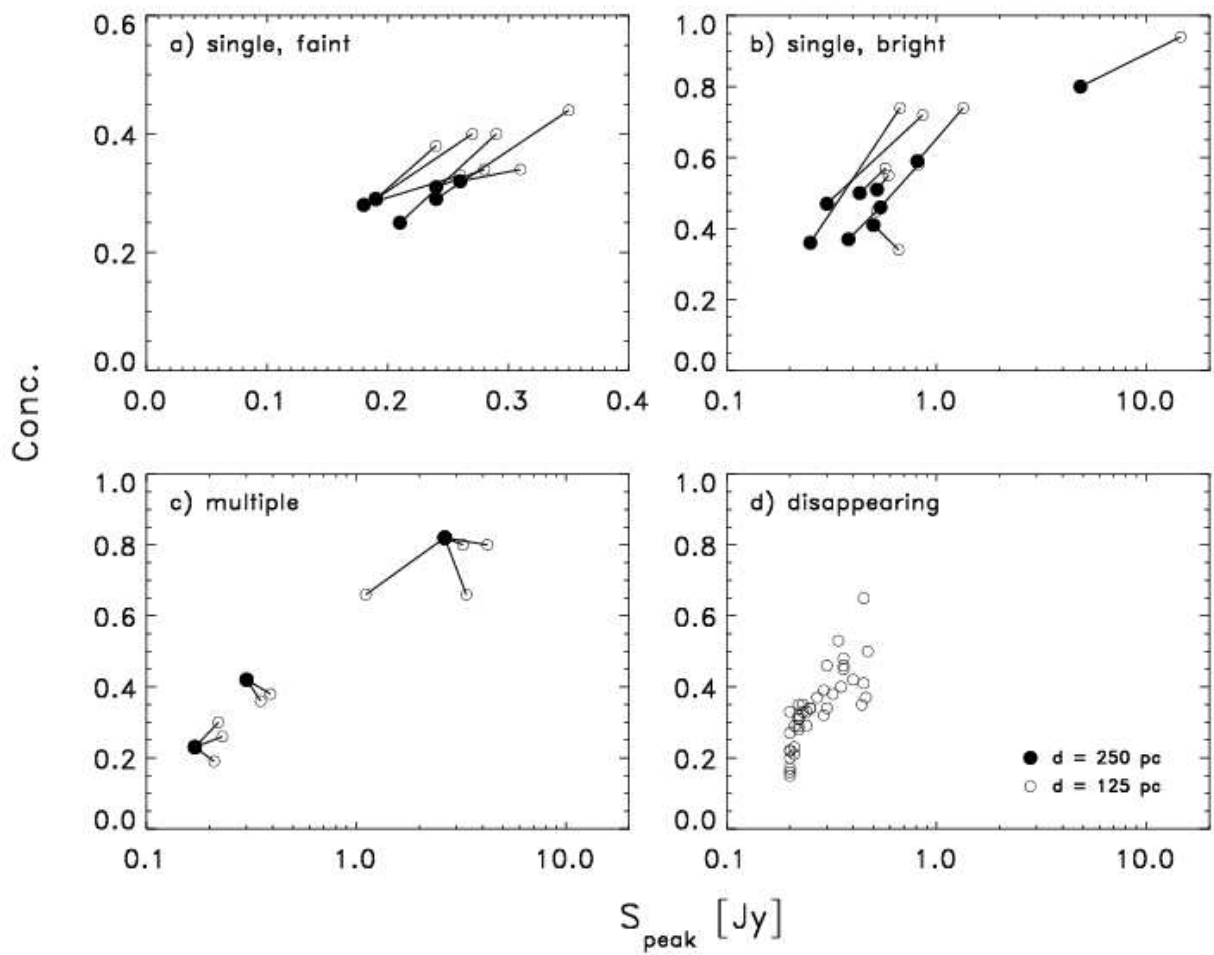}}
\caption{Change in peak fluxes and concentrations for the Ophiuchus dataset
  moved from 125 to 250~pc. In each panel open symbols represent SCUBA cores
  at 125 pc and closed symbols SCUBA cores at 250 pc. The loss in angular
  resolution makes individual core identifications ambiguous and thus all
  cores from the $d=125$~pc map which lie within one core radius of a measured
  core at $d=250$~pc are grouped together. The four panels contain: \emph{a)}
  single cores in both datasets with peak fluxes smaller than 0.5 Jy,
  \emph{b)} single cores in both datasets with peak fluxes larger than 0.5 in
  the original ($d=125$~pc) dataset, \emph{c)} systems were multiple cores are
  joined into one core in the more distant cloud and \emph{d)} and cores that
  fall below the detection limit at the larger distance.}\label{conc_change}
\end{figure}

Since many of the SCUBA cores are barely resolved by only a few beams,
resolution might affect the derived physical properties. The factor of two
better spatial resolution in Ophiuchus therefore challenges comparisons
between the physical properties of the two clouds.  Additionally, given the
gregarious nature of the submillimeter cores, nearby neighbors will be merged
into larger entities in the Perseus cloud.  To address both these issues we
simulated the case where the SCUBA map of Ophiuchus was observed at the
distance of Perseus: the Ophiuchus SCUBA maps were convolved with a beam twice
the size of the regular SCUBA beam and the map re-sampled to affect the
spatial and pixel resolution of Perseus.  Furthermore, the observed brightness
was reduced by a factor four, corresponding to the further distance.  The
Clumpfind algorithm was then applied to this new map to identify
substructures. Fig.~\ref{conc_change} compares the peak fluxes and
concentrations of the SCUBA cores in the original map with the ones in the map
simulated at twice the distance.

As expected the number of cores decreases when the cloud is moved to a larger
distance: the lower S/N causes the smallest cores to be missed and the larger
cores have their sizes reduced, since their boundaries in Clumpfind are
defined by the contour at an absolute flux level. Although the total number of
cores thereby decreases, the relative distribution of core sizes remains the
same as some of the closest cores, e.g., in the Oph~A ridge, are merged into
larger cores. For the cores with high fluxes the concentration appears to be
largely unchanged. For the fainter cores, the cores with the highest
concentrations are more likely disappearing as their peaks drop below the
detection threshold. The lower concentration faint cores in contrast have a
better chance of survival as their faint emission is more evenly distributed
and the convolution just moves emission from larger scales into the central
beam. Finally, the concentration of the cores formed by a merger of multiple
cores tends to be similar to the highest concentration core merged. Taken
together, these effects will bias against low flux, high concentration sources
at larger distances, i.e., typically more evolved sources such as embedded
objects with lower mass envelopes or disks. The SCUBA maps of Ophiuchus are
therefore expected to reveal more sources than Perseus although these will
generally be less massive and smaller. This is indeed seen when comparing the
list for Perseus \citep{kirk06} with that from Ophiuchus from this paper: the
median core radius is 7000~AU (28$''$) in Perseus vs. 3875~AU (31$''$) in
Ophiuchus and likewise the median core mass 0.90~$M_\odot$ in Perseus
vs. 0.44~$M_\odot$ in Ophiuchus.

Since, the criteria for associating a MIPS source with a SCUBA core is based
on its distance on angular and not linear scale, a natural question is whether
we instead ought to increase the radius from 15$''$ to 30$''$ for associations
according to our second criterion, ``{\bf B}''. Table~\ref{distantsources}
summarize the sources separated 15$''$--30$''$ from a SCUBA core. The numbers
in themselves argue that these are not a significant population of embedded
YSOs: only 9 MIPS sources are found in this separation range compared to the
24 within 15$''$ of the center of a SCUBA core (the latter covering a factor 3
smaller area). None of these sources have red colors fullfilling our ``{\bf
A}'' criterion or are associated with well-studied, embedded protostars. On
the other hand, it is also not reasonable to decrease the size scale in which
SCUBA cores are associated with MIPS sources in Perseus - given the intrinsic
uncertainty in the pointing of the submillimeter data (a few arcseconds) and
the size of the JCMT/SCUBA beam (15\arcsec). That most of the red MIPS sources
in fact fall within this radius (e.g., Fig.~\ref{inverse_12}-\ref{inverse_45})
and \citealt{scubaspitz}) suggests that most of scatter seen in this diagram
is due to the instrumental limitations and that the distribution in fact is
even tighter (typical separations $<$5-10$''$).

An important, and somewhat subtle, effect is that the distance difference also
skews the interpretation of the embedded protostellar sources - compared to
the pre-stellar cores. The dust continuum maps measure the optically thin
emission from dust, weighted by the temperature within the SCUBA beam. For an
optically thin envelope the temperature dependence on radius, $r$ is:
\begin{equation}
T_{\rm d}(r)=60\left(\frac{r}{2\times10^{17}~{\rm cm}}\right)^{-q}\left(\frac{L_{\rm bol}}{10^5~L_\odot}\right)^{q/2}
\label{e:chandlertemp1}
\end{equation}
where $q=2/(4+\beta)$ depends on the slope of the dust opacity law $\kappa
\propto \nu^\beta$ \citep[e.g.,][]{chandler00}. In more convenient units at a
radius corresponding to the 15$''$ size of the SCUBA beam for ISM dust with
$\beta = 2$ Eq.~\ref{e:chandlertemp1} becomes:
\begin{equation}
T_{\rm d}(r)=21\left(\frac{d}{125~{\rm pc}}\right)^{-1/3}\left(\frac{L_{\rm bol}}{1~L_\odot}\right)^{1/6}
\label{e:chandlertemp2}
\end{equation}
A protostellar envelope will therefore have a dust temperature which increases
in the central SCUBA beam as it is brought nearer (25\% higher if the distance
to the source is halved) and it will brighten significantly (by about 50\% at
half the distance). This, however, is not the case for the starless cores
which are close to isothermal (or actually decreasing in temperature toward
their centers) as the actual temperature measured in the SCUBA beam is not
varying with radius. We should therefore expect to have an easier task
detecting protostellar cores in Ophiuchus relative to Perseus - or
correspondingly, be sensitive to even lower masses of envelopes in Ophiuchus
relative to Perseus. This will bias the results in a relative sense towards a
situation with fewer low-mass protostellar sources compared to pre-stellar
cores in the more distant cloud, i.e., Perseus.

The improved sensitivity in the nearer cloud will also cause circumstellar disks
to start ``contaminating'' our sample. \cite{andrews05} surveyed a sample of
isolated sources in Taurus with SCUBA, including a number of well-known
classical T-Tauri stars, or ``Class II'' \citep{lada87}, YSOs . They found that
35\% of those more evolved sources had fluxes larger than 100~mJy (about the
3$\sigma$ sensitivity limit of our SCUBA maps) whereas less than 10\% had fluxes
larger than 500~mJy. The distance to Taurus ($d=140$) and Ophiuchus ($d=125$)
are comparable and thus we would expect to pick up some more evolved, ``Class
II'' young stellar objects without envelopes in the Ophiuchus SCUBA data. In
contrast the 10\% cut-off from Taurus would be moved to a flux of 125~mJy in
Perseus which would be very close to the sensitivity limit of our SCUBA maps.

\clearpage
{\scriptsize \begin{table}
  \caption{List of candidate embedded YSOs in Ophiuchus. \label{embeddedysolist}}
{\tiny\begin{tabular}{lllllllll} \hline\hline
RA (J2000)\tablenotemark{a} & DEC (J2000)\tablenotemark{a}   & Conc.\tablenotemark{b} & $[3.6]-[4.5]$ & $[8.0]-[24]$ & Sep.\tablenotemark{c} & SCUBA flux\tablenotemark{d} & Code\tablenotemark{e} & Common identifiers\tablenotemark{f} \\
hh mm ss.ss & dd mm ss.s   &     &  &  & [$''$] & [Jy~beam$^{-1}$] &  & \\ \hline
16 26 10.32 &  -24 20 54.6 &  0.65 &  0.81         &   2.66   &   3.4  &   \phantom{0}0.45  & -BC & $\bullet$ GSS~26  \\ 
16 26 14.63 &  -24 25 07.5 &  0.34 &  $\ldots$     & $\ldots$ &   7.8  &   \phantom{0}0.28  & AB- & $\ldots$ \\  
16 26 17.23 &  -24 23 45.1 &  0.32 &  1.20         &   3.23   &   3.1  &   \phantom{0}0.29  & -B- & $\bullet$ CRBR~12 / ISO-Oph 21 \\  
16 26 21.35 &  -24 23 04.3 &  0.58 &  1.31         & $\ldots$ &  10.2  &   \phantom{0}0.82  & AB- & $\bullet$ Elias~21 / GSS~30-IRS1\tablenotemark{g} \\  
16 26 24.07 &  -24 16 13.4 &  0.41 &  0.54         &   2.30   &   2.5  &   \phantom{0}0.45  & -B- & BKLT J162624-241616 / Elias 24 \\  
16 26 25.46 &  -24 23 01.3 &  0.66 &  1.06         &   5.33   &  35.2  &   \phantom{0}0.34  & A-- & $\bullet$ CRBR~2324.1-1619 / [GY92]~30\\
16 26 25.62 &  -24 24 28.9 &  0.80 &  2.12         &   6.77   &  11.5  &   \phantom{0}3.23  & ABC & GDS J162625.6-242429 \\     
16 26 26.42 &  -24 24 30.0 &  0.80 &  $\ldots$     &$\ldots$  &   1.0  &   \phantom{0}3.23  & ABC & $\bullet$ VLA~1623 \\  
16 26 40.46 &  -24 27 14.3 &  0.53 &  0.91         &   4.07   &   1.9  &   \phantom{0}0.34  & -B- & $\bullet$ BKLT~J162640-242715 / [GY92]~91 \\  
16 26 44.19 &  -24 34 48.4 &  0.29 & 1.72          & $\ldots$ &   3.2  &   \phantom{0}0.24  & AB- & $\bullet$ WL~12 / ISO-Oph 65 \\
16 26 45.02 &  -24 23 07.6 &  0.74 &  0.76         &   2.68   &   3.1  &   \phantom{0}0.67  & -BC & $\bullet$ BKLT J162645-242309 / Elias~27 \\
            &              &       &               &          &        &          &     &  / GSS~39 / VSSG~28 \\  
16 26 58.42 &  -24 45 31.8 &  0.42 &  0.52         &   2.53   &   6.6  &   \phantom{0}0.40  & -B- & SR~24N / Elias~28 \\
16 26 59.10 &  -24 35 03.3 &  0.55 &  1.2          &   4.9    &  38.6  &   \phantom{0}0.17  & A-- & $\bullet$ WL~22 / ISO-Oph 90 \\
16 27 05.24 &  -24 36 29.6 &  0.34 &  1.09         &   5.36   &   1.6  &   \phantom{0}0.25  & AB- & $\bullet$ CRBR~2403.7-2948 / [GY92]~197 \\  
16 27 06.75 &  -24 38 14.8 &  0.22 &  1.08         &   3.89   &   4.6  &   \phantom{0}0.20  & -B- & CRBR~2404.8-3133 / WL~17 \\  
16 27 09.40 &  -24 37 18.6 &  0.48 &$\ldots$       & $\ldots$ &   1.4  &   \phantom{0}0.36  & AB- & $\bullet$ Elias~29 / WL~15 \\
            &              &       &               &          &        &          &     &  / CRBR~2407.8-3033 \\
16 27 26.91 &  -24 40 50.7 &  0.57 &  1.22         & $\ldots$ &   3.8  &   \phantom{0}0.57  & AB- & $\bullet$ YLW15 / IRS43 / IR16244-2434 \\
16 27 27.99 &  -24 39 33.4 &  0.37 &  1.48         & $\ldots$ &   1.0  &   \phantom{0}0.27  & AB- & $\bullet$ YLW16 / IRS44 / IR16244-2432 \\
16 27 28.44 &  -24 27 21.0 &  0.34 &  0.89         &   2.81   &  12.5  &   \phantom{0}0.66  & -B- & [AMD2002]~J162728-242721 \\  
16 27 30.17 &  -24 27 43.2 &  0.50 &  1.00         &   2.20   &   9.7  &   \phantom{0}0.47  & -B- & $\bullet$ Elias~33 / VSSG~17 \\  
16 27 39.81 &  -24 43 15.0 &  0.27 &  0.70         &   3.40   &   1.9  &   \phantom{0}0.20  & -B- & $\bullet$ IRS51 / IR16246-2436 \\ 
16 28 21.61 &  -24 36 23.4 &  0.39 &  2.08         &   6.03   &   2.8  &   \phantom{0}0.29  & AB- & ($\bullet$) [SSG2006]~MMS126 / IR16253-2429\\[2.0ex] 
16 31 35.65 &  -24 01 29.3 &  0.72 &  1.01         &   3.25   &   2.6  &   \phantom{0}0.86  & -BC & GWAYL~4 / IR16285-2355 \\  
16 31 52.45 &  -24 55 36.2 &  0.40 &  1.21         &   4.59   &  30.5  &   \phantom{0}0.12  & A-- & ISO-Oph 203 \\
16 32 00.99 &  -24 56 42.6 &  0.45 &  1.46         & $\ldots$ &   1.4  &   \phantom{0}0.36  & AB- & [B96]~L1689S1~3 / ISO-Oph 209 \\
16 32 22.63 &  -24 28 31.8 &  0.94 &  $\ldots$     &$\ldots$  &   6.5  &             14.6\phantom{0}   & ABC & IR16293-2422 \\  
16 33 55.60 &  -24 42 05.0 &  0.33 &  0.15         &   4.49   &   1.7  &   \phantom{0}0.20  & -B- & RX J1633.9-2442 \\ \hline
\end{tabular}}
\tablenotetext{a}{Position of Spitzer source from c2d catalog.} 
\tablenotetext{b}{Concentration of the nearest SCUBA core.}
\tablenotetext{c}{Separation between the Spitzer source and SCUBA
core.}
\tablenotetext{d}{For sources with code ``A--'' the SCUBA flux refers to flux density
at the exact position in the maps; for the remaining sources to the peak flux
from the Clumpfind algorithm.}
\tablenotetext{e}{Identifier indicating whether the given
source obeys ({\bf A:} MIPS~24~$\mu$m source with $[3.6]-[4.5]>1$ and
$[8.0]-[24] > 4.5$, {\bf B:} separation between Spitzer source and nearest SCUBA
core less than 15$''$ and {\bf C:} Associated submillimeter core has
concentration higher than 0.6). Note only {\bf A} and {\bf B} are used to
construct this sample.} 
\tablenotetext{f}{List of common identifiers from the SIMBAD database
(IR=IRAS). Sources indicated with ``$\bullet$'' are included in the list of embedded
YSOs by \cite{ophhandbook}.} 
\tablenotetext{g}{A complex system of at least three
protostellar candidates: in addition to GSS30-IRS1 a nearby source is seen at
16:26:22.38; -24:22:52.9 at a distance of 10.4$''$ from the SCUBA core:
[AMD2002] J162622-242254 / VSSG~12 / ISO-Oph 34 / GSS~30-IRS2. A third source,
$\bullet$~GSS~30-IRS3 / LFAM~1, is at 16:26:21.72 -24:22:50.5, separated by 4.5$''$ and
with $[3.6]-[4.5] = 0.9$ and $[8.0]-[24]=5.3$ (although with low S/N in MIPS1
due to confusion with GSS30-IRS1).}
\end{table}

\begin{table}
\caption{As Table~\ref{embeddedysolist} for SCUBA cores in Ophiuchus with high concentrations and no embedded YSOs (see also Table~\ref{submmtab}).}
{\scriptsize\begin{tabular}{lllllllll} \hline\hline
Name (SMM J)& RA (J2000) & DEC (J2000)   & Conc. &   Sep.  & SCUBA flux & Common identifiers \\ \hline
162627-24233 & 16 26 27.34 &  -24 23 33.9 &  0.66 &   19.6  &    3.350  & \\ 
162628-24235 & 16 26 27.55 &  -24 23 54.9 &  0.80 &   25.8  &    4.220  & \\  
162628-24225 & 16 26 28.00 &  -24 22 54.9 &  0.66 &   35.2  &    1.110  & [MAN98] A-MM6 \\  
163229-24291 & 16 32 29.06 &  -24 29 07.2 &  0.74 &   94.7  &    1.340  & IRAS~16293E \\ \hline
\end{tabular}}
\end{table}

\begin{table}
  \caption{As Table~\ref{embeddedysolist} for interesting MIPS sources with very red colors, but no associated SCUBA flux (e.g., candidate edge-on disks).\label{disks}.}
{\scriptsize\begin{tabular}{lllllllll} \hline\hline
RA (J2000) & DEC (J2000)   & Conc. & $[3.6]-[4.5]$ & [8.0]-[24] & Sep. & SCUBA flux & Code & Common identifiers \\ \hline
16 21 45.13 &  -23 42 31.6  & 0.34 &  1.03         &   5.25  &  4256.5 &    0.033 & A-- & ... \\
16 21 59.55 &  -23 16 02.5  & 0.41 &  1.11         &   4.86  &  5122.8 &    0.007 & A-- & ... \\
16 27 38.93 &  -24 40 20.5  & 0.31 &  1.05         &   4.73  &   136.8 &    0.007 & A-- & BKLT J162738-244019 \\
16 27 48.23 &  -24 42 25.4  & 0.27 &  1.14         &   5.17  &   123.5 &    0.005 & A-- & ... \\
16 31 52.06 &  -24 57 26.0  & 0.46 &  1.22         &   4.69  &    69.3 &   -0.026 & A-- & ISO-Oph 202 \\
16 31 56.87 &  -24 54 03.2  & 0.40 &  1.36         &   4.50  &   124.2 &   -0.026 & A-- & ... \\ \hline
\end{tabular}}
\end{table}

\begin{table}
  \caption{As Table~\ref{embeddedysolist} for sources with separations of
    15$''$-30$''$ of a MIPS source.\label{distantsources}.}
{\scriptsize\begin{tabular}{lllllllll} \hline\hline
RA (J2000) & DEC (J2000)   & Conc. & $[3.6]-[4.5]$ & [8.0]-[24] & Sep. & SCUBA flux & Code & Common identifiers \\ \hline
16 26 25.99 & -24 23 40.5 & 0.66 & 1.86 & 2.47 &  19.6 & 0.57         & $\ldots$ \\
16 27 10.02 & -24 29 13.1 & 0.21 & 0.43 & 4.97 &  24.5 & 0.12         & [GY92]~218 \\
16 27 11.82 & -24 39 47.6 & 0.29 & $\ldots$ & $\ldots$ &  26.1 & 0.11 & $\ldots$ \\
16 27 15.50 & -24 30 53.6 & 0.34 & 1.03 & 2.81 &  30.0 & 0.29         & [GY92]~238 \\
16 27 30.91 & -24 27 33.2 & 0.50 & 1.64 & 3.04 &  20.7 & 0.38         & $\bullet$ [AMD2002] J162730-242734 / ISO-Oph 150\\
16 27 37.23 & -24 42 37.9 & 0.31 & 0.96 & 3.62 &  21.0 & 0.12         & [GY92]~301\\
16 27 37.88 & -24 42 10.8 & 0.31 & 1.21 & $\ldots$ &  29.1 & 0.04     & $\ldots$ \\
16 27 41.02 & -24 43 32.4 & 0.27 & $\ldots$ & $\ldots$ &  24.7 & 0.02 & $\ldots$ \\
16 31 33.52 & -24  3 34.5 & 0.35 & 0.14 & 1.38 &  23.2 & 0.12         & $\ldots$ \\
16 31 34.29 & -24  3 25.2 & 0.35 & 0.73 & 2.58 &  26.0 & 0.14         & $\ldots$ \\
16 31 52.11 & -24 56 15.8 & 0.40 & 0.90 & 2.95 &  28.1 & 0.20         & L1689-IRS5 \\ \hline
\end{tabular}}
\end{table}}

\clearpage

\subsection{Comparison to other mid-infrared and submillimeter surveys}\label{submmcomparison}
While the Spitzer observations of Ophiuchus are unprecedented at mid-infrared
wavelengths in terms of the area covered and sensitivity, a number of surveys
have covered similar areas of the cloud at (sub)millimeter
wavelengths. Besides JCMT/SCUBA observations
\citep{johnstone00maps,visser02,johnstone04,nutter06}, Ophiuchus has been
surveyed at 1.1--1.2~mm using the MAMBO bolometer on the IRAM 30~m telescope
\citep{motte98}, the SIMBA bolometer on the SEST telescope \citep{stanke06}
and Bolocam on the Caltech Submillimeter Observatory 10.4 meter telescope
\citep{young06}.

Our number of cores, 66, is somewhat lower than the 100 submillimeter cores
originally reported by \cite{johnstone04}: \citeauthor{johnstone04} went down
to a much deeper depth (0.02~Jy~beam$^{-1}$) than we have done here mainly to
confirm that their main result (the absence of cores at low $A_V$) was
not a result of a biased look at only the brightest cores. We do not include
these fainter sources here for consistency with the \cite{kirk06} and
\cite{scubaspitz} studies of Perseus.

\cite{nutter06} surveyed L1689B in the submillimeter, using a subset of the
SCUBA 850~$\mu$m data presented here. By going to a lower noise level, they
identified about twice as many dust continuum cores compared with the list
used in this paper. These ``extra'' cores are predominantly fainter emission
stretching over arcminute scales. For the brighter, more concentrated cores,
there is a direct one-to-one mapping between our sample and that of
\citeauthor{nutter06}. Likewise none of the ``extra'' cores in the list of
\citeauthor{nutter06} have associated MIPS sources within 30$''$ - consistent
with their starless nature as also found by \citeauthor{nutter06}.

\cite{ophhandbook} listed a sample of 35 embedded YSOs in L1688 based on near-
and mid-infrared surveys, in particular, the ISO survey of Ophiuchus by
\cite{bontemps01}. Of these sources 15 (16 if one includes LFAM~1 which is a
faint companion to GSS30-IRS1 in our list; see footnote to
Table~\ref{embeddedysolist}) are common with our list (22 of the objects from
our list associated with L1688). \citeauthor{ophhandbook} also mentions
[SSG2006]~MMS126 as a potential Class 0 source, although it is not included in
their list. The list also includes sources that are associated with
well-studied YSOs, including IRS~46 \citep{lahuis06} and IRS~48
\citep{geers07} that both show little submillimeter emission and have been
suggested to be disk sources. Two sources from the list are associated with
regions of strong submillimeter emission but are likely also disk candidates:
one is CRBR~2422.8-3423 modeled in detail by \cite{pontoppidan05crbr} who
suggested that it was a disk seen edge-on. It is not associated with a
separate peak in the SCUBA map (the emission being dominated by the nearby
protostar, YLW~15 or IRS~43) although with a flux of 0.39~Jy~beam$^{-1}$ at
the location of the MIPS source and colors, $[3.6]-[4.5]=1.4$ and
$[8.0]-[24]=4.45$, it just misses the criteria in this paper. ISO-Oph 150
(IRAC~30 in the list of \cite{ophhandbook}) shows similar red $[3.6]-[4.5]$
colors and likewise has a SCUBA flux of 0.37~Jy~beam$^{-1}$, but even bluer
$[8.0]-[24]$ colors than CRBR~2422.8-3423. It is possible that similar sources
are present in our list, but the only way to systematically filter such
sources would be either high angular resolution observations (at mid-infrared
or submillimeter wavelengths) - or spectroscopical
confirmation. Table~\ref{disks} lists the most prominent candidates edge-on
disk sources with no associated SCUBA emission\label{disksources}. The
remaining sources in the \cite{ophhandbook} are found to show little if any
submillimeter emission and/or $[3.6]-[4.5]$ and $[8.0]-[24]$ colors that do
not full-fill our criteria, but are bluer. This would suggest that these
sources are either only slightly embedded or perhaps more evolved sources
extincted by a large amount of foreground material.

Two of the sources in our list, but which are not included in the list of
\cite{ophhandbook}, BKLT~J162624-241616/Elias~24 and SR~24N/Elias~28, are
optically visible and have optically-determined spectral types
\citep[e.g][]{wilking05,struve49}. These sources do not show particularly red
mid-IR colors but are located within a few arcseconds of the center of SCUBA
cores, SMM~J162624-24162 and SMM~J162659-24454, respectively. Both of these
SCUBA cores have effective radii less than 15$''$ and masses less than
0.1~$M_\odot$. This suggests that these sources only have tenuous envelopes
and that the dust continuum emission largely has its origin in their
circumstellar disks.

One source in L1689 selected based on our criteria, SSTc2d J163355.6-244204,
is likely also a more evolved object. It is associated with an X-ray source,
RX~J1633.9-2422, and although it is located only 1.7$''$ from the center of a
faint SCUBA core (0.2~Jy), its colors do not obey the criteria for
embeddedness. The SCUBA core also appears unresolved, consistent with it being
submillimeter emission from a compact disk. This illustrates that the ``ABC
classification scheme'' should be taken as a gradation of the likely
embeddedness of a given source: the more criteria passed for a given source,
the greater likelihood that source is embedded.

\clearpage

\subsection{Extended cloud emission and absorption of mid-IR emission}\label{extendedemission}
As discussed by \cite{padgett08}, the MIPS 24~$\mu$m images of Ophiuchus show
large swatches of variable brightness extended emission across the cloud.  In
particular, \citeauthor{padgett08} draw attention to significant dark patches
where up to 80\% of the extended background emission is absorbed (their
Fig.~4). Recently \cite{stutz07} analyzed Spitzer images of a similar shadow
of the Bok Globule CB~190.

\begin{figure}
\resizebox{0.5\hsize}{!}{\includegraphics{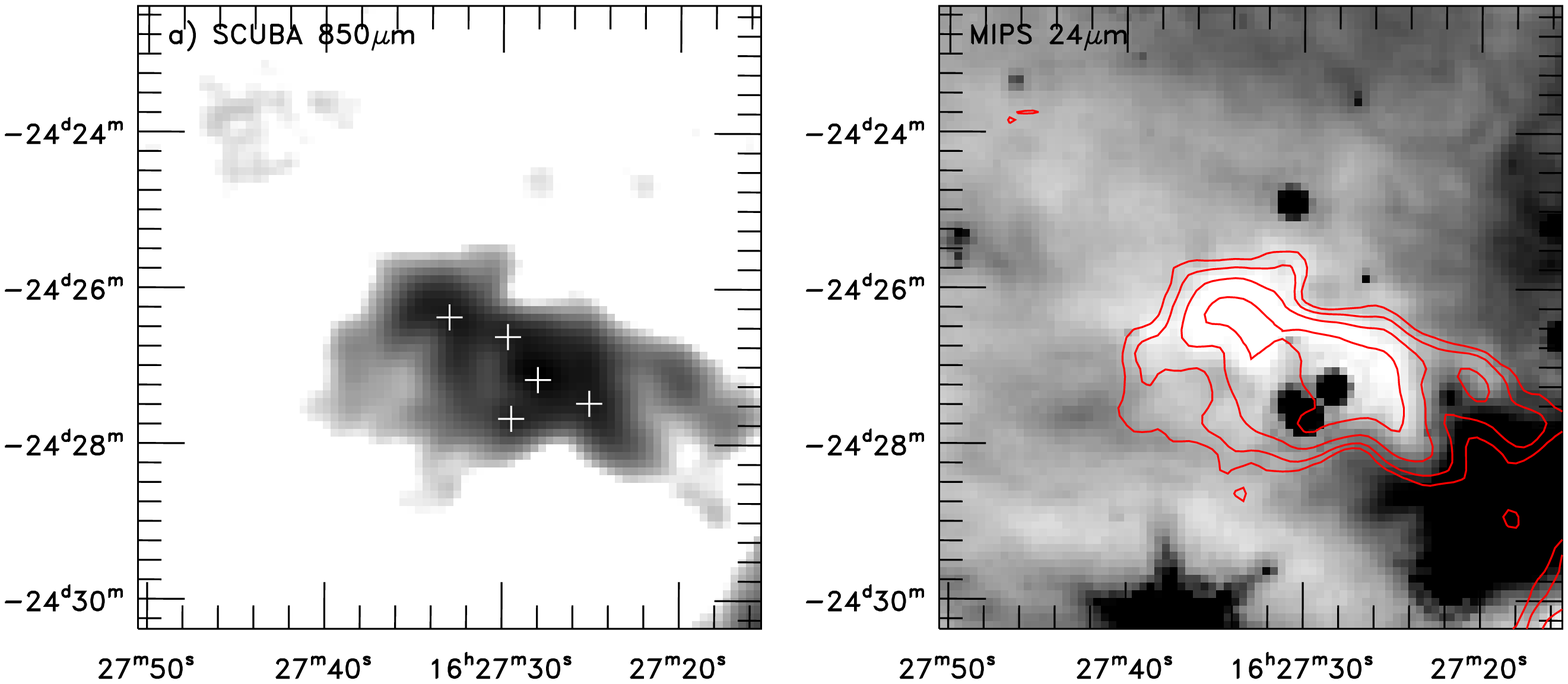}}
\resizebox{0.5\hsize}{!}{\includegraphics{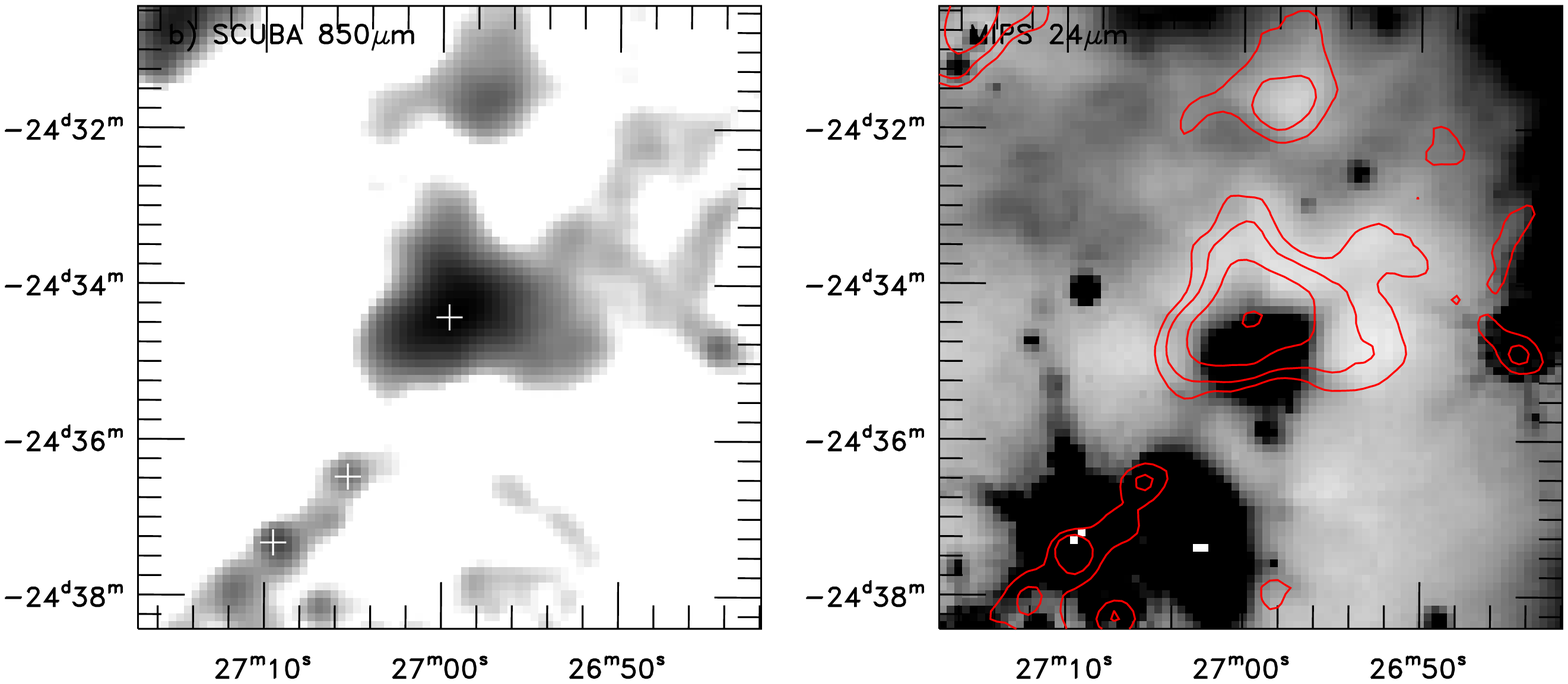}}
\resizebox{0.5\hsize}{!}{\includegraphics{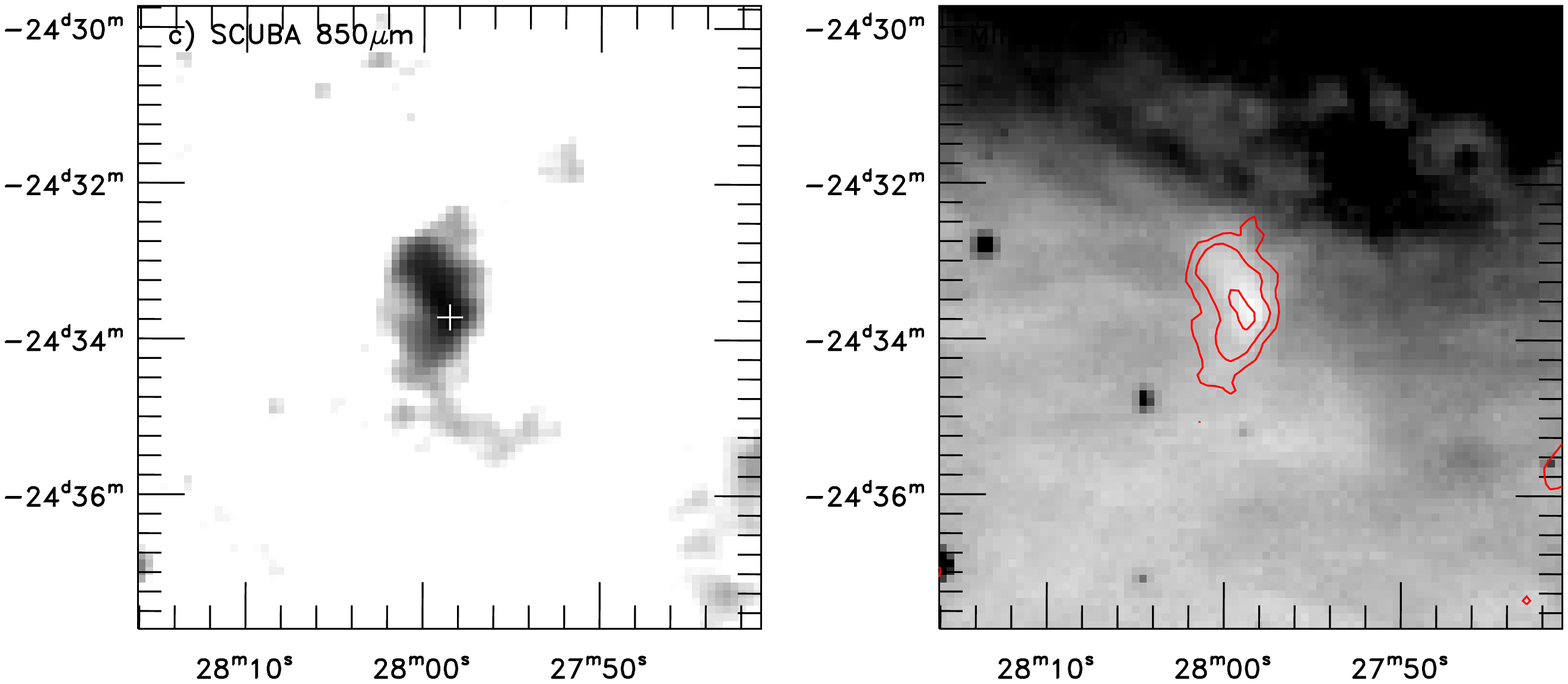}}
\resizebox{0.5\hsize}{!}{\includegraphics{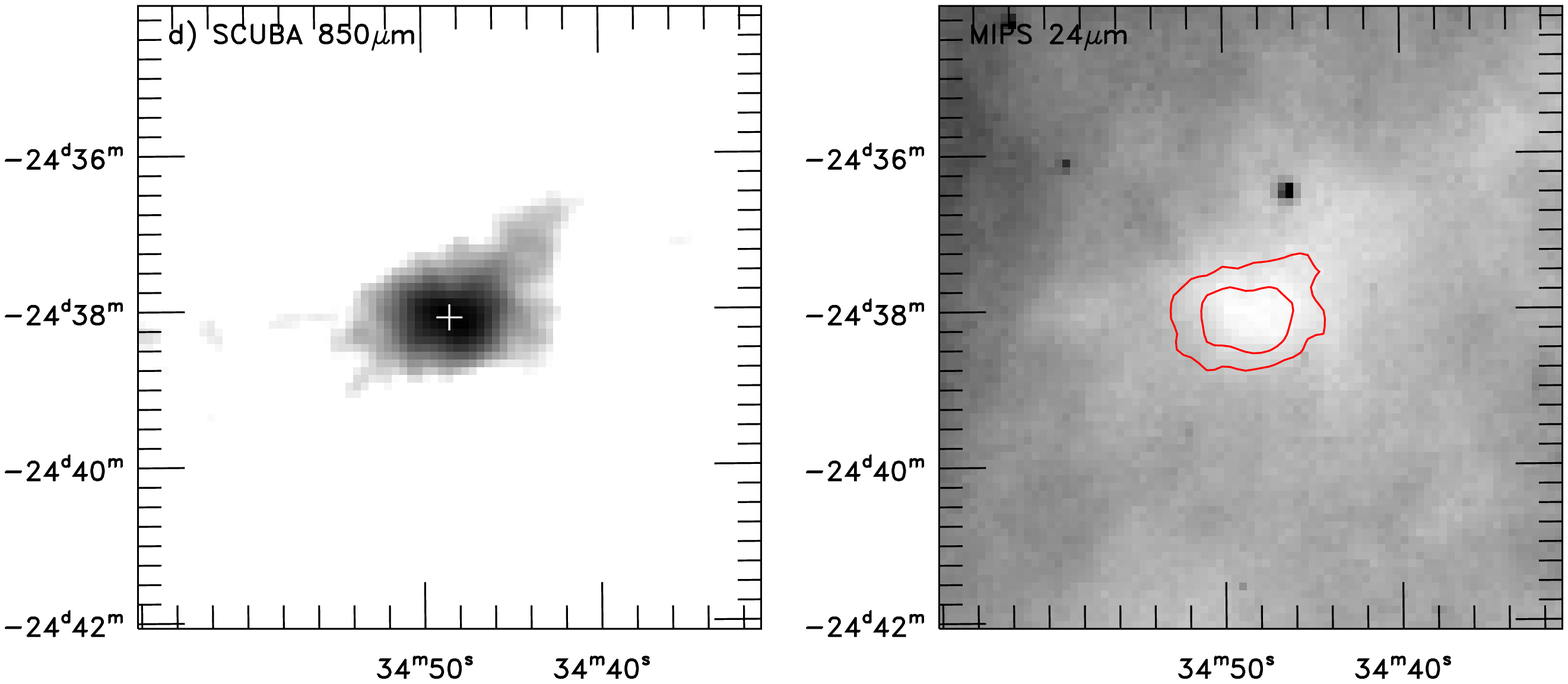}}
\caption{24~$\mu$m shadowed cores in the SCUBA 850~$\mu$m maps (left) and MIPS
  24~$\mu$m maps (right). In both panels dark is high intensity of the
  emission and white is low intensity. The contours in the righthand panels
  indicate the SCUBA 850~$\mu$m emission. The location of the identified SCUBA
  cores (Table~\ref{submmtab}) are indicated by white plus-signs in the
  righthand panels: in \emph{a)} SMM J162725-24273, J162728-24271,
  J162729-24274, J162730-24264, J162733-24262, \emph{b)} SMM J162660-24343,
  J162705-24363, J162709-24372, \emph{c)} SMM J162759-24334 and \emph{d)} SMM
  J163448-24381.}
  \label{shadows}
\end{figure}

Fig.~\ref{shadows} compare the 24~$\mu$m images to SCUBA continuum maps for
four cores seen across the field. The resemblance between the shape of the
core absorption shadows at 24~$\mu$m and their emission profiles at 850~$\mu$m
illustrate (1) the good correspondence between the resolution at these two
wavelengths (6$''$ at 24~$\mu$m and 15$''$ at 850~$\mu$m) and (2) that these
cores contain very dense material. A decrease in flux by 80\% corresponds to
an optical depth of the 24~$\mu$m shadow of 1.6 or a visual extinction $A_V =
35$ under the same assumptions as \citeauthor{stutz07} and assuming that the
cores are in the foreground to the extended emission. In reality both zodiacal
light and foreground emission (i.e., emission from cloud material in which the
core is embedded) may contribute to the observed profile - and the derived
extinction is therefore a lower limit to the actual extinction toward the core
profiles. For comparison, for a temperature of 10~K the typical SCUBA peak
850~$\mu$m fluxes of these cores of 0.2--0.8~Jy~beam$^{-1}$ correspond to an
extinction of $A_V=30-70$ averaged over the 15$''$ SCUBA beam.
\clearpage
\clearpage

\section{Characterizing the current star formation populations}\label{comparison}
\subsection{Motions of YSOs and dissipation of cores}
As discussed by, e.g., \cite{harvey07}, the Spitzer data provides large
samples of YSOs that can be classified in the traditional scheme for the
evolution based on their spectral indices at near- and mid-infrared
wavelengths, $\alpha$\footnote{The spectral index, $\alpha$, where $\nu F_\nu
\propto \nu^\alpha$ is calculated in the c2d catalogs from a least square fit
to the observations from 2.2 to 24~$\mu$m. Following \cite{greene94}, Class I
sources have $\alpha > 0.3$, ``Flat Spectrum'' sources $-0.3 < \alpha \le 0.3$
and Class II sources $-1.6 < \alpha \le -0.3$.}. This scheme does not directly
capture differences within the group of embedded sources which are typically
classified into ``Class 0'' and ``Class I'' according to their submillimeter
signatures (see discussion in \S~\ref{physorigin}), e.g., their submillimeter
relative to total luminosities. Such a distinction therefore requires fully
sampled SEDs as presented, e.g., by \cite{hatchell07} and
\cite{enoch08}. Still, since we expect almost all Class 0 sources to be
detected at 24~$\mu$m and these sources per definition are associated with
submillimeter cores, our sample of embedded YSOs is expected to be complete
with respect to these sources. For the remainder of this section we therefore
use the term ``Class~I'' in the sense of the $\alpha$-classification scheme
which thereby also encompass the more deeply embedded sources.

\begin{figure}
\resizebox{\hsize}{!}{\includegraphics{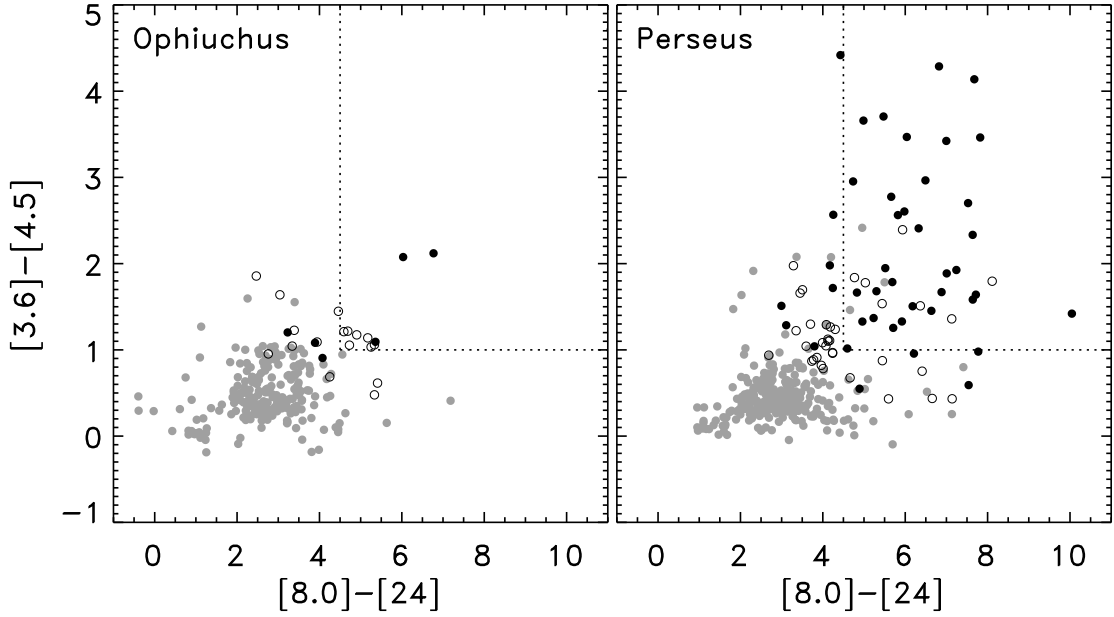}}
\caption{$[3.6]-[4.5]$ vs. $[8.0]-[24]$ color-color diagram for the YSOs
  identified by the c2d color criteria. The ``evolved'' YSOs (``Flat
  spectrum'', Class II and Class III) have been shown with
  grey symbols whereas the Class I YSOs are shown with black
  symbols. The Class I YSOs have furthermore been divided into those
  that are (filled symbols) and are not (open symbols) located within 15$''$
  of a SCUBA core. The dotted lines indicate the color criteria for embedded
  YSOs ($[3.6]-[4.5] > 1.0$ and $[8.0]-[24] > 4.5$). The Class I
  sources that are not associated with SCUBA cores are found close to the more
  evolved YSOs in this diagram, suggesting that these sources are less
  embedded and have envelopes with masses below the detection limit of the
  SCUBA survey (0.1~$M_\odot$).}\label{classinosubmm}
\end{figure}

The distribution of the more evolved YSOs identified by the Spitzer data
(e.g., ``Flat spectrum'', Class~II) compared to the SCUBA cores provides
insight into the dynamical properties of star
formation. Table~\ref{dissipation} lists the number of YSOs in each Class for
both Perseus and Ophiuchus from the classification given in the catalogs from
the final delivery of c2d data \citep{delivery4} and gives the fraction which
are located within 15$''$ and 30$''$ of a SCUBA core relative to the total
number of YSOs within the area covered by SCUBA. The fraction of embedded YSOs
associated with submillimeter cores are clearly decreasing from more than half
to less than a few percent in the Class~II stages. The numbers suggest both
that Class~II sources are genuine envelope-less sources and that about
40--50\% of the Class~I sources do not have envelopes more massive than the
sensitivity limit of the SCUBA survey, i.e., about 0.1~$M_\odot$. This is
further illustrated in Fig.~\ref{classinosubmm}, which shows the $[3.6]-[4.5]$
vs. $[8.0]-[24]$ color-color diagrams for Ophiuchus and Perseus. In both
diagrams the evolved YSO population based on their $\alpha$ (``Flat
spectrum'', Class II and Class III) has been separated from the Class I
sources. The latter group has furthermore been divided into Class I sources
that are or are not within 15$''$ from the center of a SCUBA
core. Fig.~\ref{classinosubmm} shows that the Class I sources in Perseus that
are not associated with a SCUBA core are located close to the more evolved
population of YSOs in this diagram, i.e., typically have less red colors than
those that do lie within 15$''$ of a core. This again suggests that these
sources are more evolved and although they may be associated with envelopes
providing their increasing SEDs in the mid-infrared, those envelopes have
masses smaller than the sensitivity limit of the SCUBA survey. On the other
hand, the population of red sources associated with SCUBA cores seen
prominently in the Perseus panel is absent in the Ophiuchus panel, arguing in
favor of the more evolved/less embedded nature of the Class I sources in
Ophiuchus.

Naturally per definition most (89\%) of the sources in the sample constructed
in this paper (Table~\ref{embeddedysolist}) are associated with SCUBA cores
with the few exceptions being red sources in region of crowded submillimeter
emission. This sample is expected to contain both Class 0 and I protostellar
sources.

\clearpage
\clearpage

\begin{table}
\caption{Number of mid-infrared classified YSOs within 15$''$ or 30$''$ within SCUBA cores and in total.}\label{dissipation}
\begin{tabular}{lccc@{\extracolsep{0.5cm}}ccc} \hline \hline
      & \multicolumn{3}{c}{Ophiuchus (265 total)}   & \multicolumn{3}{c}{Perseus (353 total)} \\\cline{2-4}\cline{5-7}
Class & Within 15$''$ & Within 30$''$ & Total & Within 15$''$ & Within 30$''$ & Total \\ \hline
I     &   15    (47\%) & 17    (53\%) &  32 & 49    (58\%) & 55    (66\%) &  84 \\
Flat spectrum  &    4   (9.1\%) &  8    (18\%) &  44 &  4   (9.5\%) &  6    (14\%) &  42 \\
II    &    5   (3.1\%) &  5   (3.1\%) & 163 &  2   (1.0\%) &  6   (3.0\%) & 199 \\
III   &    0   (0.0\%) &  0   (0.0\%) &  26 &  0   (0.0\%) &  1   (3.6\%) &  28 \\[2.0ex]
\emph{This paper\tablenotemark{a}}    &    24    (89\%) & 24    (89\%) &  27 & 50    (94\%) & 51    (96\%) &  53 \\\hline
\end{tabular}
\tablenotetext{a}{Samples of embedded objects from this analysis.}
\end{table}
\clearpage

The decrease in fraction of sources associated with submillimeter cores in the
later YSO classes can be interpreted in two ways: either the envelopes around
the embedded sources dissipate over the duration of the Class I stage or the
YSOs themselves disperse (i.e., move away from their parental cores through
their evolution). However, although the number of red YSOs in Ophiuchus is
lower than in Perseus these are still strongly clustered within 15$''$
(\S\ref{midirsubmm} and Fig.~\ref{clustplot}). As discussed in \cite{perspitz}
this in itself suggests that very little dispersal of the YSOs take place
during their evolution through their embedded stages. In fact, this is likely
to be expected given the required escape velocity from centers of the SCUBA
cores: the median mass of $M_0=$0.44~$M_\odot$ and radius of $r_0=$31$''$
(3875~AU) of the cores from Table~\ref{submmtab} corresponds to an escape
speed from the center of a constant density core to infinity, $v=\sqrt{3 G M /
r_0}$, of 0.5~km~s$^{-1}$ - about 5 times the sound speed in a molecular cloud
with a temperature of 10--15~K. As the density profile of the SCUBA cores are
centrally condensed (and the escape velocity thus higher) it therefore shows
that highly supersonic motions would be required at the birth of the central
protostar in order for it to escape its parental core during the embedded
protostellar stages.

\subsection{Nearest neighbor surface and volume density maps}
\begin{figure}
\resizebox{0.3\hsize}{!}{\includegraphics{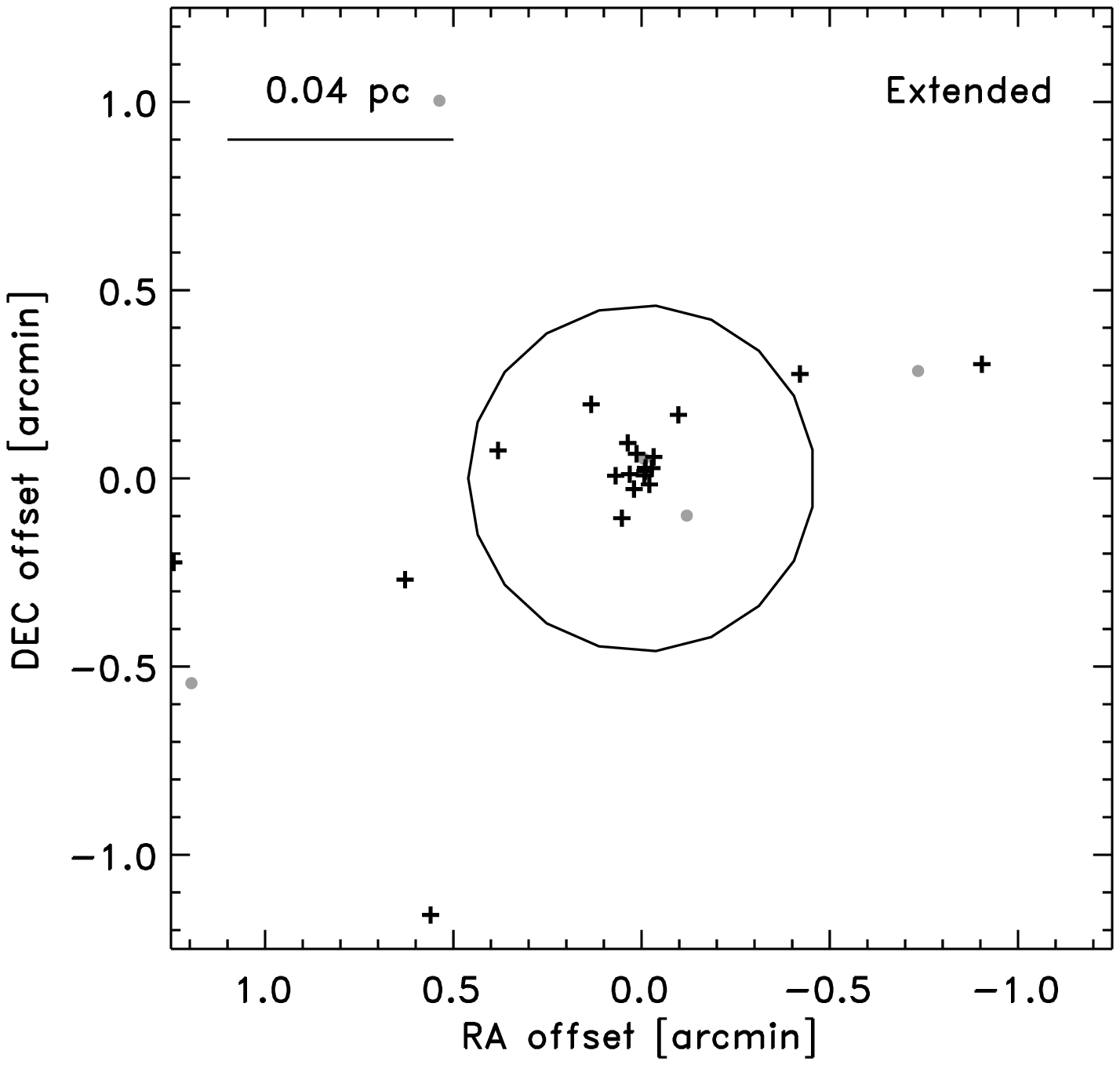}}
\resizebox{0.3\hsize}{!}{\includegraphics{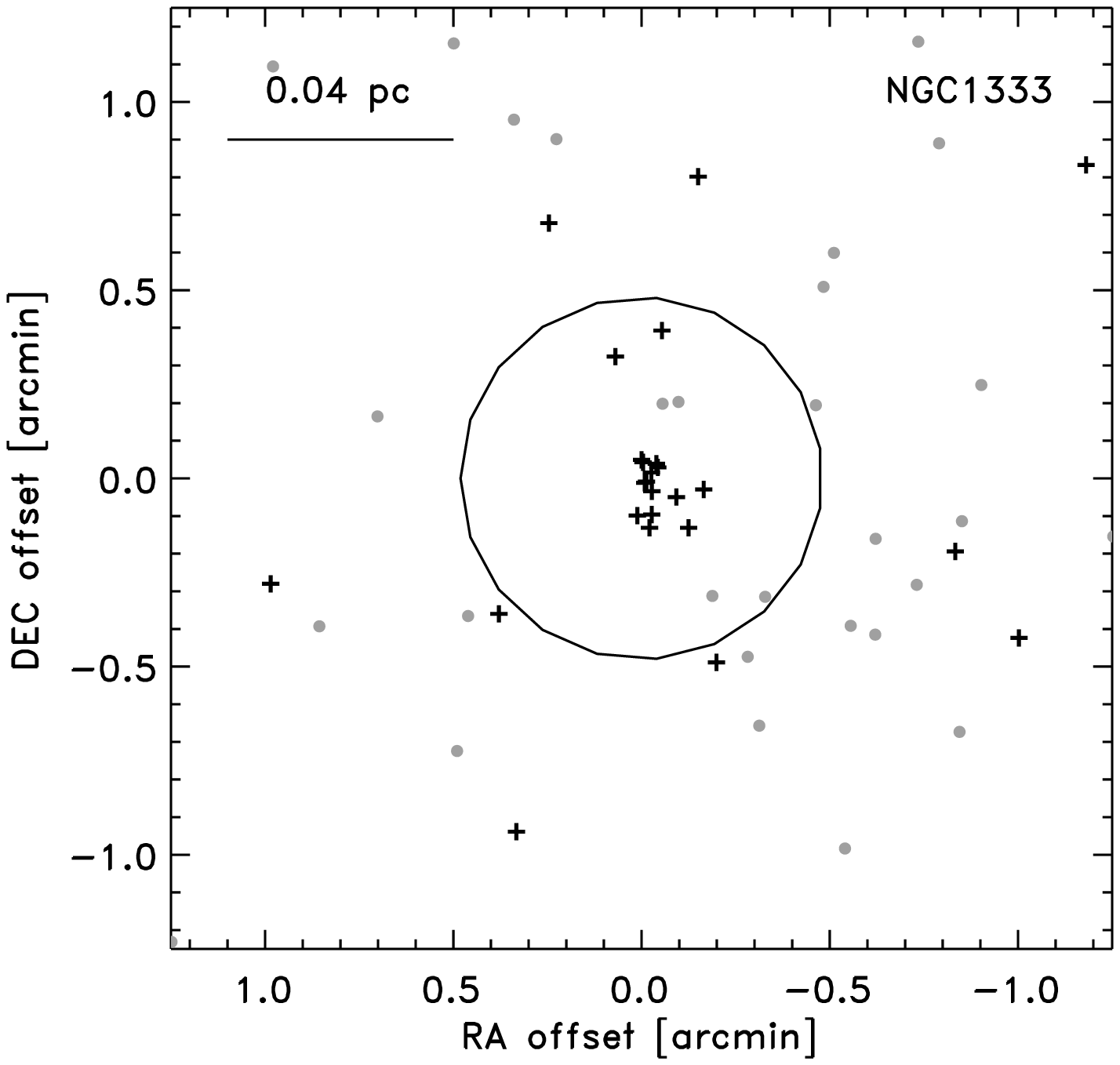}}
\resizebox{0.3\hsize}{!}{\includegraphics{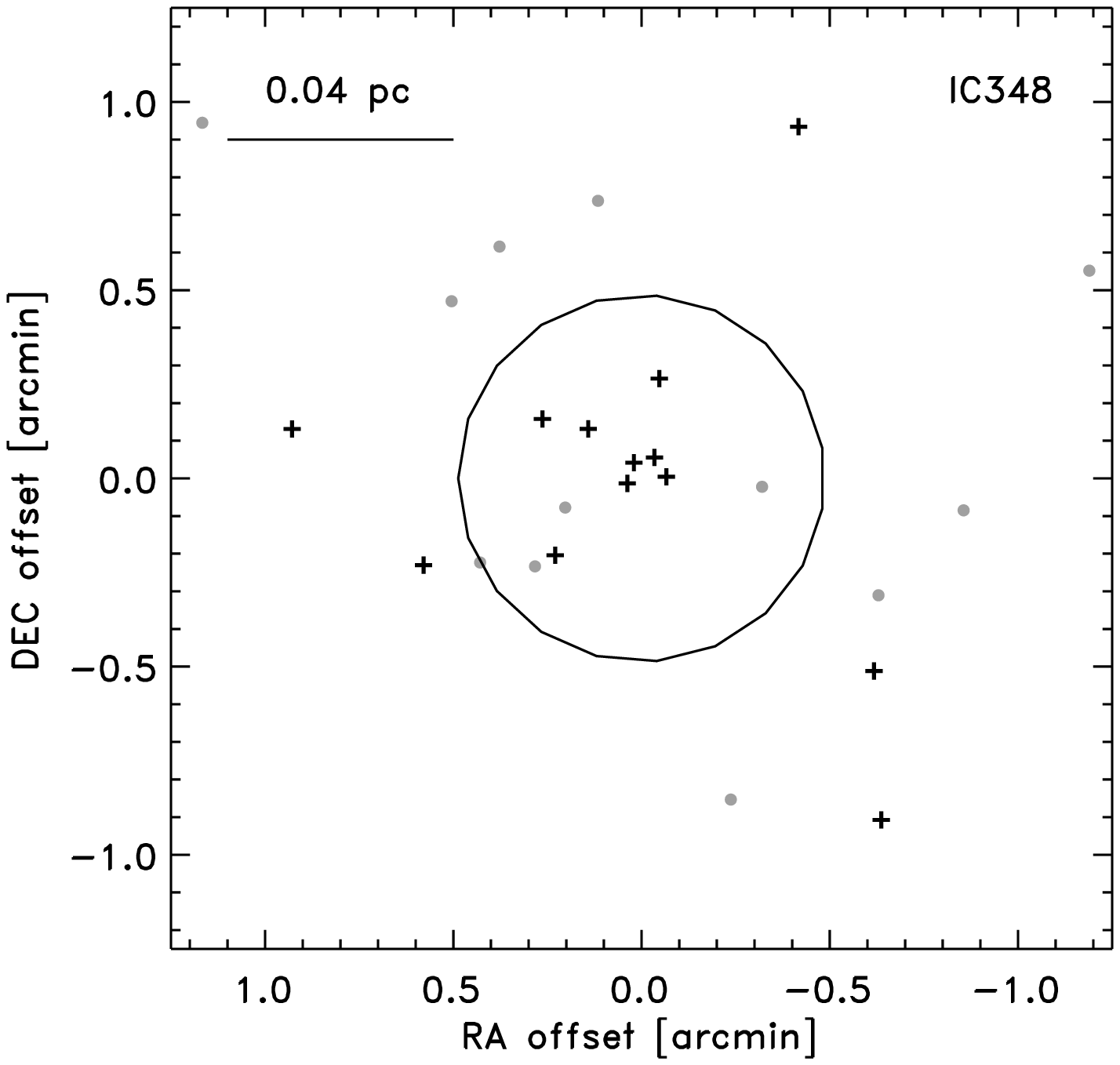}}
\caption{As Fig.~\ref{clustplot} showing the distribution of MIPS sources
  around each SCUBA core (shifted to the same center) for subregions in
  Perseus: the extended cloud, NGC~1333 and IC~348. The concentration of red
  MIPS sources (black plus signs) are seen to be significantly higher in the
  two former regions than in IC~348 suggesting differences in their star
  formation histories.}\label{regioncompare}
\end{figure}

As discussed in \cite{perspitz} and \cite{padgett08}, on large scales the
Perseus and Ophiuchus clouds are highly non-uniform. Significant star
formation occurs in well-defined regions, e.g., the two main clusters, IC~348
and NGC~1333, of Perseus or L1688 in Ophiuchus. On the other hand significant
star formation was also found to occur outside of these clusters in Perseus,
e.g., in the smaller groups surrounding B1, L1448 and
L1455. Fig.~\ref{regioncompare} compares the distribution of red MIPS sources
around the cores in NGC~1333, IC~348 and the rest of the cloud (using the
distinction between the two clusters and the remainder of Perseus from
\cite{perspitz}). As seen in this figure the ``extended'' region which
predominantly covers SCUBA cores in the smaller groups, B1, L1448 and L1455,
shows the most concentrated distribution of YSOs around the center of the
SCUBA cores whereas IC~348 in the other extreme shows the most loose
distribution of YSOs. This in itself suggests that the smaller groups are
still in the early star forming stages, whereas the region around IC~348
predominantly contains more evolved YSOs that are no longer associated with
SCUBA cores.

\clearpage
\clearpage

Many quantifiable properties about individual star-forming regions depend on
the choice of boundary used to describe the region. For example, the star
formation efficiency depends on the gas mass enclosed within a given region
surrounding YSOs. It is thus important to have objective criteria for defining
the boundaries of subregions within clouds and determining the memberships of
these groups. To apply this in a systematic fashion we utilize a
nearest-neighbor algorithm similar to that discussed by \cite{gutermuth05}
(see also \citealt{casertano85}). Across the cloud we calculate the local
surface density, $\sigma$, at grid position $(i,j)$ from
\begin{eqnarray}
\sigma & = & \frac{n-1}{\pi r_n^2(i,j)} \label{sigmadens}
\end{eqnarray}
where $r_n(i,j)$ is the distance to the $n$'th nearest neighbor (following
\citeauthor{gutermuth05}, $n=5$). Assuming local spherical symmetry the
volume density, $\rho$, can likewise be defined as:
\begin{eqnarray}
\rho   & = & \frac{n-1}{\frac{4}{3} \pi r_n^3(i,j)} \label{rhodens}
\end{eqnarray}

From Eq.~\ref{sigmadens} and \ref{rhodens} it is easily seen that
$\rho = \frac{3}{4}\sqrt{\frac{\pi}{n-1}}\,\sigma^{3/2}$ under these
assumptions. In case the distribution of sources are flattened along
the line of sight, the derived volume density is a lower limit to the
actual volume density. 

As discussed by \citeauthor{gutermuth05} and others this is just one of the
many different ways to quantify the substructure of clouds and we do not
attempt to address fundamental questions such as what fraction of stars form
in isolation and clusters and whether the star formation process is
hierarchical or fractal in nature. We also note that significant substructure
exists within the regions defined in this manner (see, e.g., study of NGC~1333
by \citeauthor{gutermuth08}). A more exhaustive analysis of the clustering
substructure in these clouds is outside the scope of this work.

\begin{figure}
\resizebox{0.9\hsize}{!}{\includegraphics{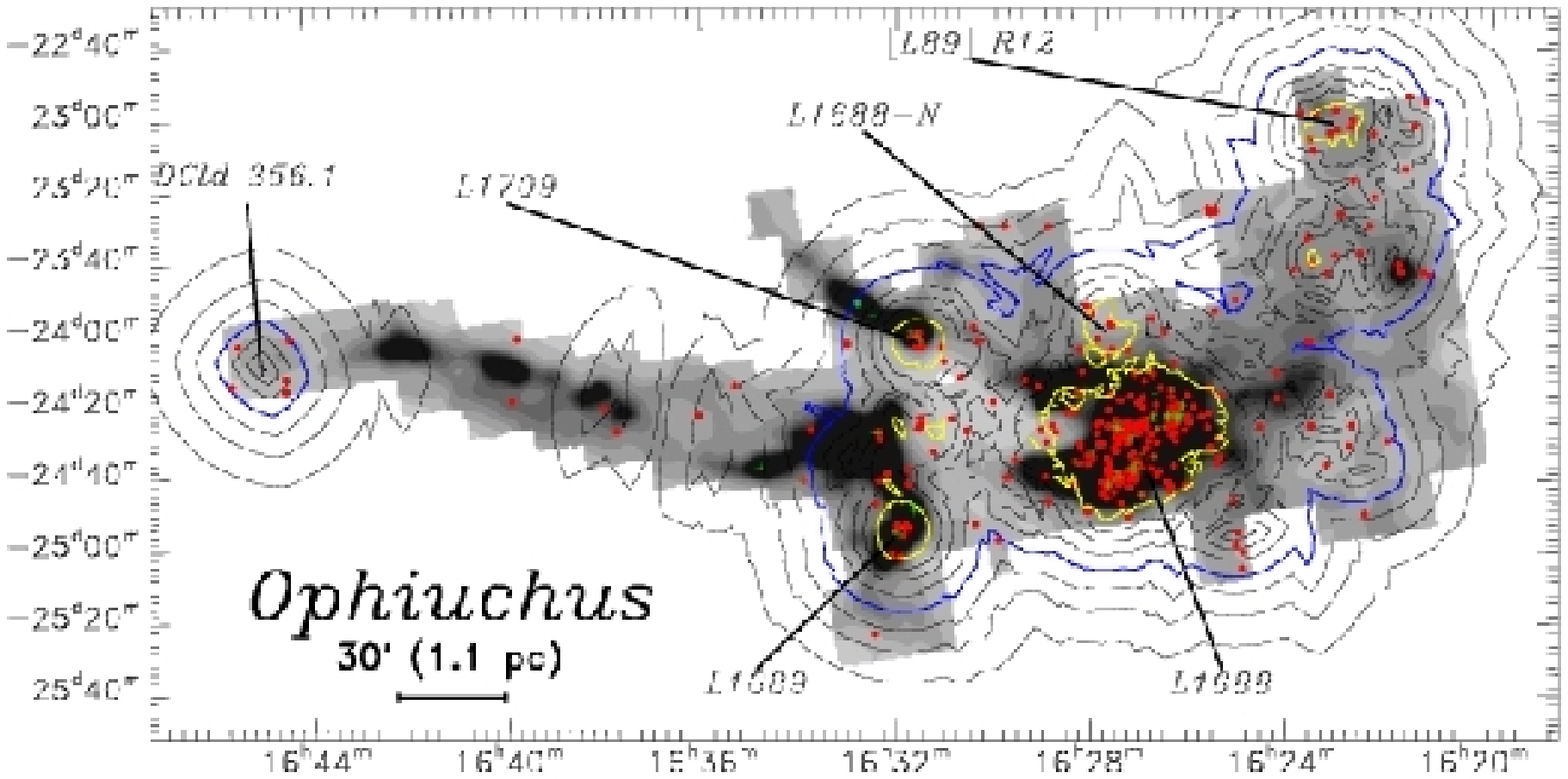}}
\resizebox{0.9\hsize}{!}{\includegraphics{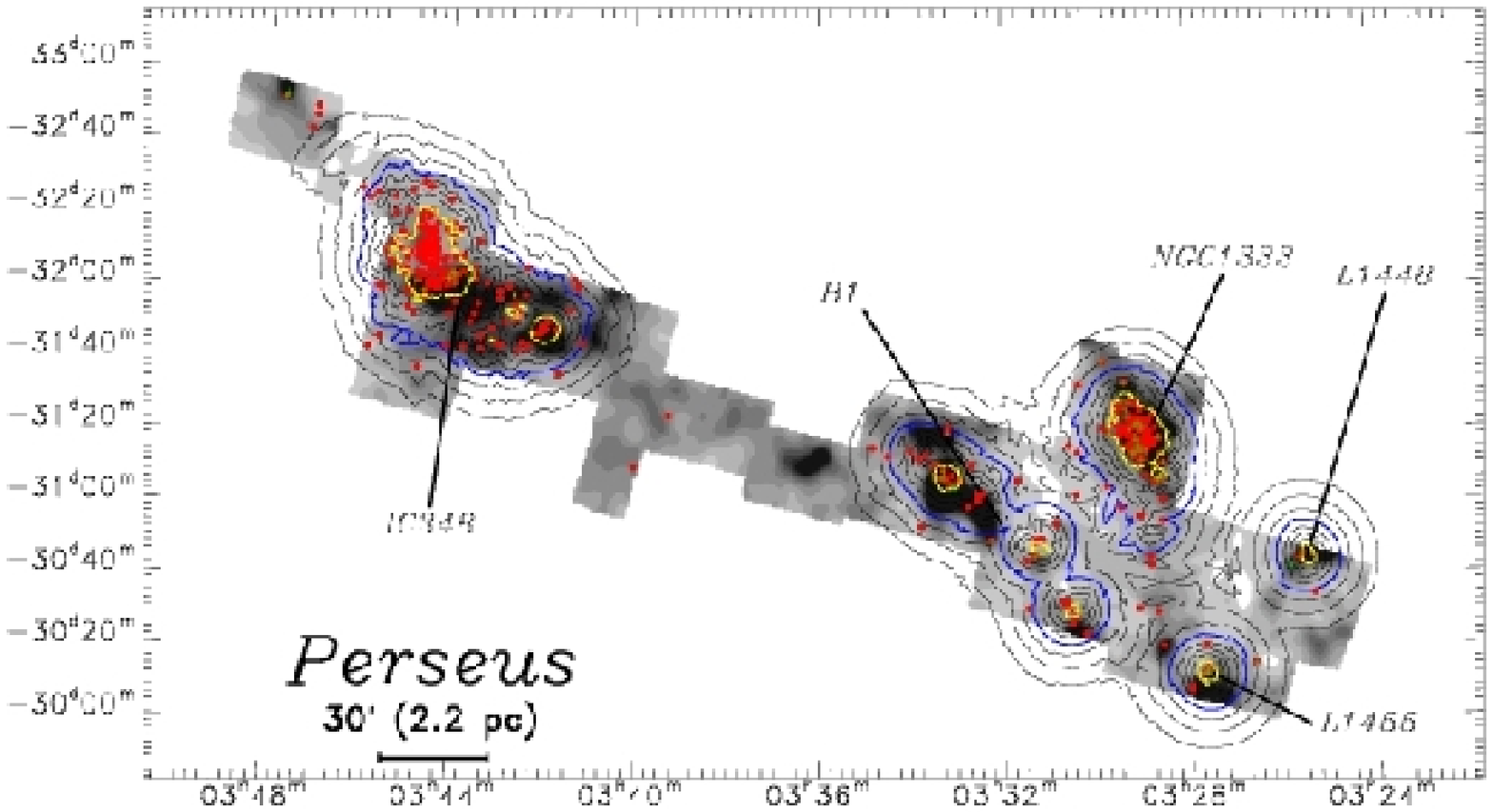}}
\caption{Volume density contours of the YSOs in Ophiuchus (upper panel),
  Perseus (lower panel) shown on top of extinction maps (grey-scale) based on
  the c2d data \citep{delivery4}. In both panels the black contours indicate
  the volume densities corresponding to 0.125, 0.25, 0.50, 2.0 and
  4.0~$M_\odot$~pc$^{-3}$. The blue contours show volume densities of
  1~$M_\odot$~pc$^{-3}$, corresponding to the criterion (1$\times$LL03 in the
  text) for identifying clusters suggested by \cite{lada03}, and the yellow
  contours to volume densities of 25~$M_\odot$~pc$^{-3}$ (25$\times$LL03 in
  the text). The red dots show the locations of the YSOs and the green plus
  signs the locations of the SCUBA cores in the two
  clouds.}\label{surfdensfig}
\end{figure}

As input for this analysis we use the list of YSO candidates from the catalogs
included in the final delivery of c2d data \citep{delivery4}. These lists of
YSOs include sources identified using their mid-infrared colors
\citep[e.g.,][]{harvey07} as well as the deeply embedded protostars from this
work. These lists are therefore restricted to YSOs with a mid-infrared excess,
i.e., predominantly YSOs the Class 0--II evolutionary stages. The results of
this analysis is shown in Fig.~\ref{surfdensfig} for Ophiuchus and Perseus.

\citeauthor{lada03} (2003; LL03 in the following) compiled a catalog of
clusters, defining those as being groups of stars stable against tidal
disruption by the galaxy or passing interstellar clouds. This criterion
translates to a mass density of 1.0~$M_\odot$~pc$^{-3}$ (or 2
sources~pc$^{-3}$ for an IMF with a mean mass of 0.5~$M_\odot$ or a surface
density of 2.1~sources~pc$^{-2}$ with the assumption of local spherical
symmetry). In addition, LL03 required clusters to have more than 35 members to
ensure that its evaporation time was larger than 10$^8$~years. The volume
density level of LL03 is shown in both panels of Fig.~\ref{surfdensfig} with
the blue contour while the black contours correspond to volume densities
0.125, 0.25, 0.5, 2, 4 and 8 times this level. It is clear that the volume
density contour suggested by the stability arguments of LL03 does not capture
the substructure of the YSO distributions, but rather that a higher contour
level is required. A similar issue is seen when applying the algorithm to the
samples of in Serpens \citep{harvey07}: there a much higher surface density
encompasses the entire surveyed region. It thus appears that the LL03
criterion is too loose to capture the substructure of the regions. This
becomes even more aggravated by the fact that our YSO catalogs do not include
the most evolved YSOs, those with little mid-infrared excesses. If these were
included in the lists of YSOs, the groups would be found to cover even larger
areas and volumes. Judging by eye, a volume density twenty-five times higher
than the LL03 level (yellow contour in Fig.~\ref{surfdensfig}) appears to
better capture the structure of the YSO distributions seen in these clouds.

We thus use the following criteria to determine the subgroupings of YSOs
(``associations'') within the c2d clouds: we define associations of YSOs as
loose or tight depending on whether they have a minimum volume density of
1~$M_\odot$~pc$^{-3}$ (1$\times$LL03) or 25~$M_\odot$~pc$^{-3}$
(25$\times$LL03).  The associations are furthermore split into ``clusters''
(greater than 35 members) and ``groups'' (less than 35 members). Due to the
definition of surface/volume density from the fifth nearest neighbor, the lower
limit to the number of members in a given group is 5.
\clearpage
\clearpage

\subsubsection{Substructure of Ophiuchus}
At the 1$\times$LL03 level most of the YSOs in Ophiuchus are included in one
large region, ``e-Oph'', encompassing the main associations of YSOs around L1688,
L1709 and L1689. The only other loose group is located close to DCld 356.1+13.7
\citep{feitzinger84} in the northwestern part of the Ophiuchus streamer. This
loose group is located slightly to the south of a $^{13}$CO core, [L89]~R76, in
the maps of \cite{loren89a}. This region was not covered by the SCUBA maps
(although inside the c2d coverage - and thus the extinction maps in
Fig.~\ref{surfdensfig}). The same is true for the two extinction cores
associated with L1712 and L1729 [seen in the c2d maps at (16:40;-24:10) and
(16:42;-24:10)] which do not contain any YSOs.

At the higher 25$\times$LL03 level, the main loose cluster, ``e-Oph'', is
broken into five separate tight regions. The dominating tight concentration is
associated with the L1688 dark cloud which contains 171 members (61\% of those
in the larger loose cluster) and is the only ``cluster'' in Ophiuchus
according to our definition. It also dominates in terms of the number of SCUBA
cores, with 71\% of the SCUBA cores (83\% of those associated with MIPS
sources).

Four additional tight regions have 5--10 members each and are therefore
classified as ``groups''. Only two of these regions have SCUBA cores within
their boundaries; those associated with the dark clouds L1689 and L1709. L1689
and L1709 show extended dust extinction but only small regions containing
embedded YSOs. In L1709 a small group of starless cores are identified at the
DCO$^+$ peak, L1709A, from the maps of \cite{loren90} (see also
Fig.~\ref{findingchart}). The small group of YSOs at the peak of the L1709
filament is associated with the IRAS point source, IRAS~16285-2355. In L1689
the YSO density peaks at the southern part of the extinction core associated
with L1689S \citep{loren89a}. A few additional isolated starless cores and
YSOs are found in the north, surrounded by the extinction core associated with
L1689N.

In addition to L1689 and L1709, two tight groups of YSOs are found that do not
have associated SCUBA cores. One, L1688-N, is found just north of the L1688
cluster, but is not connected directly to the main cluster through strong
extinction or $^{13}$CO emission in the maps of \cite{loren89a}. The other
association is located toward the west of the cloud complex where
\cite{loren89a} found a number of smaller CO clouds. The YSO volume density in
this region peaks at the location of the [L89]~R12 core: \citeauthor{loren89a}
found that this particular core together with another, [L89]~R7, showed
spatially compact $^{13}$CO emission and strong $^{13}$CO peaks suggesting that
star formation is ongoing.  Whereas [L89]~R7 does not show up as a separate
group in the YSO density maps, it is associated with a high level of extinction
$A_V > 10$ and contains a small number of Spitzer identified YSOs. Both the
[L89]~R12 and L1688-N regions have very low levels of extinction, $\langle A_V
\rangle \approx 4$ compared to L1709, L1689 and L1688 where the average
extinction is $\langle A_V\rangle = 10-15$.

Of the remaining 9 SCUBA cores located outside the boundaries of the
L1688, L1689 and L1709 tight groups, seven are associated with the
larger loose ``e-Oph'' association and  two are located within 
the extended cloud where the YSO volume density is lower 
than 1~$M_\odot$~pc$^{-3}$.

\subsubsection{Substructure of Perseus}
Perseus shows a larger variety of structure than Ophiuchus with the regions
around the two main clusters, NGC~1333 and IC~348, containing about 80\% of the
YSOs but only about 55\% of those associated with SCUBA cores. This in itself
suggests that significant current star formation is going on outside the two
main clusters. In contrast to Ophiuchus, the YSO associations in Perseus can be
traced to lower levels of volume densities. At the 1$\times$LL03 level five
loose associations of YSOs are identified, connected to the two main clusters,
NGC~1333 and IC~348, and the three small groups, B1, L1455 and L1448.

At the 1$\times$LL03 level the B1 group extends south along the main
ridge seen in the extinction and (sub)millimeter emission maps, and
encompasses the regions around the deeply embedded YSOs,
IRAS~03285+3035 and IRAS~03292+3039 as well as the aggregate of YSOs
in the Per~6 region \citep{rebull07}.

Each of the loose regions can also be traced to the higher 25$\times$LL03
level. At this level only e-IC~348 breaks up into multiple tight associations:
the main cluster, IC~348, and the smaller group to the southwest, IC~348-SW
\citeauthor{cambresy06}). The main IC~348 tight cluster contains all of the
SCUBA cores within the loose boundary. As pointed out by \cite{muench07} these
cores are all located toward the southwest, where the extinction peaks. The
association of YSOs in L1448 only has three members at the 25$\times$LL03 level
and is therefore not counted as a tight group.

The other regions each contain one tight cluster or group within their
1$\times$LL03 levels. NGC~1333 contains close to half of the SCUBA
cores in the Perseus cloud, with 85\% of the cores within the loose cluster
also within the boundaries of the tight cluster. The number of YSOs
within the tight groups of IC~348 and NGC~1333 are almost the same
(121 and 102, respectively) as are the masses from the extinction maps
(318 and 319~$M_\odot$). There is, however,  more than a factor 3 
difference between the number of Class I objects in these regions, 
as also noted by \cite{perspitz}.

\subsection{Key numbers for Perseus and Ophiuchus}\label{numbers_peroph}
Table~\ref{keynumbers} summarizes the key numbers for each of the regions in
Ophiuchus and Perseus. We are here concerned with properties directly related to
the current/ongoing star formation status of each region and refer to future
papers for further discussion about the more evolved YSOs and overviews of the
full YSO populations. In addition to the number of cores, the total core mass,
and the number of embedded YSOs from this paper within each region, we also
tabulate the number of YSOs from the c2d catalogs \citep{delivery4} together
with the number of Class I sources. We also calculate two estimates of the star
formation efficiency, $SFE=\frac{M_{\rm YSO}}{M_{\rm gas+dust}+M_{\rm YSO}}$,
where $M_{\rm YSO}$ is the total mass of YSOs and $M_{\rm gas+dust}$ is the mass
of gas and dust in the considered region. The ``core SFE'' ($SFE_{\rm core}$) is
calculated on basis of the number of embedded YSOs from this paper and where
$M_{\rm gas+dust}$ is the total mass in SCUBA cores. The other estimate, the
``cloud SFE'' ($SFE_{\rm cloud}$), is calculated using the total number of YSOs
from the c2d list (i.e., including the more evolved YSOs as well as the embedded
YSOs from this paper) and with $M_{\rm gas+dust}$ being the mass within the
considered region (e.g., within a given volume density contour) based on the c2d
extinction maps \citep{delivery4} and using the conversion between extinction
and hydrogen column density, $N_{{\rm H}_2} =9.4\times 10^{20} {\rm cm}^{-2}
(A_V / {\rm mag})$, from the Copernicus results of \cite{bohlin78}. The ``core
SFE'' is a measure of how efficient stars form from existing cores with central
densities higher than about $5\times 10^4$cm~$^{-3}$ corresponding to the
detection limit of the SCUBA data, but does not take into account the YSOs that
have already been formed and additional cores which may still form within the
cloud. In contrast, the cloud SFE measures how efficient the cloud has been in
turning dust and gas into YSOs up to this point in time. For both measures of
the SFE we assume a mean stellar mass of 0.5~$M_\odot$.

\clearpage
\thispagestyle{empty}
\setlength{\voffset}{40mm}
{\rotate
\begin{table}[!htb]
\newcommand{\tn}[1]{\tiny{#1}}
\newcommand{\an}[1]{\tiny{#1}}
\newcommand{\nn}[1]{\tiny{#1}}
\newcommand{\Nn}[1]{\scriptsize{\bf #1}}
\newcommand{\An}[1]{\footnotesize{\bf #1}}
\caption{Key numbers for Ophiuchus and Perseus.}\label{keynumbers}
\scriptsize
\begin{tabular}{llccccccccccc}\hline\hline
\multicolumn{2}{l}{}           & \#YSOs & \#Class I\tablenotemark{a}   & \#Cores & \#Cores w/MIPS & $\langle A_V \rangle$\tablenotemark{b} & Volume     & $M_{\rm cloud}$\tablenotemark{c}  & $M_{\rm core}$\tablenotemark{d} & $M_{\rm starless}$ & SFE$_{\rm core}$ & SFE$_{\rm cloud}$ \\
\multicolumn{2}{l}{}           &          &             &         &             & [mag]                 & [pc$^{3}$] & [$M_\odot$] & [$M_\odot$]         & [$M_\odot$]       &             &            \\\hline
\multicolumn{13}{c}{\An{Ophiuchus}}\\[1.5ex]
\multicolumn{2}{l}{\An{Total}}      & \An{290} & \An{32 (11\%)}   & \An{66} & \An{23 (35\%)} & \an{$\ldots$} & \an{$\ldots$} & \An{3586} & \An{74  } & \An{45      } & \An{13\%    } & \An{3.9\% } \\[1.5ex]
\multicolumn{2}{l}{\Nn{e-Oph}}      & \Nn{266} & \Nn{30 (11\%)}   & \Nn{64} & \Nn{22 (34\%)} & \Nn{5.7     } & \Nn{75      } & \Nn{2570} & \Nn{74  } & \Nn{45      } & \Nn{13\%    } & \Nn{4.9\% } \\
              & \tn{L1688}          & \tn{154} & \tn{20 (13\%)}   & \tn{47} & \tn{19 (40\%)} & \tn{18.1    } & \tn{2.1     } & \tn{ 742} & \tn{57  } & \tn{36      } & \tn{14\%    } & \tn{9.4\% } \\
              & \tn{L1709}          & \tn{  8} & \tn{ 0 (0.0\%)}  & \tn{ 3} & \tn{ 1 (33\%)} & \tn{9.6     } & \tn{0.071   } & \tn{  42} & \tn{0.62} & \tn{0.36    } & \tn{45\%    } & \tn{8.7\% } \\
              & \tn{L1689}          & \tn{ 11} & \tn{ 3 (27\%)}   & \tn{ 7} & \tn{ 1 (14\%)} & \tn{14.9    } & \tn{0.091   } & \tn{  77} & \tn{4.0 } & \tn{3.9     } & \tn{11\%    } & \tn{6.7\% } \\
              & \tn{[L89] R12}      & \tn{  7} & \tn{ 1 (14\%)}   & \tn{ 0} & \tn{$\ldots$ } & \tn{4.5     } & \tn{0.054   } & \tn{  16} & \tn{0.0 } & \tn{$\ldots$} & \tn{$\ldots$} & \tn{18\%  } \\
              & \tn{L1688-N}        & \tn{  7} & \tn{ 0 (0.0\%)}  & \tn{ 0} & \tn{$\ldots$ } & \tn{4.1     } & \tn{0.067   } & \tn{  17} & \tn{0.0 } & \tn{$\ldots$} & \tn{$\ldots$} & \tn{17\%  } \\
              & \tn{other}          & \tn{ 79} & \tn{ 6 (6.6\%)}  & \tn{ 7} & \tn{ 1 (14\%)} & \tn{$\ldots$} & \tn{$\ldots$} & \tn{1686} & \tn{12  } & \tn{4.6     } & \tn{4.0\%   } & \tn{2.2\% } \\
\multicolumn{2}{l}{\Nn{DCld 356.1}} & \Nn{  5} & \Nn{ 2 (40\%)}   & \Nn{ 0} & \nn{$\ldots$ } & \Nn{2.4     } & \Nn{0.36    } & \Nn{  31} & \Nn{0.0 } & \nn{$\ldots$} & \nn{$\ldots$} & \Nn{7.5\% } \\
\multicolumn{2}{l}{\Nn{Extended}}   & \Nn{ 19} & \Nn{ 0 (0.0\%)}  & \Nn{ 2} & \Nn{ 1 (50\%)} & \nn{$\ldots$} & \nn{$\ldots$} & \Nn{ 985} & \Nn{0.69} & \Nn{0.66    } & \Nn{7.2\%   } & \Nn{0.96\% }\\ \hline
\multicolumn{13}{c}{\An{Perseus}}\\[1.5ex]					             	       	      	       	     	          	      	     			     	       	             
\multicolumn{2}{l}{\An{Total}}      & \An{385} & \An{89 (23\%)}   & \An{72} & \An{42 (58\%)} & \an{$\ldots$} & \an{$\ldots$} & \An{7080} & \An{106 } & \An{24      } & \An{17\%    } & \An{2.6\% } \\[1.5ex]
\multicolumn{2}{l}{\Nn{e-NGC1333}}  & \Nn{115} & \Nn{41 (36\%)}   & \Nn{33} & \Nn{20 (61\%)} & \Nn{6.2     } & \Nn{12      } & \Nn{ 822} & \Nn{ 55 } & \Nn{11      } & \Nn{15\%    } & \Nn{6.5\% } \\
              & \tn{NGC1333}        & \tn{102} & \tn{35 (34\%)}   & \tn{28} & \tn{16 (57\%)} & \tn{12.7    } & \tn{0.98    } & \tn{ 319} & \tn{ 53 } & \tn{10      } & \tn{13\%    } & \tn{14\%  } \\
              & \tn{other}          & \tn{ 13} & \tn{ 6 (46\%)}   & \tn{ 5} & \tn{4  (80\%)} & \tn{$\ldots$} & \tn{$\ldots$} & \tn{ 502} & \tn{1.8 } & \tn{0.54    } & \tn{53\%    } & \tn{1.3\% } \\
\multicolumn{2}{l}{\Nn{e-IC348}}    & \Nn{189} & \Nn{15 (7.9\%)}  & \Nn{12} & \Nn{4  (33\%)} & \Nn{6.2     } & \Nn{33      } & \Nn{1611} & \Nn{ 14 } & \Nn{5.0     } & \Nn{13\%    } & \Nn{5.5\% } \\
              & \tn{IC348}          & \tn{121} & \tn{11 (9.1\%})  & \tn{12} & \tn{4  (33\%)} & \tn{8.1     } & \tn{1.9     } & \tn{ 318} & \tn{ 14 } & \tn{5.0     } & \tn{13\%    } & \tn{16\%  } \\
              & \tn{IC348-SW}       & \tn{ 10} & \tn{ 2 (20\%)}   & \tn{ 0} & \tn{$\ldots$ } & \tn{11.8    } & \tn{0.086   } & \tn{  59} & \tn{0.0 } & \tn{$\ldots$} & \tn{$\ldots$} & \tn{7.8\% } \\
              & \tn{other}          & \tn{ 58} & \tn{ 2 (3.4\%)}  & \tn{ 0} & \tn{$\ldots$ } & \tn{$\ldots$} & \tn{$\ldots$} & \tn{1234} & \tn{0.0 } & \tn{$\ldots$} & \tn{$\ldots$} & \tn{2.3\% } \\
\multicolumn{2}{l}{\Nn{e-B1}}       & \Nn{ 34} & \Nn{14 (41\%)}   & \Nn{10} & \Nn{7  (70\%)} & \Nn{5.9     } & \Nn{15      } & \Nn{ 919} & \Nn{ 15 } & \Nn{2.0     } & \Nn{25\%    } & \Nn{1.8\% } \\
              & \tn{B1}             & \tn{  9} & \tn{ 8 (89\%)}   & \tn{ 7} & \tn{5  (71\%)} & \tn{17.5    } & \tn{0.075   } & \tn{  79} & \tn{ 13 } & \tn{0.84    } & \tn{16\%    } & \tn{5.4\% } \\
              & \tn{other}          & \tn{ 25} & \tn{ 6 (24\%)}   & \tn{ 3} & \tn{2  (67\%)} & \tn{$\ldots$} & \tn{$\ldots$} & \tn{ 840} & \tn{2.4 } & \tn{1.0     } & \tn{30\%    } & \tn{1.5\% } \\
\multicolumn{2}{l}{\Nn{e-L1455}}    & \Nn{  8} & \Nn{ 6 (75\%)}   & \Nn{ 5} & \Nn{5  (80\%)} & \Nn{5.3     } & \Nn{1.9     } & \Nn{ 211} & \Nn{3.1 } & \Nn{0.55    } & \Nn{39\%    } & \Nn{1.8\% } \\
              & \tn{L1455}          & \tn{  5} & \tn{ 5 (100\%)}  & \tn{ 5} & \tn{5  (80\%)} & \tn{8.1     } & \tn{0.035   } & \tn{  22} & \tn{3.1 } & \tn{0.55    } & \tn{39\%    } & \tn{10\%  } \\
              & \tn{other}          & \tn{  3} & \tn{ 1 (33\%)}   & \tn{ 0} & \tn{$\ldots$ } & \tn{$\ldots$} & \tn{$\ldots$} & \tn{ 189} & \tn{0.0 } & \tn{$\ldots$} & \tn{$\ldots$} & \tn{0.79\%} \\
\multicolumn{2}{l}{\Nn{L1448}}      & \Nn{  5} & \Nn{ 5 (100\%)}  & \Nn{ 6} & \Nn{3  (50\%)} & \Nn{4.6     } & \Nn{1.4     } & \Nn{ 147} & \Nn{16  } & \Nn{5.4     } & \Nn{8.5\%   } & \Nn{1.7\% } \\
\multicolumn{2}{l}{\Nn{Extended}}   & \Nn{ 34} & \Nn{ 8 (24\%)}   & \Nn{ 6} & \Nn{4  (67\%)} & \nn{$\ldots$} & \nn{$\ldots$} & \Nn{3371} & \Nn{3.6 } & \Nn{0.55    } & \Nn{36\%    } & \Nn{0.52\%} \\ \hline
\end{tabular}
\tablenotetext{a}{Number of Class I YSOs and fraction of the total number of YSOs in the given region.}
\tablenotetext{b}{Average $A_V$ from the extinction maps within the volume density contour used to define the given region.}
\tablenotetext{c}{Total mass within the given volume density contour from the c2d extinction map.}
\tablenotetext{d}{Total mass of the SCUBA cores located within the given volume density contour.}
\end{table}
}
\clearpage
\setlength{\voffset}{0mm}

\subsection{Perseus vs. Ophiuchus: fraction of embedded sources and SFE}
In contrast to Ophiuchus, the Perseus regions show large variety in the
relative numbers of embedded YSOs - both when defined as being associated with
submillimeter cores or based on their mid-infrared spectral index (i.e., the
classical definition of Class~I YSOs). The difference between IC~348 and the
collection of groups including NGC~1333, L1455, L1448 and B1 was also
discussed by \cite{perspitz}. The smaller groups contain only 20\% of the YSOs
in the entire Perseus complex but about 40\% of both the Class I sources and
the MIPS sources associated with submillimeter cores. In the central tight,
B1, L1455 and L1448, groups, the ratios of Class I sources to total YSOs are
in fact 90--100\% compared to 34\% in NGC~1333 and 10--20\% in the two tight
clusters/groups in IC~348. In the more extended regions around these tight
groups the ratios of Class Is to total YSOs are, in contrast, much lower.  For
example, in B1 the loose group further south in the extinction ridge and the
Per~6 aggregate have about 25\% Class I sources.  The ratio of Class I YSOs in
the Ophiuchus groups is more similar to that of IC~348 than those of the
smaller groups and NGC~1333.

Another property which is interesting to compare is the star formation
efficiency, $SFE$ (see \S\ref{numbers_peroph}). As discussed in
\cite{scubaspitz} it is convenient to distinguish between the ``core'' star
formation efficiency, $SFE_{\rm core}$, and ``cloud'' star formation
efficiency, $SFE_{\rm cloud}$: $SFE_{\rm core}$ is based on the masses of the
embedded YSOs in the cloud compared to the mass of the dust and gas already
collected in dense cores. It is therefore a good measure of the star formation
efficiency for the current generation of dense cores but does not take into
account those stars which have evolved away from the embedded stages or those
cores which may still form from gas within the cloud. $SFE_{\rm cloud}$ in
contrast is based on the masses of all (identified) YSOs in the studied region
and the total mass based on extinction maps (i.e., mass in the extended cloud
at lower column densities) and therefore captures the star formation
efficiency over the larger time-span corresponding to the evolution of the
YSOs from the formation of the cores through the YSO Class 0--II stages. The
biggest caveats about these definitions are (1) the completeness of the YSO
samples (large number of diskless YSOs may for example remain undefined by the
photometric surveys because they have too little dust emission to show
significant infrared excesses) and (2) that dust and gas in the cloud might
disperse over the timescale during which the YSOs evolve.

For the regions in Perseus and Ophiuchus, values of $SFE_{\rm core}$ are
10--30\% - with no significant differences between the two clouds. For
$SFE_{\rm cloud}$, the average number for Perseus (2.6\%) is only slightly
lower than in Ophiuchus (4.1\%). When individual regions are considered,
however, a more pronounced difference is found: the two clusters and the
Ophiuchus region show $SFE_{\rm cloud}$ values of about 6\% whereas the more
isolated groups show lower efficiencies of about 2\% (at the 1$\times LL03$
level). For the tight groups and clusters, the regions in Ophiuchus - except
L1689 - and NGC~1333 and IC~348, also show higher efficiencies than do the
smaller regions of Perseus. These differences can be interpreted in two ways:
one possibility is that there is a continuous replenishment of material in the
dense cores and the higher $SFE_{\rm cloud}$ of the larger regions simply
reflect that these are more effective in turning dust and gas into new
stars. A different explanation is that the smaller regions in Perseus only
recently have started forming stars and therefore show a lower $SFE_{\rm
cloud}$ than the larger clouds. We will return to this discussion in
\S\ref{discuss}.

\section{Discussion: Evolution of embedded protostars}\label{discuss}
The differences between Perseus and Ophiuchus in the numbers of embedded
protostars are interesting in the context of the evolution of YSOs through
their earliest stages - and suggests evolutionary differences between the two
clouds related to differences either in the evolution of individual YSOs or
the star formation complex as a whole. To reiterate, compared to Perseus, the
regions in Ophiuchus show:
\begin{enumerate}
 \item a lower fraction of Class I sources compared to the total number of YSOs
 \item a lower fraction of MIPS sources associated with SCUBA cores
 \item fewer ``red'' ($[3.6]-[4.5] > 1$ and $[8.0]-[24] > 4.5$) sources from
the comparison of the SCUBA cores and MIPS sources from this paper and
\cite{scubaspitz}
\end{enumerate}
The differences between the number of sources in these different classes raise
concerns about timescales for the evolution of prestellar cores and protostars
derived on basis of for example the relative numbers of starless and
starforming cores or the number of Class I vs. II sources. In terms of star
formation efficiencies, the central Ophiuchus region(s) and the two clusters
in Perseus, IC~348 and NGC~1333, show higher cloud star formation efficiencies
than L1689 in Ophiuchus and the smaller L1455, L1448 and B1 groups in Perseus.

\subsection{Difference in evolutionary stages?}
One explanation for the differences between the number of embedded sources in
Ophiuchus and Perseus might be that the clouds have different ``ages''. If
star formation has initiated (e.g., been triggered) at a given time in the
recent past of the cloud history and the embedded YSOs evolve on similar
timescales the lower number of embedded sources in Ophiuchus suggest that this
cloud is more evolved compared to Perseus. A similar effect has been observed
within Perseus with clear differences in the numbers of embedded YSOs in
IC~348 and NGC~1333 \citep{perspitz} in line with the different age estimates
for the two clusters of about 2~Myr \citep{luhman03} and $<$~1~Myr
\citep{lada96,wilking04}, respectively. The difference in $SFE_{\rm cloud}$
between Ophiuchus and Perseus could then simply reflect that a smaller amount
of the dust and gas has turned into stars at this point of the evolution of
Perseus. In principle, there is no reason why star formation in Perseus and
Ophiuchus should have initiated at the same time, so from that perspective
this is definitely a plausible scenario.

What does complicate the picture of a pure evolutionary effect are the
relative numbers of SCUBA cores with and without associated MIPS sources (or
protostellar and starless cores, respectively). As pointed out above, the
small groups in Perseus show very high numbers of Class I sources
corresponding to 90--100\% of the YSOs in the groups, but these regions do not
have the same high fractions of associated MIPS sources. This is naturally
explained if B1, L1448 and L1455 have all only recently started forming stars
and therefore not all of their cores have yet collapsed. The issue with this
interpretation is that Ophiuchus also shows a low fraction of cores with
embedded protostars relative to the total number of cores. In a scenario where
the time for the initiation star formation differs, the relatively high number
of starless cores in Ophiuchus would therefore mean that these remaining
starless cores either are ``fizzles'' (i.e., did not ``manage'' to collapse) -
or that a second generation of star formation is about to occur in this
cloud. \cite{nutter06} argued that the differences seen between L1689 and
L1688 - the latter having a more active star formation history - could be
attributed to sequential star formation triggered by a small group of young
high mass stars, $\sigma$Sco, in the Sco-Cen association.

\subsection{Difference in star formation physics?}
Another possible explanation for the differences between the regions in
Perseus and Ophiuchus is that the physics regulating the star formation
between these regions differ, e.g., the time-scale for evolution through the
embedded stages is different in the dense clusters in Perseus and in Ophiuchus
compared to the more isolated groups. If the collapse of an ensemble of cores
occurs stochastically over a time-scale which is long compared to the collapse
time for an individual core, the ratio of sources in different evolutionary
stages can be used as a statistical tool for measuring the relative durations
of star formation phases. Thus, if the collapse of cores occurs slower (or the
rate is lower) or if the timescale for evolution through the embedded stages
is shorter - more Class 0/I sources would have time to evolve away from the
embedded stages skewing the statistics toward more evolved (Class II)
YSOs. Such a solution could explain the undercounting of embedded sources in
Ophiuchus as a whole but does not explain the higher $SFE_{\rm cloud}$ in
Ophiuchus and NGC~1333 and IC~348.

What could be the physical explanations for such a difference? As pointed out
by \cite{jayawardhana01} pre-stellar cores and protostars formed in more dense
star-forming regions are likely to have high masses on small scales, which in
turn will imply high accretion rates, e.g., compared to objects in less dense
regions such as Taurus. If this is the case, one would therefore expect that
the Ophiuchus protostars were found in regions with higher extinction (and
thus further evolved) than those in Perseus. If one compares the major
clusters in the two clouds, there does seem to be such a trend with L1688 and
L1689 having an average extinction of 15--17 compared to the 8--13 for IC~348
and NGC~1333. This would in turn require that the current population of
embedded protostars in Perseus have lower accretion rates than those few
sources observed in Ophiuchus. Other factors, such as differences in the
global and local magnetic field strengths between Perseus and Ophiuchus, could
of course also be important in this context, although no strong observational
constraints exist.

\subsection{Difference in definitions?}\label{physorigin}
A final explanation for the differences between Ophiuchus and Perseus could
simply be that the empirical evolutionary classification scheme does not apply
well to comparisons across clouds. For example, it is important to realize the
difference between the mid-infrared and submillimeter observations: where the
latter measures the dust (plus gas) mass, $M$, and core temperature, the
former measures the extinction or line of sight column density, $N$, - and
thus mid-infrared ``redness'' - of the central star+disk system. If one just
takes a spherical symmetric envelope with a density profile, $n\propto r^{-p}$
and inner and outer radii, $R_{\rm in}$ and $R_{\rm out}$, it is easily seen
that:
\begin{eqnarray}
M \propto & \int_{R_{\rm in}}^{R_{\rm out}} n(r) 4\pi r^2\, {\rm d}r \propto \int_{R_{\rm in}}^{R_{\rm out}} r^{2-p}\,{\rm d}r \propto R_{\rm out}^{3-p}-R_{\rm in}^{3-p} & \approx R_{\rm out}^{3-p} \\
N \propto & \int_{R_{\rm in}}^{R_{\rm out}} n(r)\, {\rm d}r \propto \int_{R_{\rm in}}^{R_{\rm out}} r^{-p}\,{\rm d}r \propto R_{\rm out}^{1-p}-R_{\rm in}^{1-p} & \approx R_{\rm in}^{1-p}
\end{eqnarray}
when $R_{\rm out} >> R_{\rm in}$ and $1.0 < p < 3.0$. That is, a larger inner
radius decreases the extinction of a given protostellar source (e.g., making
it visible at shorter wavelengths \citep{iras16293letter} or reducing its
``redness'') while not affecting the envelope mass and vice versa: a larger
outer radius only increases the envelope mass and thus submillimeter flux, but
does not affect the mid-infrared signature of the source. If the inner radius
or flattening of the protostellar envelopes in Ophiuchus is larger than that
in Perseus the protostars there will appear more evolved if classified based
on their mid-infrared colors or $\alpha$. This could for example be the case
if the angular momenta of the pre-stellar cores had been different, resulting
in larger centrifugal radii for the sources in Ophiuchus. The difference in
the maximum redness in Perseus and Ophiuchus (\S\ref{midirsubmm}) could be
taken as evidence that such an effect is in play.

The bottom line is that based on the current data it is difficult to say which
scenario is most likely; indeed, a combination very likely applies. Regardless,
challenges are brought to the definition of the most deeply embedded stages of
low-mass protostars, i.e., Class 0 stage.  It is here useful to remind of the
three definitions of Class 0 sources \citep[see][]{andre93,andreppiv}: the
conceptional definition of the Class 0 stage is that the protostar has accreted
less than half its final mass, but since it is difficult to determine the masses
of the central protostars, this definition is not useful for practical
purposes. Two main observational definitions are in play: most often that the
object emit most of its luminosity at submillimeter wavelength (either
quantified by a ratio of the submillimeter luminosity to total luminosity,
$L_{\lambda\, \ge\, 350 \mu{\rm m}} / L_{\rm bol} > 0.5$\% or a bolometric
temperature lower than 70--100~K) - or that it has a high mass accretion rate or
equivalently drives a powerful outflow.

The main issue is that taken separately, the Ophiuchus and Perseus data
provide very different arguments for the life-time of the Class 0
stage. Assuming a constant birthrate, the relatively large number of embedded
YSOs in Perseus compared to more evolved YSOs suggests that the Class 0 stage
is not simply a short-lived, high accretion stage. This is in direct contrast
to the conclusion derived from the Ophiuchus YSO sample, where the deeply
embedded sources are significantly outnumbered, requiring a very short-lived
Class 0 phase. A reasonable solution to this discrepancy may be that
differences between the appearance of the YSOs reflects strongly their
environments (physical conditions).  In this case, Class 0 protostars may not
represent a clearly defined early stage in the evolution of protostars but
rather, an environmentally affected, continuum into Class I protostars (as
also discussed by \cite{jayawardhana01}). An alternative explanation, that
star formation occurs in bursts, could explain the differences if the most
recent burst in Perseus has occurred more recently than that in Ophiuchus. In
this case, however, the relative number of embedded versus more evolved YSOs
does not reveal their relative time-scales and thus do not provide the
evidence that the time-scale for the evolution through the most deeply
embedded Class~0 stage is significantly shorter than the evolution through the
Class~I stage.

\section{Summary}
We have combined SCUBA submillimeter dust continuum and Spitzer mid-IR surveys
of the current star formation in the Ophiuchus cloud complex. The results are
compared to our previous study of Perseus. The main conclusions of this study
are as follows:
\begin{itemize}
\item We have applied a method developed for identifying embedded protostars
  in Perseus \citep{scubaspitz} directly to the Ophiuchus datasets.
\item Little dispersal of the protostars with respect to the SCUBA cores are
  seen similar to the situation in Perseus. Fewer red sources are found in
  Ophiuchus compared to Perseus - but also a higher fraction of starless SCUBA
  cores.
\item Although distance does not directly play into the definition of SCUBA
  cores, the Ophiuchus surveys are more sensitive to dust emission from disks,
  representing a small, but non-negligible fraction of the sources seen in the
  SCUBA survey.
\item As was the case in Perseus most of the SCUBA cores in Ophiuchus with high
  concentrations also have associated MIPS sources, likely reflecting the
  heating of the dust from the central source making them appear more
  peaked. Four high concentration SCUBA cores without associated YSOs are
  identified and are found to be the likely result of the impact of outflow
  and the structure of the ambient environment.
\item Using a nearest-neighbor algorithm we calculate surface and volume
  densities of the YSOs in Perseus and Ophiuchus and from those identify
  groups and clusters in each of the clouds. For each of these regions we
  calculate key numbers such as masses, star formation efficiencies and the
  fraction of embedded objects.
\item The star formation efficiency varies across and within clouds. Perseus
  has a high efficiency in turning material from the SCUBA cores into embedded
  protostars (a high $SFE_{\rm core}$), driven primarily by the efficiency in
  its smaller groups (B1, L1455 and L1448). The large clusters in Perseus
  (NGC~1333 and IC~348) have lower core SFEs, similar to those found across
  Ophiuchus. The cloud SFE measured on basis of the larger scale dust
  extinction and the overall population of YSOs, however, is somewhat higher
  in Ophiuchus than in Perseus primarily because these same small groups in
  Perseus (B1, L1455 and L1448) have a larger fraction of their mass still not
  turned into cores or stars. Generally, though the cloud SFE is lower than
  the core SFE, suggesting that the star formation mechanism is more
  inefficient on larger relative to smaller scales. The fraction of Class I
  sources relative to YSOs also vary from about 10\% in Ophiuchus and IC~348
  to 35\% in NGC~1333 and more than 90\% in the inner regions of B1, L1455 and
  L1448. Finally, the fraction of starless cores in Ophiuchus is found to be
  high compared to in Perseus.
\item We discuss possible explanations for the differences between the regions
  in Perseus and Ophiuchus, such as (i) that the YSOs have evolved over
  different timescales, (ii) that the physics regulating the star formation,
  e.g., the accretion in the two clouds, differ, or (iii) that the empirical
  evolutionary classification scheme does not apply well when comparing
  samples of YSOs from different clouds. Based on the existing samples it is
  not possible to rule out any of these scenarios, and likely all three may be
  relevant for the full picture of the star formation properties of the two
  clouds. In any case, the results do raise questions about the standard
  picture of the Class 0 stage being a short-lived precursor to the Class I
  stage in the evolution of embedded YSOs.
\end{itemize}

The results presented here provide a direct comparison of the global star
formation properties of the larger scale clouds and also provides important
constraints on - and raises questions about - the general low-mass star
formation scenario. It demonstrates the importance of the systematic approach
using the large scale maps which are currently appearing from Spitzer and the
new submillimeter cameras, LABOCA on APEX and SCUBA-2 on the JCMT. Together
these surveys can address whether differences between the number of sources in
the embedded stages are generally seen or, e.g., whether Perseus or Ophiuchus
contain YSO populations ``more representative'' for other regions currently
forming stars. Other complementary observations (e.g., large scale
submillimeter line maps) can address whether differences in the initial
conditions (e.g., pressure or turbulence) could be responsible for the
observed differences in the YSO populations. At the same time these
large-scale surveys are providing important unbiased samples of embedded
low-mass protostars that will be obvious targets for detailed observations
with upcoming facilities such as ALMA.

\acknowledgments 
We are grateful to Rob Gutermuth, Neal Evans, Melissa Enoch, and Jens Kauffmann
for helpful discussions and comments about the paper. We also thank the
referee, Bruce Wilking, for insightful comments that improved the
paper. D.J. is supported by a Natural Sciences and Engineering Research
Council of Canada grant. H.K. is supported by a Natural Sciences and
Engineering Research Council of Canada CGS Award and a National Research
Council of Canada GSSSP Award. This paper has made use of the SIMBAD database,
operated at CDS, Strasbourg, France.

\appendix

\section{Properties of submillimeter cores.}
We here present the list of submillimeter cores in Ophiuchus. See
\cite{johnstone04} for further details about the observations and
\cite{kirk06} for further discussion about the algorithm for identifying cores
and their properties.

\clearpage
\pagestyle{empty}
\begin{deluxetable}{lccccccccccc}
\rotate
\tabletypesize{\footnotesize}
\tablewidth{0pt}
\tablecaption{Properties of submillimeter cores in Ophiuchus\label{submmtab}}
\tablehead{
\colhead{Name\tablenotemark{a}} & 
\colhead{R.A.\tablenotemark{b}} & 
\colhead{Decl.\tablenotemark{b}} & 
\colhead{$f_0^c$} & 
\colhead{$S_{850}$\tablenotemark{c}} & 
\colhead{$R_{\rm eff}$\tablenotemark{c}} & 
\colhead{Mass\tablenotemark{d}} & 
\colhead{Concentration} & 
\colhead{Temperature\tablenotemark{e}} & 
\colhead{$M_{\rm BE}$\tablenotemark{e}} & 
\colhead{$\log n_{\rm cent}$\tablenotemark{e}} & 
\colhead{$\log P_{\rm ext}/k$\tablenotemark{e}} \\
\colhead{(SMM J)} & 
\colhead{(J2000)} & 
\colhead{(J2000)} & 
\colhead{(Jy beam$^{-1}$)} & 
\colhead{(Jy)} & 
\colhead{(arcsec)} & 
\colhead{($M_\odot$)} &
\colhead{ }  & 
\colhead{(K)} & 
\colhead{($M_\odot$)} & 
\colhead{(cm$^{-3}$)} & 
\colhead{(cm$^{-3}$~K$^{-1}$)}
}
\startdata
162608-24202 & 16  26  08.0 & -24  20  24.5 &  0.31 &  3.32 & 38   &  0.76 & 0.34 & 24   & 0.40 & 4.8 & 6.1 \\ 
162610-24195 & 16  26  09.6 & -24  19  45.5 &  0.30 &  2.11 & 31   &  0.48 & 0.34 & 21   & 0.29 & 5.0 & 6.2 \\ 
162610-24231 & 16  26  10.2 & -24  23  09.5 &  0.29 &  1.94 & 31   &  0.44 & 0.40 & 16   & 0.40 & 5.1 & 6.2 \\ 
162610-24206 & 16  26  10.4 & -24  20  57.5 &  0.45 &  3.50 & 44   &  0.80 & 0.65 & 11   & 1.3  & 5.7 & 5.8 \\ 
162614-24232 & 16  26  13.9 & -24  23  24.6 &  0.20 &  0.58 & 17   &  0.13 & 0.15 & 16   & 0.13 & 5.3 & 6.5 \\ 
162614-24234 & 16  26  13.9 & -24  23  42.6 &  0.20 &  1.10 & 25   &  0.25 & 0.22 & 17   & 0.20 & 5.1 & 6.2 \\ 
162614-24250 & 16  26  14.4 & -24  25  00.6 &  0.28 &  5.93 & 53   &  1.4  & 0.34 & 25   & 0.62 & 4.6 & 5.9 \\ 
162615-24231 & 16  26  14.6 & -24  23  06.6 &  0.20 &  2.16 & 34   &  0.49 & 0.17 & 20   & 0.31 & 4.9 & 6.1 \\ 
162616-24235 & 16  26  15.7 & -24  23  54.7 &  0.20 &  1.03 & 24   &  0.23 & 0.20 & 17   & 0.19 & 5.1 & 6.3 \\ 
162617-24235 & 16  26  17.5 & -24  23  45.7 &  0.29 &  3.93 & 42   &  0.89 & 0.32 & 23   & 0.45 & 4.8 & 6.0 \\ 
162622-24225 & 16  26  21.6 & -24  22  54.8 &  0.82 & 13.2  & 58   &  3.0  & 0.58 & 17   & 2.4  & 5.4 & 6.0 \\ 
162624-24162 & 16  26  24.1 & -24  16  15.9 &  0.45 &  0.27 & 10   &  0.06 & 0.41 & 12   & 0.10 & 6.1 & 7.0 \\ 
162626-24243 & 16  26  26.5 & -24  24  30.9 &  3.23 & 32.2  & 66   &  7.3  & 0.80 & 22   & 4.1  & 5.8 & 6.0 \\ 
162627-24233 & 16  26  27.3 & -24  23  33.9 &  3.35 & 13.3  & 32   &  3.0  & 0.66 & 21   & 1.8  & 6.2 & 6.7 \\
162628-24235 & 16  26  27.5 & -24  23  54.9 &  4.22 & 33.3  & 59   &  7.6  & 0.80 & 23   & 3.9  & 6.0 & 6.2 \\ 
162628-24225 & 16  26  28.0 & -24  22  54.9 &  1.11 & 21.0  & 70   &  4.8  & 0.66 & 18   & 3.4  & 5.5 & 5.9 \\ 
162633-24261 & 16  26  32.8 & -24  26  13.0 &  0.45 & 30.0  & 100  &  6.8  & 0.41 & 29   & 2.6  & 4.4 & 5.7 \\ 
162641-24272 & 16  26  40.5 & -24  27  16.1 &  0.34 &  3.72 & 45   &  0.85 & 0.53 & 13   & 1.1  & 5.3 & 5.9 \\ 
162644-24173 & 16  26  43.6 & -24  17  28.2 &  0.35 &  0.95 & 20   &  0.22 & 0.40 & 14   & 0.24 & 5.5 & 6.5 \\ 
162644-24345 & 16  26  44.4 & -24  34  49.2 &  0.24 &  0.19 & 10   &  0.04 & 0.29 & 13   & 0.06 & 5.8 & 6.8 \\
162644-24253 & 16  26  44.5 & -24  25  28.2 &  0.21 &  1.81 & 31   &  0.41 & 0.23 & 19   & 0.28 & 4.9 & 6.1 \\ 
162645-24231 & 16  26  45.1 & -24  23  10.2 &  0.67 &  3.06 & 39   &  0.70 & 0.74 & 11   & 1.2  & 6.0 & 5.9 \\ 
162646-24242 & 16  26  46.0 & -24  24  19.2 &  0.20 &  1.74 & 30   &  0.40 & 0.16 & 19   & 0.27 & 5.0 & 6.1 \\ 
162648-24236 & 16  26  47.5 & -24  23  55.2 &  0.23 &  2.25 & 36   &  0.51 & 0.33 & 20   & 0.33 & 4.8 & 6.0 \\ 
162659-24454 & 16  26  58.7 & -24  45  37.3 &  0.40 &  0.35 & 12   &  0.08 & 0.42 & 12   & 0.13 & 6.0 & 6.8 \\ 
162660-24343 & 16  26  59.6 & -24  34  25.3 &  0.59 & 16.0  & 73   &  3.6  & 0.55 & 18   & 2.8  & 5.1 & 5.8 \\ 
162705-24363 & 16  27  05.3 & -24  36  28.4 &  0.25 &  0.37 & 14   &  0.08 & 0.34 & 14   & 0.09 & 5.5 & 6.6 \\ 
162705-24391 & 16  27  05.3 & -24  39  13.4 &  0.35 &  2.31 & 32   &  0.53 & 0.44 & 14   & 0.58 & 5.3 & 6.2 \\ 
162707-24381 & 16  27  07.1 & -24  38  13.4 &  0.20 &  0.18 & 10   &  0.04 & 0.22 & 12   & 0.06 & 5.7 & 6.7 \\ 
162709-24372 & 16  27  09.5 & -24  37  19.4 &  0.36 &  0.83 & 20   &  0.19 & 0.48 & 11   & 0.34 & 5.8 & 6.5 \\ 
162711-24393 & 16  27  10.8 & -24  39  25.4 &  0.22 &  0.90 & 23   &  0.21 & 0.29 & 17   & 0.17 & 5.2 & 6.3 \\ 
162712-24290 & 16  27  11.7 & -24  29  04.4 &  0.21 &  0.50 & 16   &  0.11 & 0.21 & 15   & 0.11 & 5.4 & 6.5 \\ 
162712-24380 & 16  27  12.1 & -24  38  01.4 &  0.22 &  0.53 & 18   &  0.12 & 0.31 & 15   & 0.12 & 5.3 & 6.4 \\ 
162713-24295 & 16  27  12.6 & -24  29  49.4 &  0.32 &  1.28 & 24   &  0.29 & 0.38 & 18   & 0.23 & 5.2 & 6.3 \\ 
162715-24303 & 16  27  14.8 & -24  30  25.4 &  0.25 &  0.99 & 23   &  0.22 & 0.34 & 17   & 0.18 & 5.2 & 6.3 \\ 
162725-24273 & 16  27  25.1 & -24  27  28.4 &  0.46 &  3.09 & 31   &  0.70 & 0.37 & 24   & 0.34 & 5.1 & 6.3 \\ 
162727-24405 & 16  27  26.7 & -24  40  52.3 &  0.57 & 16.0  & 76   &  3.6  & 0.57 & 17   & 3.0  & 5.1 & 5.7 \\ 
162728-24271 & 16  27  28.0 & -24  27  10.3 &  0.66 &  4.26 & 30   &  0.97 & 0.34 & 28   & 0.38 & 5.2 & 6.5 \\ 
162728-24393 & 16  27  28.0 & -24  39  34.3 &  0.27 &  0.74 & 20   &  0.17 & 0.37 & 16   & 0.15 & 5.3 & 6.4 \\ 
162729-24274 & 16  27  29.5 & -24  27  40.3 &  0.47 &  2.50 & 31   &  0.57 & 0.50 & 13   & 0.70 & 5.6 & 6.3 \\ 
162730-24264 & 16  27  29.7 & -24  26  37.3 &  0.44 &  2.23 & 26   &  0.51 & 0.35 & 22   & 0.27 & 5.2 & 6.4 \\ 
162730-24415 & 16  27  30.2 & -24  41  49.3 &  0.22 &  2.69 & 40   &  0.61 & 0.32 & 20   & 0.38 & 4.8 & 6.0 \\ 
162733-24262 & 16  27  33.0 & -24  26  22.3 &  0.52 &  4.68 & 38   &  1.1  & 0.45 & 17   & 0.87 & 5.3 & 6.2 \\ 
162739-24424 & 16  27  38.8 & -24  42  37.2 &  0.22 &  1.64 & 31   &  0.37 & 0.31 & 19   & 0.26 & 4.9 & 6.1 \\ 
162740-24431 & 16  27  39.9 & -24  43  13.2 &  0.20 &  0.56 & 18   &  0.13 & 0.27 & 15   & 0.13 & 5.3 & 6.4 \\ 
162759-24334 & 16  27  58.5 & -24  33  43.0 &  0.36 &  2.78 & 35   &  0.63 & 0.46 & 14   & 0.69 & 5.3 & 6.1 \\ 
162821-24362 & 16  28  21.4 & -24  36  24.5 &  0.29 &  0.73 & 19   &  0.17 & 0.39 & 14   & 0.18 & 5.4 & 6.4 \\
163133-24032 & 16  31  32.5 & -24  03  15.9 &  0.23 &  1.63 & 31   &  0.23 & 0.35 & 17   & 0.19 & 5.1 & 6.2 \\ 
163136-24013 & 16  31  35.8 & -24  01  28.0 &  0.86 &  1.85 & 26   &  0.26 & 0.72 & 10   & 0.57 & 6.5 & 6.3 \\ 
163138-24495 & 16  31  38.2 & -24  49  47.7 &  0.39 &  4.89 & 42   &  1.1  & 0.38 & 24   & 0.56 & 4.9 & 6.1 \\ 
163139-24506 & 16  31  38.9 & -24  50  59.7 &  0.22 &  2.11 & 36   &  0.48 & 0.35 & 19   & 0.32 & 4.8 & 6.0 \\ 
163140-24485 & 16  31  40.5 & -24  48  53.8 &  0.21 &  1.03 & 25   &  0.24 & 0.29 & 17   & 0.19 & 5.1 & 6.2 \\ 
163141-24495 & 16  31  41.3 & -24  49  47.8 &  0.35 &  3.93 & 40   &  0.89 & 0.36 & 24   & 0.44 & 4.8 & 6.1 \\ 
163143-24003 & 16  31  43.3 & -24  00  28.2 &  0.24 &  0.94 & 23   &  0.13 & 0.33 & 15   & 0.13 & 5.3 & 6.4 \\ 
163154-24560 & 16  31  53.8 & -24  56  00.2 &  0.27 &  1.87 & 32   &  0.43 & 0.40 & 15   & 0.41 & 5.1 & 6.1 \\ 
163157-24572 & 16  31  57.1 & -24  57  18.4 &  0.30 &  3.43 & 43   &  0.78 & 0.46 & 14   & 0.87 & 5.2 & 6.0 \\ 
163201-24564 & 16  32  00.9 & -24  56  42.5 &  0.36 &  0.33 & 12   &  0.08 & 0.45 & 10   & 0.16 & 6.0 & 6.8 \\ 
163223-24284 & 16  32  22.9 & -24  28  37.1 & 14.57 & 33.27 & 56   &  7.6  & 0.94 & 24   & 3.8  & 6.0 & 6.2 \\ 
163229-24291 & 16  32  29.1 & -24  29  07.2 &  1.34 & 16.19 & 64   &  3.7  & 0.74 & 17   & 3.1  & 5.7 & 5.8 \\ 
163230-23556 & 16  32  29.7 & -23  55  55.6 &  0.26 &  1.46 & 27   &  0.20 & 0.33 & 17   & 0.17 & 5.2 & 6.4 \\ 
163247-23524 & 16  32  47.0 & -23  52  43.5 &  0.23 &  2.36 & 35   &  0.33 & 0.26 & 19   & 0.23 & 5.0 & 6.2 \\ 
163248-23524 & 16  32  48.0 & -23  52  40.5 &  0.22 &  0.61 & 19   &  0.08 & 0.30 & 14   & 0.09 & 5.5 & 6.5 \\ 
163248-23523 & 16  32  48.3 & -23  52  25.5 &  0.21 &  0.53 & 17   &  0.07 & 0.19 & 14   & 0.08 & 5.6 & 6.6 \\ 
163249-23521 & 16  32  49.3 & -23  52  13.5 &  0.22 &  0.99 & 24   &  0.14 & 0.28 & 16   & 0.13 & 5.3 & 6.4 \\ 
163356-24420 & 16  33  55.7 & -24  42  04.3 &  0.20 &  0.12 &  9   &  0.03 & 0.33 & 11   & 0.05 & 5.8 & 6.8 \\ 
163448-24381 & 16  34  48.5 & -24  38  05.8 &  0.24 &  2.91 & 42   &  0.66 & 0.38 & 19   & 0.47 & 4.8 & 6.0 
\enddata
\tablecomments{Units of right ascension are hours, minutes, and seconds, and units of declination are degrees, arcminutes, and arcseconds.}
\tablenotetext{a}{Name formed from J2000.0 positions.}
\tablenotetext{b}{Position of peak flux within clump (accurate to 3$''$; the pixel size in the SCUBA maps and the approximate pointing accuracy of the JCMT).}
\tablenotetext{c}{Peak flux, total flux, and outer radius derived from Clumfind \citep{williams94}. The peak and total fluxes have uncertainties of about 20\%, mostly due to uncertainties in the absolute flux calibration. The radius has not been deconvolved from the telescope beam.}
\tablenotetext{d}{Mass derived from the total flux assuming $T_d = 15$~K and 850~cm$^2$~g$^{-1}$, $d=125$~pc.}
\tablenotetext{e}{Concentration, temperature, mass, central number density, and external pressure derived from Bonnor-Ebert modeling (see text).}
\end{deluxetable}

\clearpage
\pagestyle{plaintop}

\end{document}